\documentclass[preprint,3p,times]{elsarticle}

\usepackage{amssymb}
\usepackage{amsmath}
\usepackage{subcaption}

\usepackage{lineno}
\usepackage{booktabs} 
\usepackage{tabularx}
\usepackage{multirow}
\usepackage{makecell}

\journal{Nuclear Instruments and Methods in Physics Research Section A: Accelerators, Spectrometers, Detectors and Associated Equipment}

\begin{document}
\begin{frontmatter}

\title{A muon scattering tomography system based on high spatial resolution scintillating detector}

\author[ustc]{Zheng Liang} 
\author[ustc]{Zebo Tang\corref{cor1}} 
\author[imp]{Xin Li\corref{cor1}} 
\cortext[cor1]{Corresponding authors: Zebo Tang (zbtang@ustc.edu.cn), Xin Li (lixin@impcas.ac.cn).}
\author[ustc]{Jiacheng He}
\author[dsel]{Kun Jiang} 
\author[ustc]{Cheng Li} 
\author[ustc]{Baiyu Liu}
\author[imp]{Ye Tian} 
\author[ustc]{Yonggang Wang} 
\author[ustc]{Zeyu Wang} 
\author[ustc]{Yishuang Zhang}

\affiliation[ustc]{organization={Department of Modern Physics, University of Science and Technology of China},  
            city={Hefei},
            postcode={230026},
            state={Anhui},
            country={China}}
\affiliation[imp]{organization={Institute of Modern Physics, Chinese Academy of Sciences},
            city={Lanzhou},
            postcode={730000},
            state={Gansu},
            country={China}}
\affiliation[dsel]{organization={Institute of Deep Space Sciences, Deep Space Exploration Laboratory},
            city={Hefei},
            postcode={230026},
            country={China}}

\begin{abstract}

Cosmic ray muon scattering tomography is an imaging technique that utilizes muon scattering in matter to inspect high-Z materials non-destructively, without requiring an artificial radiation source. This method is particularly suitable for scanning large and dense volumes such as cargo containers, industrial furnaces, and nuclear storage casks, where conventional radiographic methods are ineffective. It therefore holds significant potential for critical applications in border security, industrial monitoring, and long-term nuclear safeguards.
In this study, we developed a high-precision plastic‑scintillator‑based position‑sensitive detector that achieves a spatial resolution of 1~mm, corresponding to 0.09 times the strip pitch.
A fully functional, full‑scale imaging system was then constructed using four layers of such XY position‑sensitive detectors, each with an effective area of 53~cm~$\times$~53~cm.
The imaging capability of the system was experimentally validated, yielding a signal‑to‑noise ratio of 7.1 in the reconstructed image of a 2~$\times$~2~$\times$~2~cm$^3$ tungsten block, demonstrating excellent imaging quality.
This paper details the following key contributions: the Geant4‑simulated design and optimization of the imaging system, the fabrication, assembly, and testing of the detectors, and an evaluation of the imaging performance of the completed system.

\end{abstract}


\end{frontmatter}

\section{Introduction}

Cosmic ray muons, serving as natural and highly penetrating probes, enable two principal techniques for nondestructive imaging: muon scattering tomography and muon transmission radiography.
Muon scattering tomography (MST), first proposed in 2003 \cite{borozdinRadiographicImagingCosmicray2003a}, reconstructs the multiple Coulomb scattering angles of muons, a process that confers high sensitivity to atomic number (Z) \cite{SCHULTZ2004687,4271541}.
This sensitivity makes the technique particularly promising for detecting high‑Z materials, driving its development for several critical applications \cite{bonomiApplicationsCosmicrayMuons2020}. 
Notable examples include customs and border security, where dedicated portals scan cargo for concealed heavy metals or nuclear materials \cite{antonuccioMuonPortalProject2017b}; industrial monitoring, such as identifying shielded radioactive sources in scrap metal \cite{clarksonCharacterisingEncapsulatedNuclear2015} or mapping the internal burden of operating blast furnaces \cite{Cohu_2023}; and nuclear safeguards, for verifying the contents of spent nuclear fuel casks \cite{poulsonCosmicRayMuon2017}. 
In contrast, muon transmission radiography images density structures by measuring the attenuation of muon flux.
This method has been successfully applied to large scale geological and archaeological targets, such as volcanoes \cite{D'Errico_2020,VARGA2020162236} and ancient structures \cite{morishimaDiscoveryBigVoid2017b}.

Although they have different principles and applications, both muographic techniques face a shared fundamental challenge: building a robust, large-area tracker with high spatial resolution. 
The efficiency and quality of the final image directly depend on the performance of these tracking detectors.
In the field of particle detection and imaging, the selection of detector technology often involves a trade-off among performance, cost and deployability. 

Among the available options, plastic scintillator detectors read out by silicon photomultipliers (SiPMs) offer a compelling solution for practical field deployment, primarily due to their exceptional operational stability and environmental robustness. 
Unlike gaseous detectors such as drift chambers \cite{MORRIS15102008,morrisObtainingMaterialIdentification2012}, Micromegas \cite{9658560}, or resistive plate chambers (RPCs) \cite{coxDetectorRequirementsCosmic2008a, panExperimentalValidationMaterial2019}, they do not require gas circulation systems or high-voltage infrastructure.
However, conventional plastic scintillator detectors face two principal challenges when scaling up, a relatively high cost for large-area deployment and a spatial resolution that is lower than that of gaseous detectors.
These inherent challenges motivate the central objective of this work: to develop a large-area plastic scintillator detector that simultaneously achieves reduced cost and improved spatial resolution, thereby enhancing its competitiveness for widespread applications in both muon scattering tomography and transmission radiography.

In this study, we have designed and fabricated a modular, cost‑effective, position‑sensitive plastic scintillator detector with high spatial resolution, specifically for cosmic‑ray muon scattering imaging. 
The rest of this paper is structured as follows. 
Section~2 presents the Geant4‑based simulation and optimization of the overall system and detector design.
Section~3 details the assembly and testing of the detector super‑layers, along with the associated electronics.
Section~4 describes the integration of the full MST system and evaluates its imaging performance.
Finally, Section~5 provides a conclusion and outlook on future work.

\section{System design}

\subsection{System layout}
\begin{figure}
    \centering
    \includegraphics[width=0.8\linewidth]{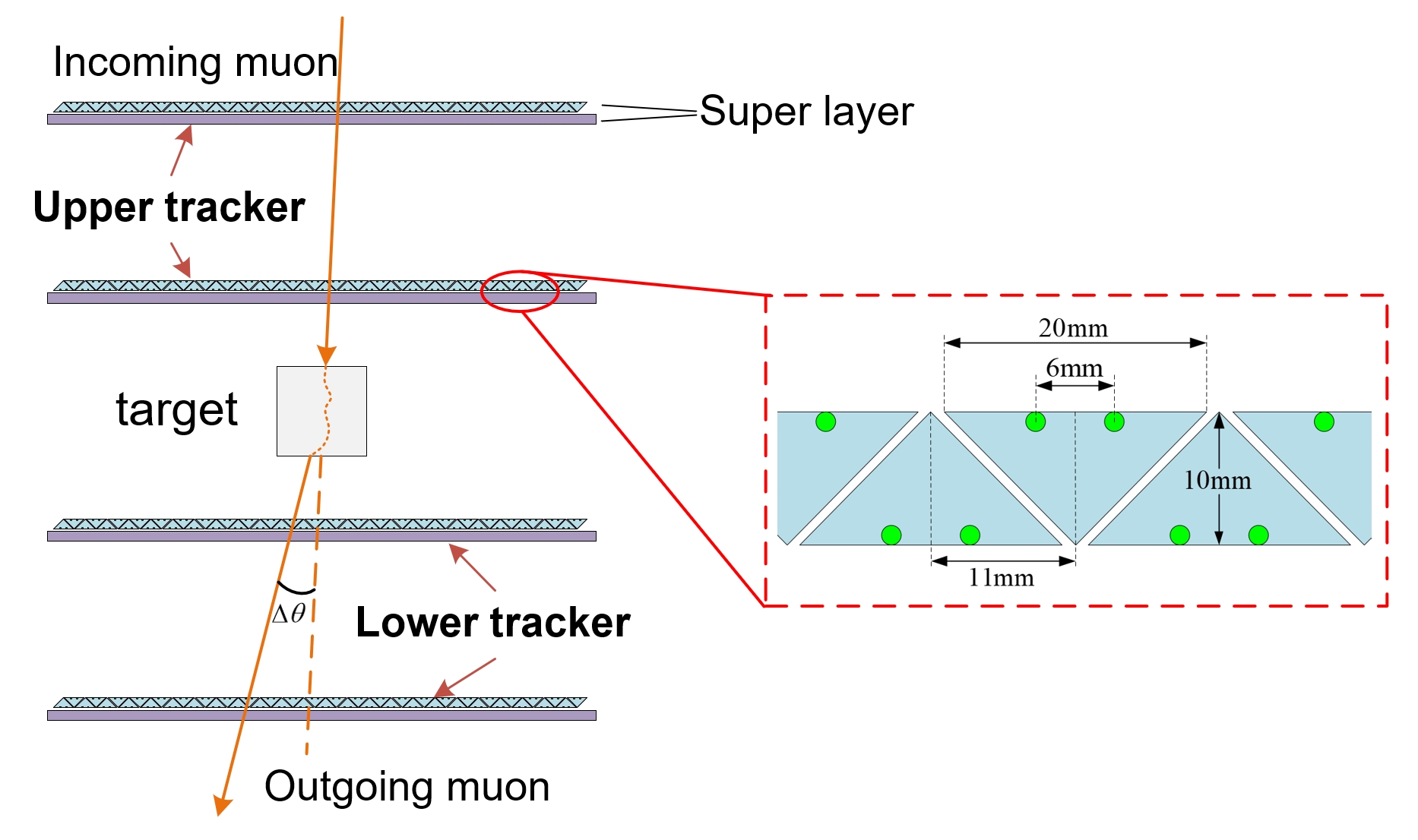}
    \caption{Structural diagram of the MST system. (Left) Four Super Layers, formed by orthogonally stacking two detection planes, provides the incoming and outgoing muon tracks. (Right) Internal grooving of the scintillator bar with embedded wavelength shifting fiber.}
    \label{fig:System-Overview}
\end{figure}

A typical muon scattering imaging system comprises at least four layers of two‑dimensional position‑sensitive detectors, arranged with two layers above and two below the inspected object.
As illustrated in the left panel of figure~\ref{fig:System-Overview}, our MST system follows this configuration. 
Both the upper tracker (recording incident muon trajectories) and the lower tracker (recording outgoing trajectories) consist of two “Super Layers,” each of which provides two‑dimensional position information. 
By measuring the scattering angles of muons that penetrate the target, the scattering density distribution within the inspection volume can be reconstructed \cite{4271541}.

A super layer consists of two orthogonally oriented one‑dimensional detection planes, each providing a coordinate measurement. 
Each plane is formed by an array of parallel scintillator bars; the fired bars in an event determine the hit position. 
Every scintillator bar is wrapped in reflective material and contains an internal groove that houses a Wavelength‑Shifting (WLS) fiber.
When a cosmic‑ray muon traverses the scintillator, it deposits energy and generates fluorescent photons. 
These photons are absorbed by the WLS fiber and re‑emitted.
A fraction of the re‑emitted light is trapped by total internal reflection and guided to the fiber end, where it is converted into an electrical signal by a SiPM.

The cross‑sectional geometry and key dimensions of the scintillator bar used in this work are presented in the right panel of figure~\ref{fig:System-Overview}.
Earlier work showed that scintillator bars with a triangular cross-section provide improved spatial resolution while maintaining the pitch \cite{Liang_2020}, we have therefore used this geometry in our current detector to obtain higher resolution.

\subsection{Simulation}
While it is widely acknowledged that enhanced detector resolution improves MST image quality, the exact quantitative relationship is not yet fully established.
To quantify this relationship, we conducted a Geant4 simulation to determine the required detector spatial resolution for the imaging system.

\begin{figure}
    \centering
    \begin{subfigure}{0.55\textwidth}
        \includegraphics[width=\textwidth]{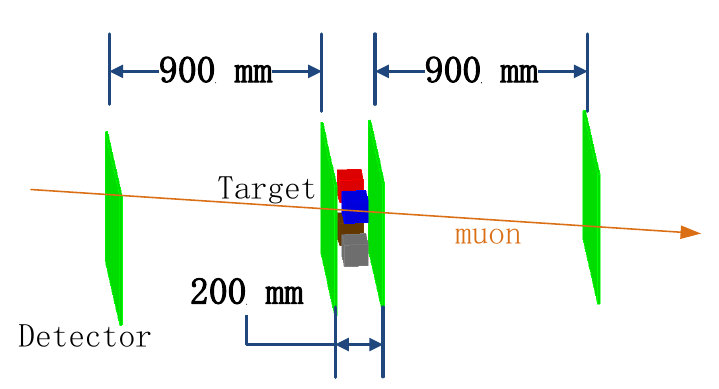}
        \caption{}
    \end{subfigure}
    \begin{subfigure}{0.35\textwidth}
        \includegraphics[width=\textwidth]{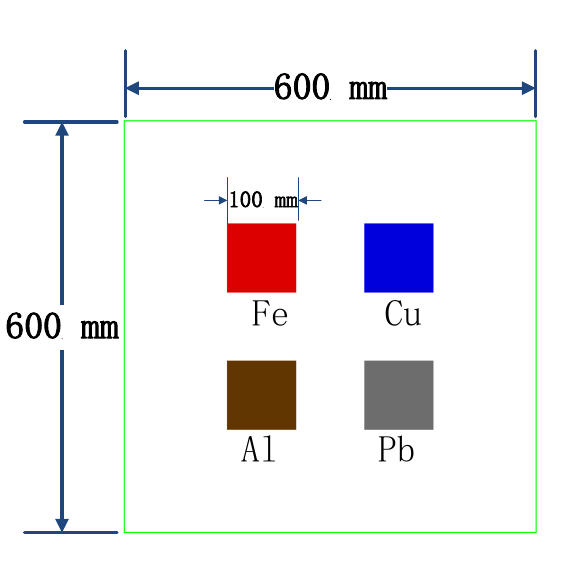}
        \caption{}
    \end{subfigure}
    \caption{Geant4 simulation geometry. (a) Layout of detectors and target. (b) Placement of four cubic targets.
    }
    \label{fig:Geant4-ImageSimu}
\end{figure}

\paragraph{Simulation configuration}
The Geant4 simulation geometry is illustrated in figure~\ref{fig:Geant4-ImageSimu}. 
Due to the limited height of the imaging system, the inter-layer spacing within both the upper and lower trackers confined to 90~cm.
A 20~cm gap was reserved between the upper and lower tracker modules to accommodate the imaging target, which is composed of 10-cm cubes made of iron (Fe), aluminum (Al), copper (Cu) and lead (Pb).
The four thin green planes in the diagram represent the detector planes.
In this model, the detectors are simplified as homogeneous plastic slabs, omitting internal structure details and scintillation photon production.
This approach was adopted to speedup the simulation without compromising the accuracy in modeling muon scattering through key components, such as plastic scintillator and imaging targets.

\begin{figure}
    \centering
    \begin{subfigure}{0.49\textwidth}
        \includegraphics[width=\textwidth]{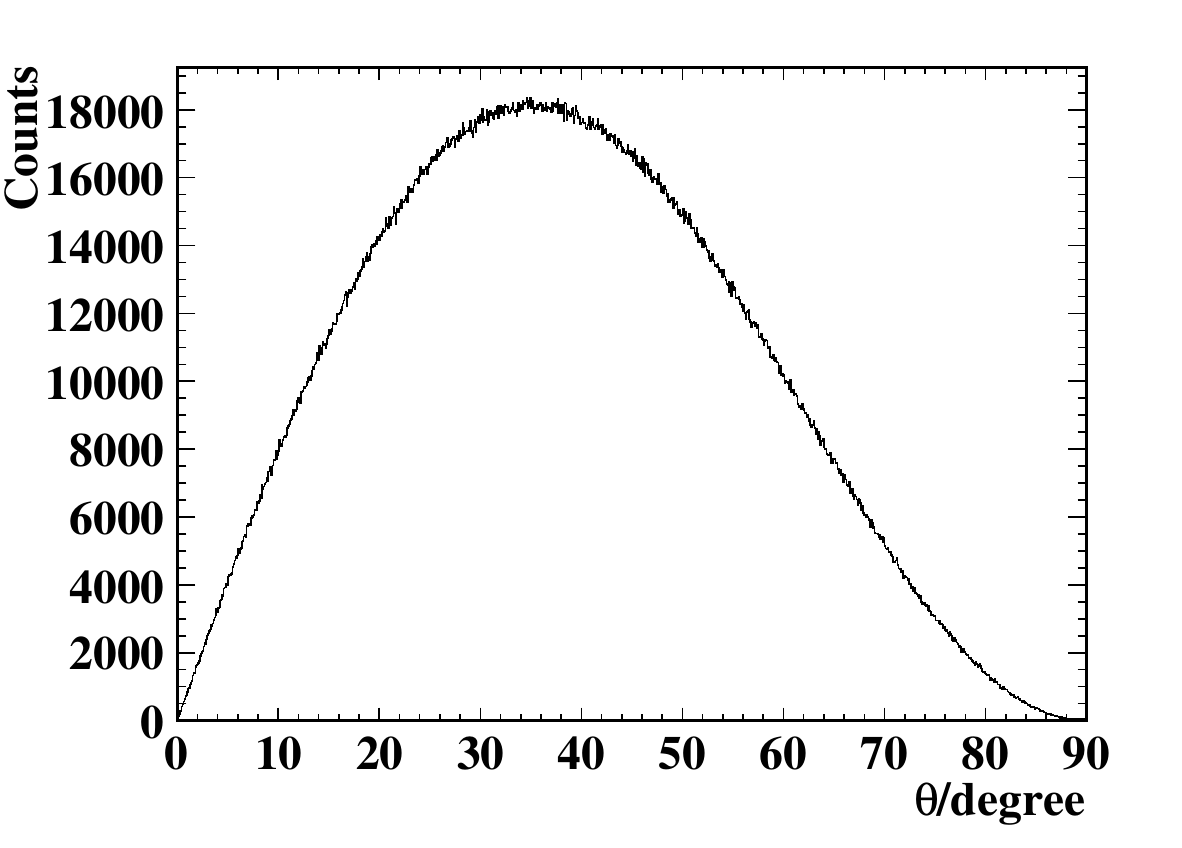}
        \caption{}
    \end{subfigure}
    \begin{subfigure}{0.49\textwidth}
        \includegraphics[width=\textwidth]{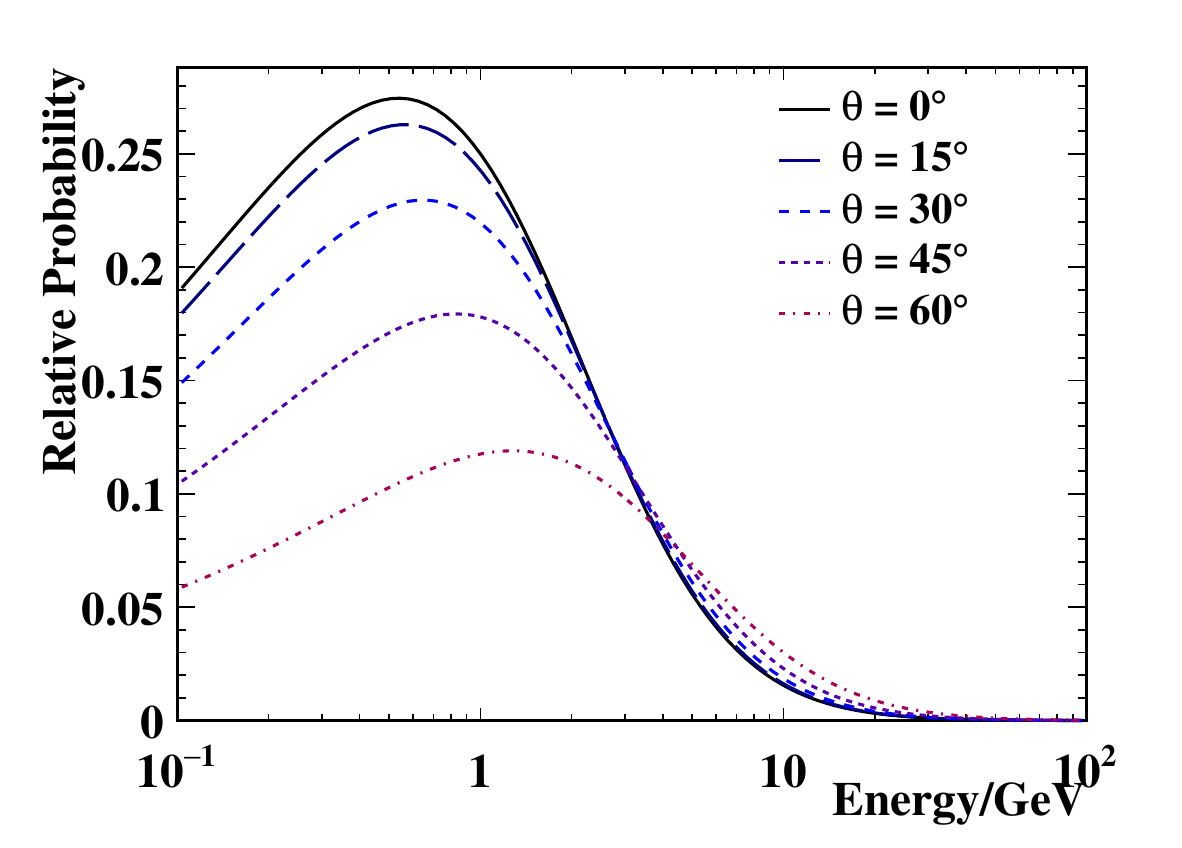}
        \caption{}
    \end{subfigure}
    \caption{Sampling of muon incident angle and energy in Geant4 simulations. (a) Distribution of zenith angle $\theta$. (b) Energy distributions for different zenith angles.}
    \label{fig:Geant4-muon-sampling}
\end{figure}

Muons were generated with initial position, incident angle and kinetic energy sampled according to the Chatzidakis model \cite{CHATZIDAKIS201533}, and propagated from left to right in the simulation geometry.
The initial positions are uniformly distributed across the incident surface, while the zenith angle probability density is given by $f(\theta) \propto \sin\theta \cos^2\theta$ and the azimuth angle $\phi$ is uniformly distributed between $0$ and $2\pi$.
After determining the $\theta$ and $\phi$, the muon energy is sampled according to the muon energy spectrum. 
The results of the zenith angle and energy sampling are shown in figure~\ref{fig:Geant4-muon-sampling}.
The same particle generator is employed both for the MST system simulation described herein and for the detector simulations that follow.

In order to model the finite spatial resolution of the detectors, the exact muon hit positions recorded by Geant4 were smeared by adding a random number drawn from a Gaussian distribution. 
The standard deviation $\sigma$ of this distribution is defined as the spatial resolution of the detector.
To systematically quantify the effect of detector spatial resolution on the reconstructed image, we performed image reconstruction across a range of $\sigma$ values from 0~mm to 5~mm.
Only muons that traverse the four detection planes are used for imaging analysis.

\paragraph{Imaging analysis}

The most common and simplest algorithm for muon tomography is the Point of Closest Approach (PoCA) algorithm \cite{SCHULTZ2004687}.
The PoCA algorithm identifies the common perpendicular between the incident and outgoing trajectories and takes its midpoint as the point of closest approach.
The scattering signal for each muon is calculated as 
\begin{equation*}
    S_i = \frac{\Delta\theta_{xi}^2 + \Delta\theta_{yi}^2}{2L_i}
\end{equation*}
where $i$ denotes the event index, $\Delta\theta_x$ and $\Delta\theta_y$ are the scattering angles in the two orthogonal planes, and $L_i$ is the total path length of the muon through the object volume (i.e., the distance the muon travels within the material).
This signal is then assigned exclusively to the voxel containing the PoCA point, while no signal is assigned to the voxels along either the incident or outgoing trajectories, or any other voxels along the common perpendicular path.
By averaging the contributions from many muons, a three-dimensional map of the scattering strength is reconstructed.

In our reconstruction, however, we've made a few modifications.
Due to the poor resolution along the Z-axis, we restrict the image reconstruction to the X-Y plane. 
The imaging signal is defined as:
\begin{equation*}
    S'_i = \sqrt{(\Delta\theta_{xi}^2 + \Delta\theta_{yi}^2) \cdot  \frac{L_0}{L_i}}=\sqrt{(\Delta\theta_{xi}^2 + \Delta\theta_{yi}^2) \cdot \cos\theta_i }
\end{equation*}
where  $L_0 = 10$~cm and $L_i = 10~\text{cm} / \cos\theta_i$, where $\theta_i$ is the incident zenith angle. 
Therefore, the reconstructed images effectively represent the mean scattering angle of muons for a material thickness of $L_0$, rather than the mean square scattering angle. 

\begin{figure}
    \centering
    \begin{subfigure}{0.4\textwidth}
        \centering
        \includegraphics[width=\textwidth]{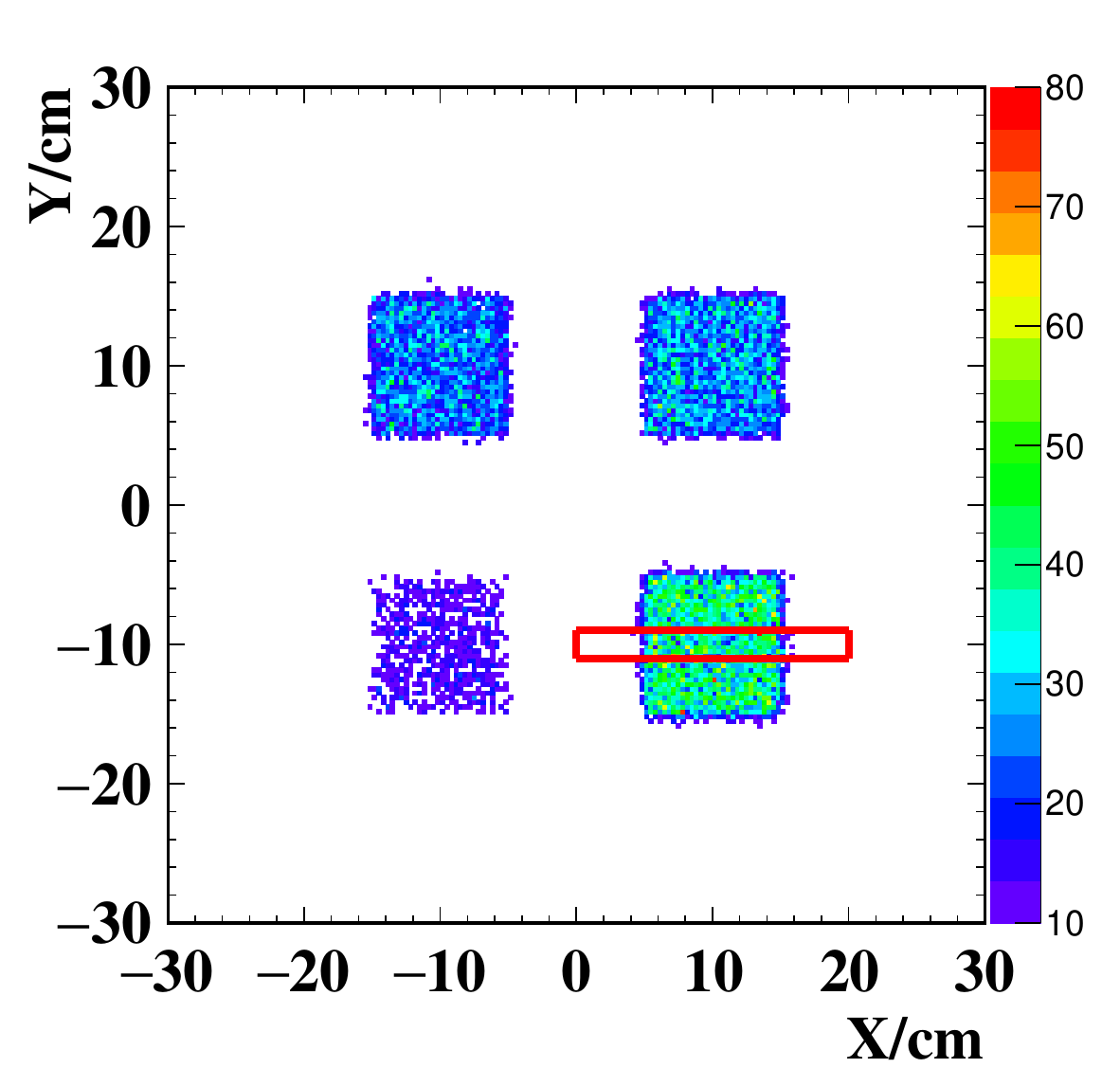}
        \caption{}
        \label{fig:poca-simu-result-lu}
    \end{subfigure}
    \begin{subfigure}{0.4\textwidth}
        \centering
        \includegraphics[width=\textwidth]{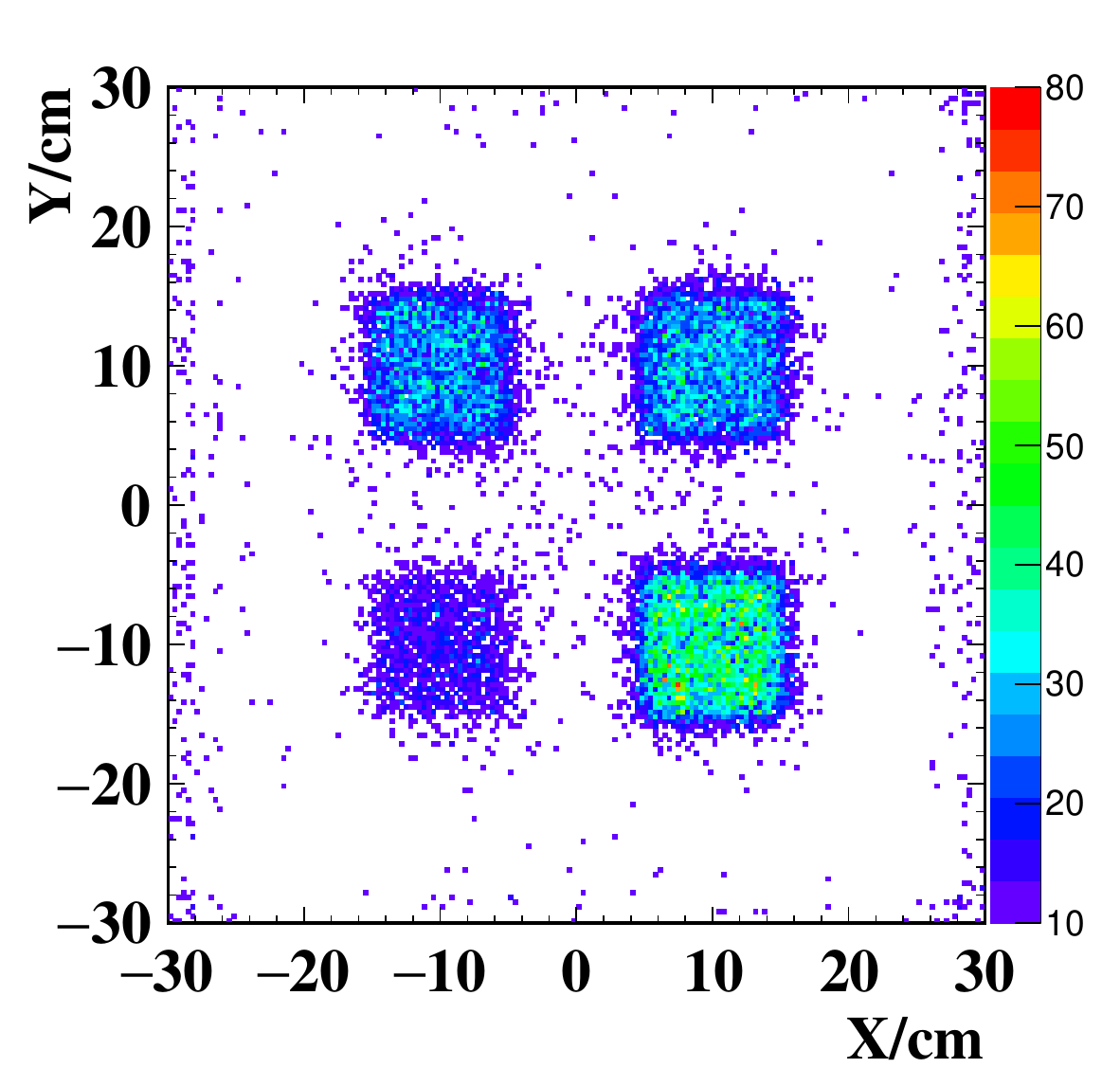}
        \caption{}
        \label{fig:poca-simu-result-ru}
    \end{subfigure}\\
    \begin{subfigure}{0.4\textwidth}
        \centering
        \includegraphics[width=\textwidth]{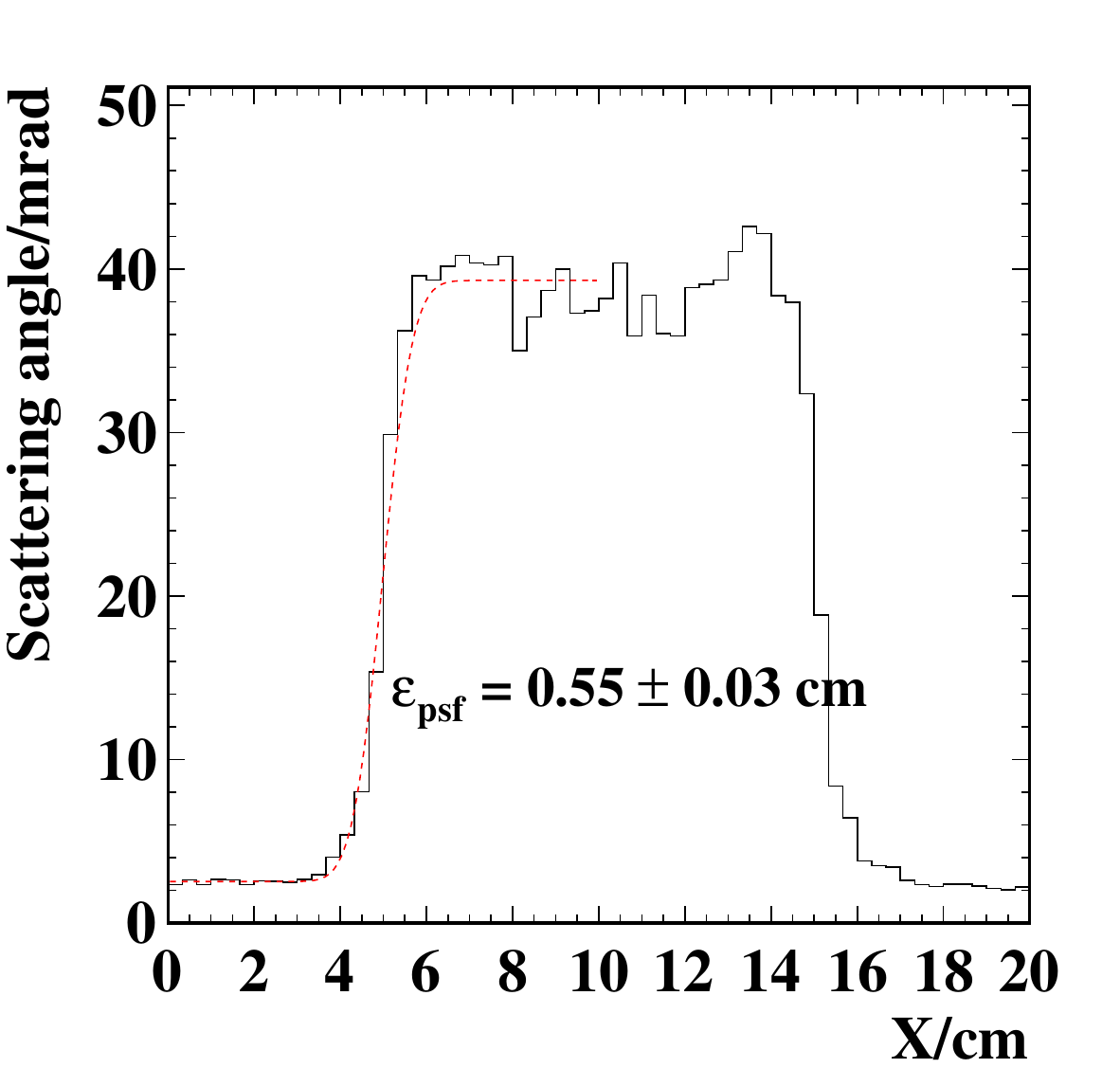}
        \caption{}
        \label{fig:poca-simu-result-ld}
    \end{subfigure}
    \begin{subfigure}{0.4\textwidth}
        \centering
        \includegraphics[width=\textwidth]{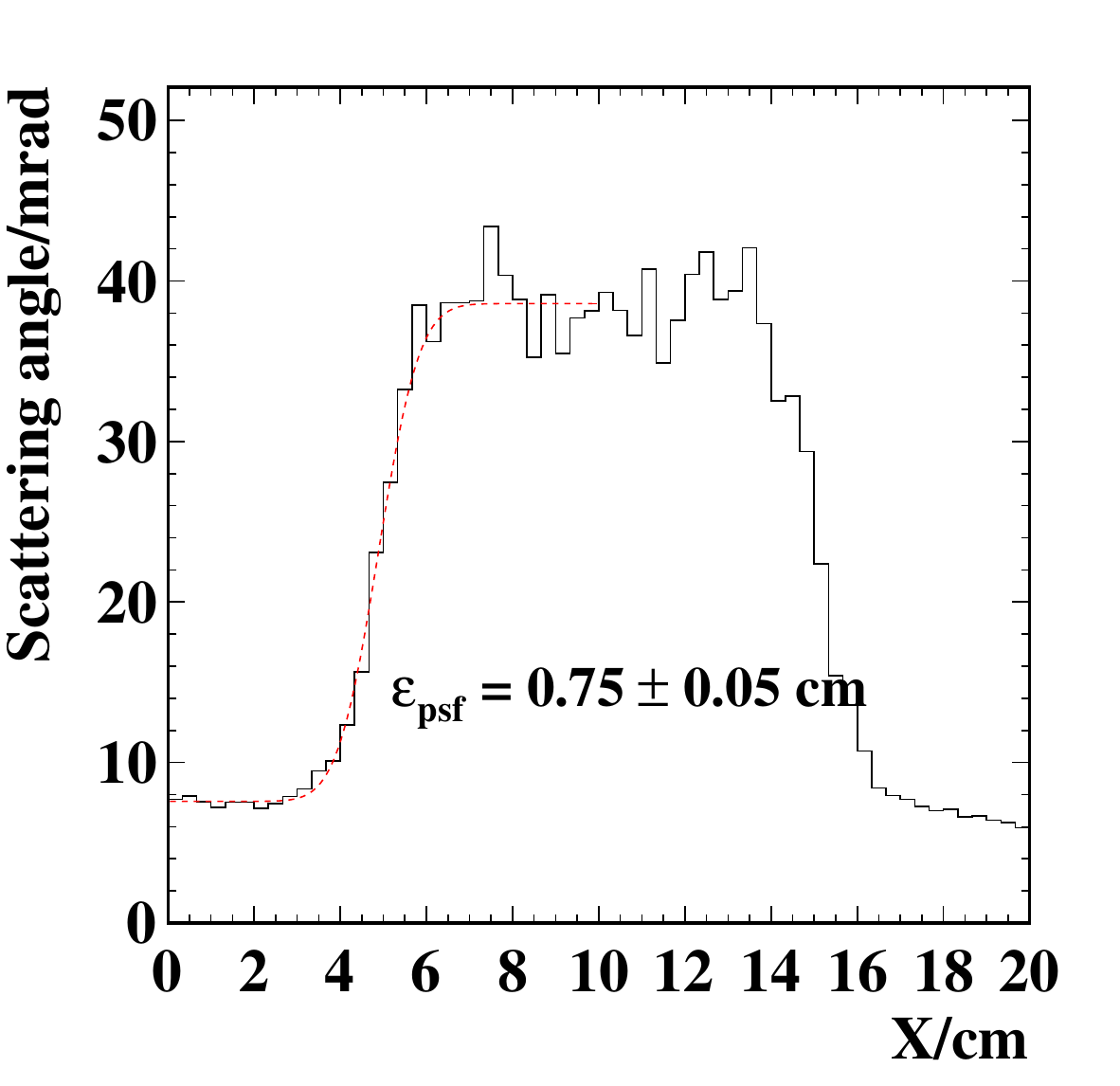}
        \caption{}
        \label{fig:poca-simu-result-rd}
    \end{subfigure}
    \caption{
    PoCA‑based reconstruction from Geant4 simulations. (a, b) Images obtained with detector spatial resolutions $\sigma_x$ of 1~mm and 3~mm, respectively. The color axis represents the mean scattering angle (mrad), normalized to a penetration length of 10~cm.  (c, d) Corresponding X‑axis intensity profiles across the lead‑brick region (5~mm < $x$ < 15~mm); the profiles are fitted with an error function to characterize the edge sharpness.
    }
    \label{fig:poca-simu-result}
\end{figure}

Figure~\ref{fig:poca-simu-result} presents the imgaing results using the PoCA algorithm under two different detector spatial resolutions ($\sigma_x$ = 1~mm and 3~mm). 
It can be observed that the impact of detector resolution on the imaging performance is mainly manifested in two aspects: the contrast between the signal and the background, and the sharpness of the object boundaries.

In optical systems, the Point Spread Function (PSF) is defined as the response of the system to an ideal point target \cite{CORLE19961}.
This concept can be extended to muon scattering tomography to quantify the blurring introduced by the finite detector resolution and the reconstruction algorithm itself.
The PSF of muon imaging system can be modeled as a Gaussian function $G(x,\varepsilon_\text{psf})$ with a resolution parameter $\varepsilon_\text{psf}$.
The input to the imaging system, which is the spatial density distribution of the object under inspection, can be approximated as a step function, $H(x)$.
Consequently, the resulting output image is mathematically described as the convolution of this step function with the Gaussian PSF, which is a error function:
\begin{equation}
    f(x)=(H\ast G)=\frac{1}{2}\left[1+\text{erf}\left(\frac{x}{\sqrt{2} \varepsilon_\text{psf} }\right)\right] \text{.}
    \label{equ:error-function}
\end{equation}
Therefore, in figures~\ref{fig:poca-simu-result-ld}~and~\ref{fig:poca-simu-result-rd}, the image edges are fitted with an error function to quantitatively analyze the contrast and blurring.
The image contrast is characterized by the signal-to-background ratio (S/B), defined as the ratio of the amplitude in the region-of-interest to that in the background, which is $A_\text{sig}/A_\text{bkg}$ \cite{CT_contrast}.
The sharpness of the boundaries in the reconstructed image is characterized by the parameter $\varepsilon_\text{psf}$ in equation~\ref{equ:error-function}.

\begin{figure}
    \centering
    \begin{subfigure}{0.49\textwidth}
        \centering
        \includegraphics[width=\textwidth]{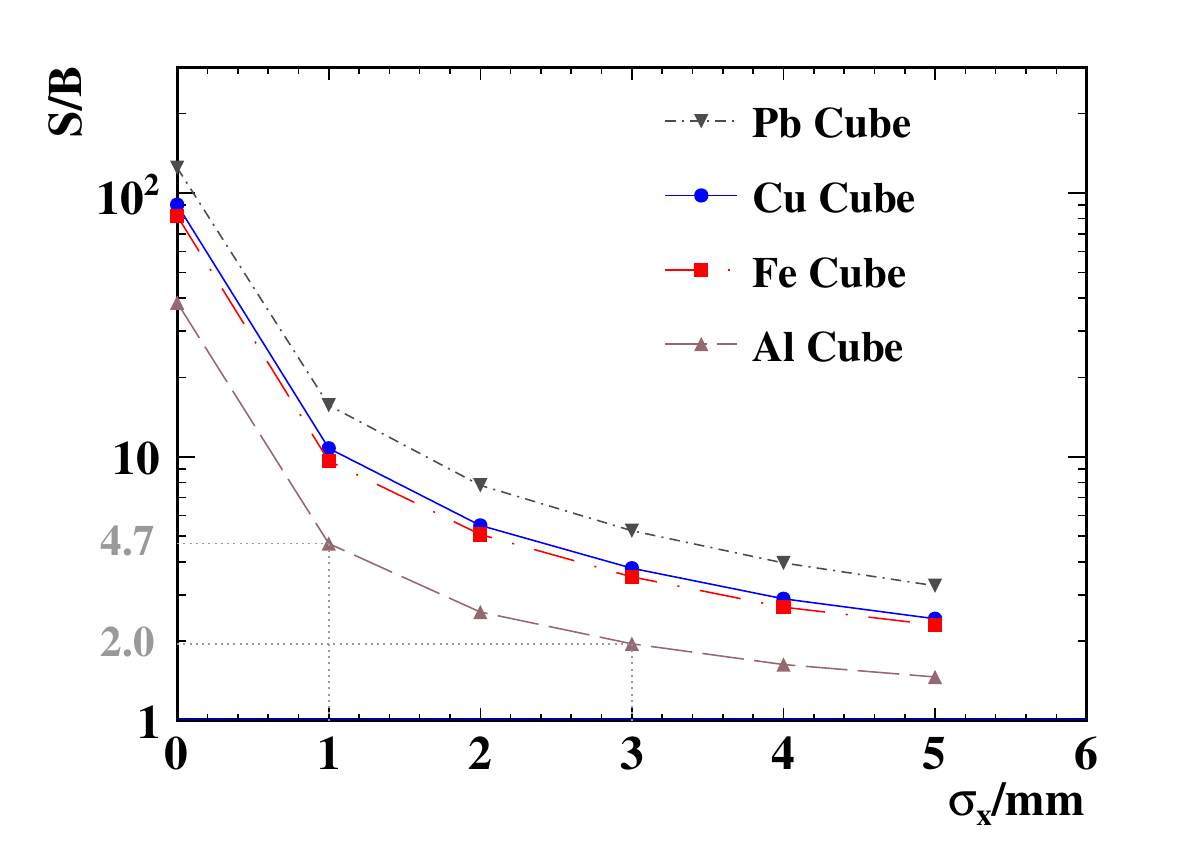}
        \caption{}
    \end{subfigure}
    \begin{subfigure}{0.49\textwidth}
        \centering
        \includegraphics[width=\textwidth]{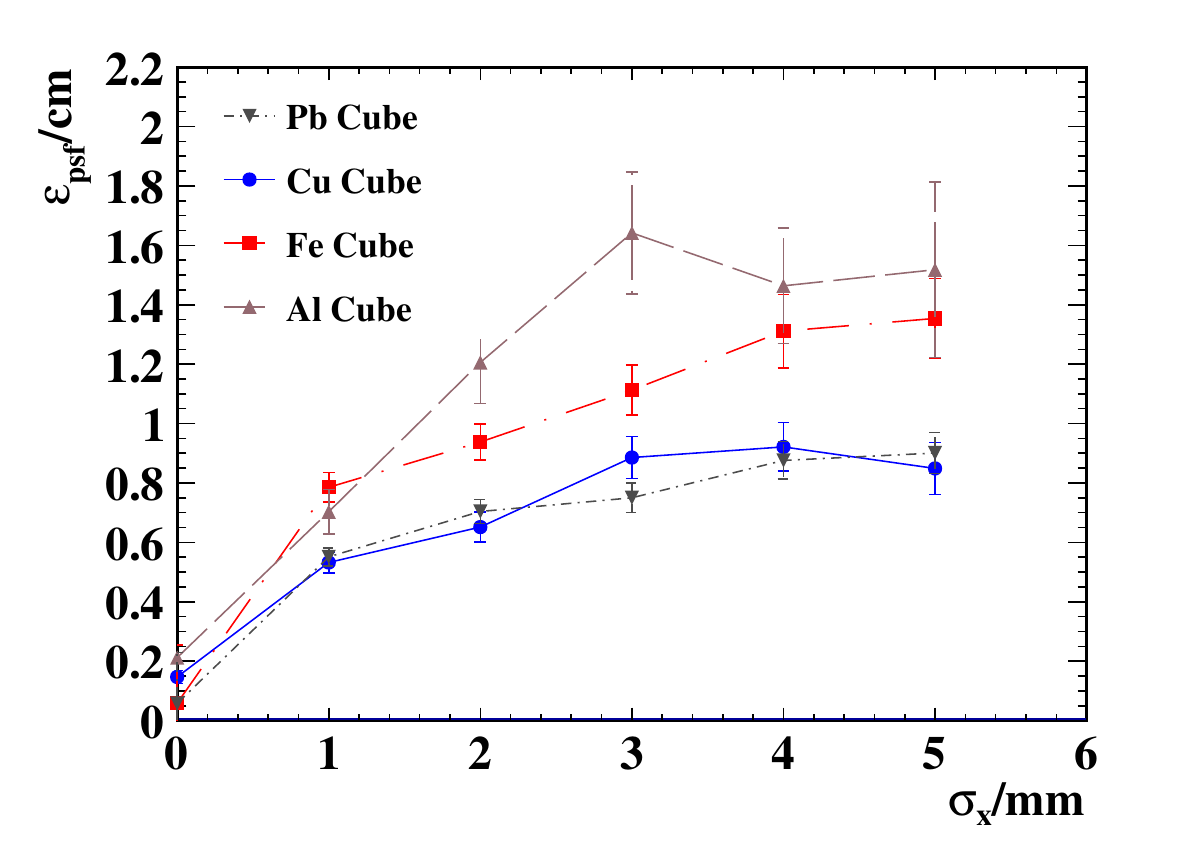}
        \caption{}
    \end{subfigure}
    \caption{
    Impact of detector spatial resolution $\sigma_x$ on image‑quality metrics. (a) Signal‑to‑background ratio (S/B). (b) Sharpness parameter $\varepsilon_\text{psf}$ derived from the system point spread function.
    }
    \label{fig:poca-SNR-sigma}
\end{figure}

Figure~\ref{fig:poca-SNR-sigma} shows the performance of imaging systems employing detectors with varying spatial resolutions.
As shown, improved spatial resolution significantly enhances both the image S/B and the boundary sharpness. 
This is demonstrated by the results for a resolution ($\sigma_x$) improvement from 3~mm to 1~mm: the S/B increased from 2.0 to 4.7 for Al and from 5.2 to 16 for Pb, while $\varepsilon_\text{psf}$ decreased from 1.6~cm to 0.70~cm for Al and from 0.75~cm to 0.55~cm for Pb, indicating sharper boundaries.

MST systems based on scintillator detectors typically achieve a spatial resolution around 3~mm \cite{ANTONUCCIO2017322}.
The Geant4 simulation results demonstrate that while a 3 mm spatial resolution can successfully image low-Z elements like aluminum, the resulting image quality is substantially limited.
To achieve high-quality reconstructions across a broader range of materials, a superior resolution is essential. 
This provides a critical reference for our detector design specifications: the spatial resolution should be aimed for as close to 1~mm as possible.

\subsection{Scintillating detector}
\subsubsection{Design and optimization}
\label{sec:detector-design}
A conventional position-sensitive scintillator detector is typically fabricated by assembling long scintillator bars into a detection plane.
The cross-sectional geometry of the scintillator bars strongly affects the detector's spatial resolution.
In our prior research, we developed and tested functional prototypes for different scintillator detector structures, each with an active area of $15\times15\text{~cm}^2$ \cite{Liang_2020}.
Results show that the detector with triangular cross-section bars achieved a spatial resolution of 1.8~mm with a pitch of 1~cm, a significant improvement from the approximately 3~mm resolution of square cross-section bars.

A practical MST system requires large-area detectors exceeding $50\times50\text{~cm}^2$.
Extending the scintillator length introduces several engineering challenges. 
These include signal readout over large areas, a modular architecture for scalability, cost-effective fabrication, and straightforward installation.
In contrast, the prototype detector's design does not satisfy these requirements.
First, scintillator bars are read out by SiPMs at both ends, and the coincidence logic between the two SiPMs is required. 
Consequently, signals from both SiPMs must be routed to the same readout module, forcing the long-distance transmission of unamplified analog signals. 
This not only complicates installation but also introduces significant noise, which in turn degrades the spatial resolution.
Another critical flaw in the prototype detector is the lack of an encoded signal readout scheme. 
This leads to a high channel count, making the readout electronics a major cost driver.

\begin{figure}
    \centering
    \begin{subfigure}{0.5\textwidth}
        \centering
        \includegraphics[width=\textwidth]{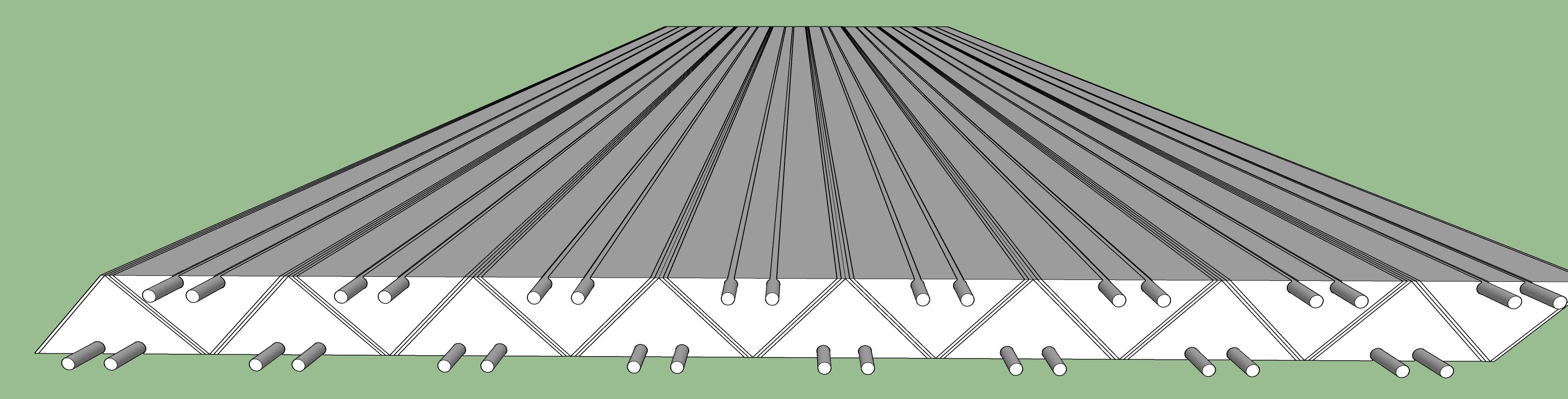}
    \end{subfigure}\\
    \begin{subfigure}{0.5\textwidth}
        \centering
        \includegraphics[width=\textwidth]{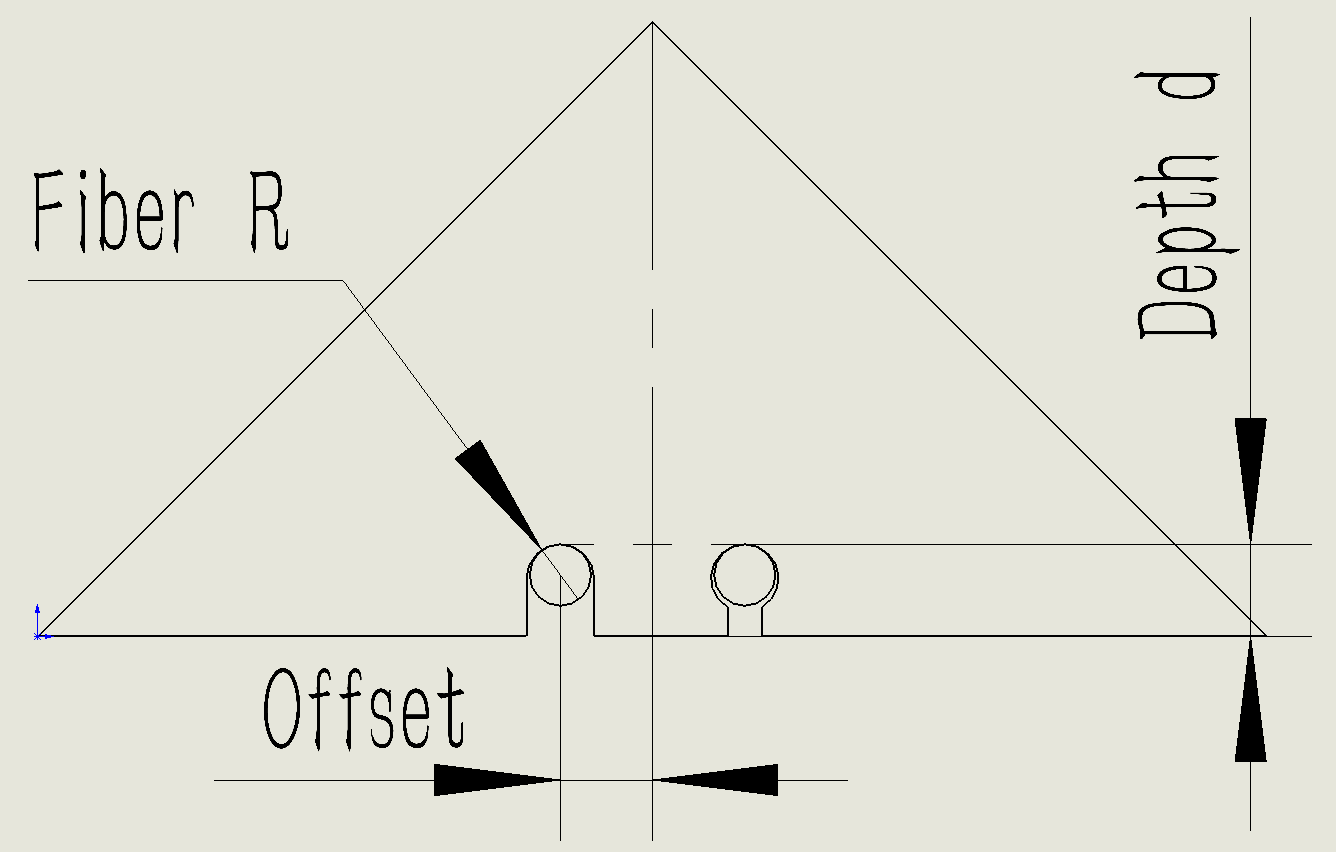}
    \end{subfigure}

    \caption{
    Illustration of scintillating detector structure.  
    Scintillator strips with a base length of 20~mm and a height of 10~mm are arranged sequentially with a pitch of 11~mm. 
    (Top) Arrangement of scintillator bar array. 
    (Bottom) Cross-section of a single bar, showing key geometric parameters for optimization (fiber radius ($R$), groove depth ($d$), and groove offset) with depictions of the two groove profiles: semicircular (left) and $\Omega$-shaped (right).
    }
    \label{fig:Detector-structure}
\end{figure}

Accordingly, the detection plane structure depicted in the figure~\ref{fig:Detector-structure} was implemented in our current design.
Scintillator strips with a base length of 20~mm and a height of 10~mm are arranged sequentially. 
Due to the reflective coating wrapped around each scintillator and the requirements of detector assembly, the center-to-center pitch between adjacent strips is set to 11~mm, resulting in a gap of 1~mm. 
Two WLS fibers are embedded within a single scintillator bar. 
Each fiber is read out by a SiPM on only one end, with the opposite end mirrored to enhance total photon collection efficiency. 
All SiPMs are positioned on the same side of the scintillator array. This optimized structure simultaneously satisfies the requirement for coincidence measurements and resolves the design flaws present in the detector prototype.

Although the use of WLS fibers raises manufacturing costs moderately, this is compensated for by their higher photon transmission efficiency, which enhances performance uniformity along the bar and allows for a larger detection area.
Moreover, the fiber-based design permits the active area to be extended cost-effectively by splicing together multiple shorter scintillator bars, adding virtually no extra cost.

The Geant4 simulation framework allows for a systematic study of how detector performance is influenced by key scintillator design parameters, ranging from the bar length and geometric features --- such as fiber radius $R$, groove depth $d$, groove offset, and groove shape (semicircular and $\Omega$-shaped) ---  to the physical coupling method at the scintillator-fiber interface (e.g., air coupling or optical grease coupling).

The optical properties of the detector materials were modeled according to the specifications of the selected components. 
For this purpose, the EJ-200 scintillator  \cite{eljen_ej200}, Kuraray Y11 WLS fibers \cite{kuraray_y11_fiber}, and Hamamatsu S13360-3075 SiPMs \cite{hamamatsu_s13360_3075pe} were implemented in the simulation.
The EJ-200 scintillator exhibits a light yield of approximately 10,000 photons per MeV, with a rise time of 0.9~ns and a decay time of 2.1~ns and its emission spectrum peaks at a wavelength of 425~nm.
The Kuraray Y11 fiber is a double-cladding fiber with an absorption band that matches the EJ-200 emission spectrum, and its emission band is well matched to the detection spectrum of the SiPM.
The Hamamatsu S13360-3075PE SiPM features a pixel pitch of 75~\textmu m and an active area of 3~mm $\times$ 3~mm.
Its peak sensitivity wavelength is 450 nm, with a corresponding photon detection efficiency (PDE) of 50\%. 
The dark count rate is approximately 500~kcps.

\begin{figure}
    \centering
    \begin{subfigure}{0.49\textwidth}
        \centering
        \includegraphics[width=\textwidth]{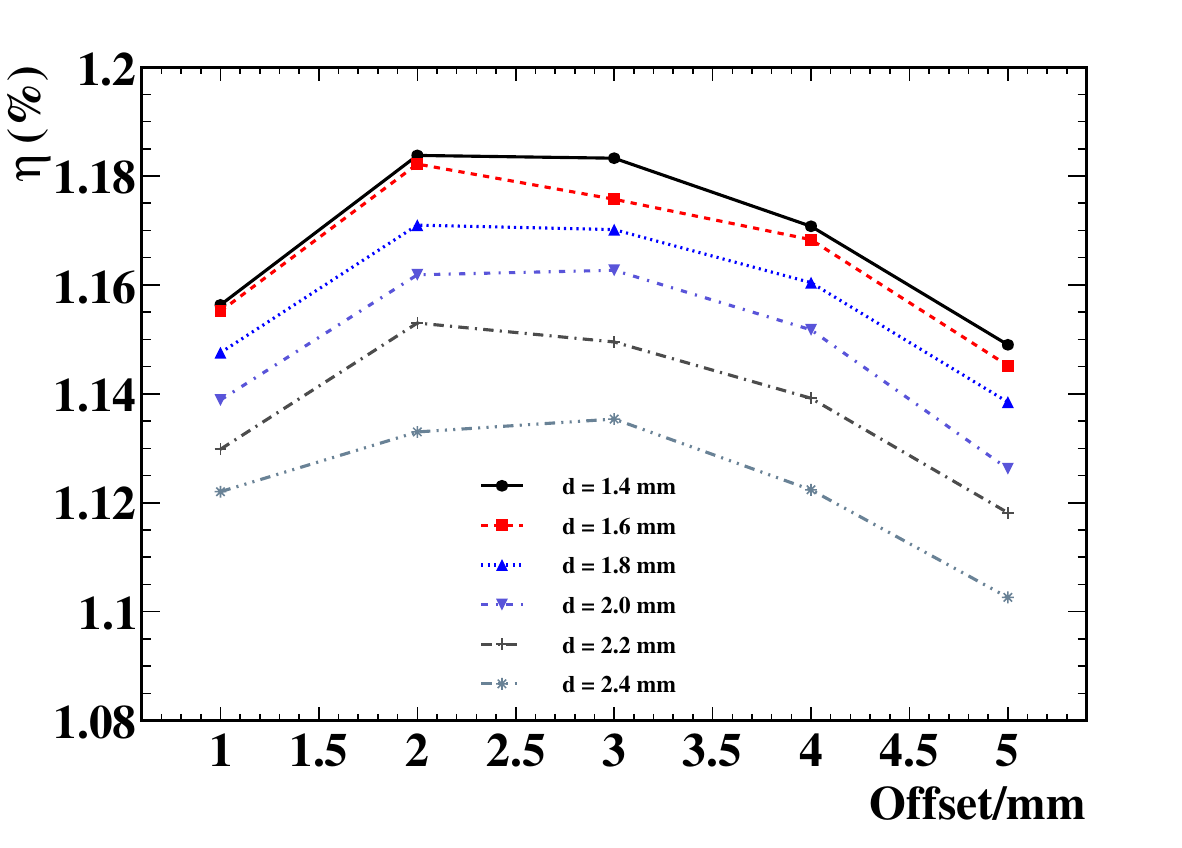}
        \caption{}
        \label{fig:DepthOffset}
    \end{subfigure}
    \begin{subfigure}{0.49\textwidth}
        \centering
        \includegraphics[width=\textwidth]{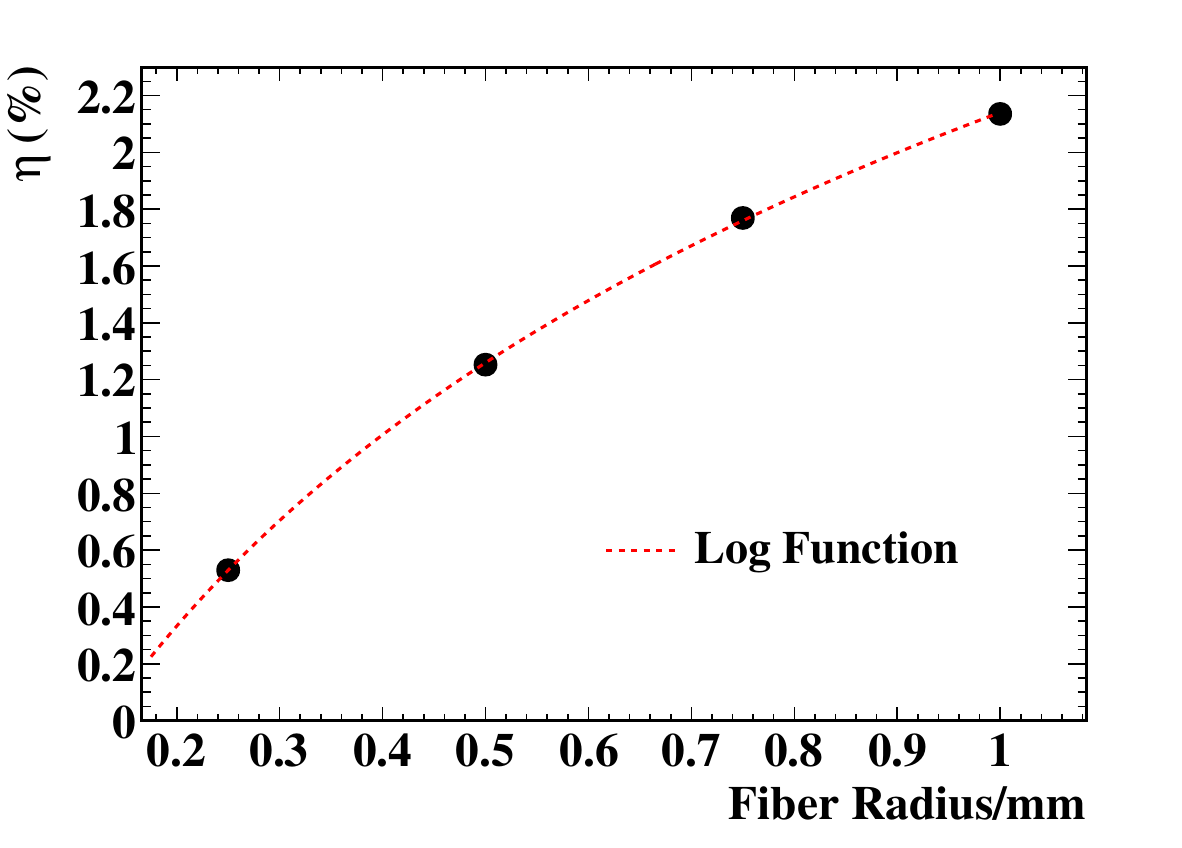}
        \caption{}
        \label{fig:FiberR}
    \end{subfigure}\\
    \begin{subfigure}{0.49\textwidth}
        \centering
        \includegraphics[width=\textwidth]{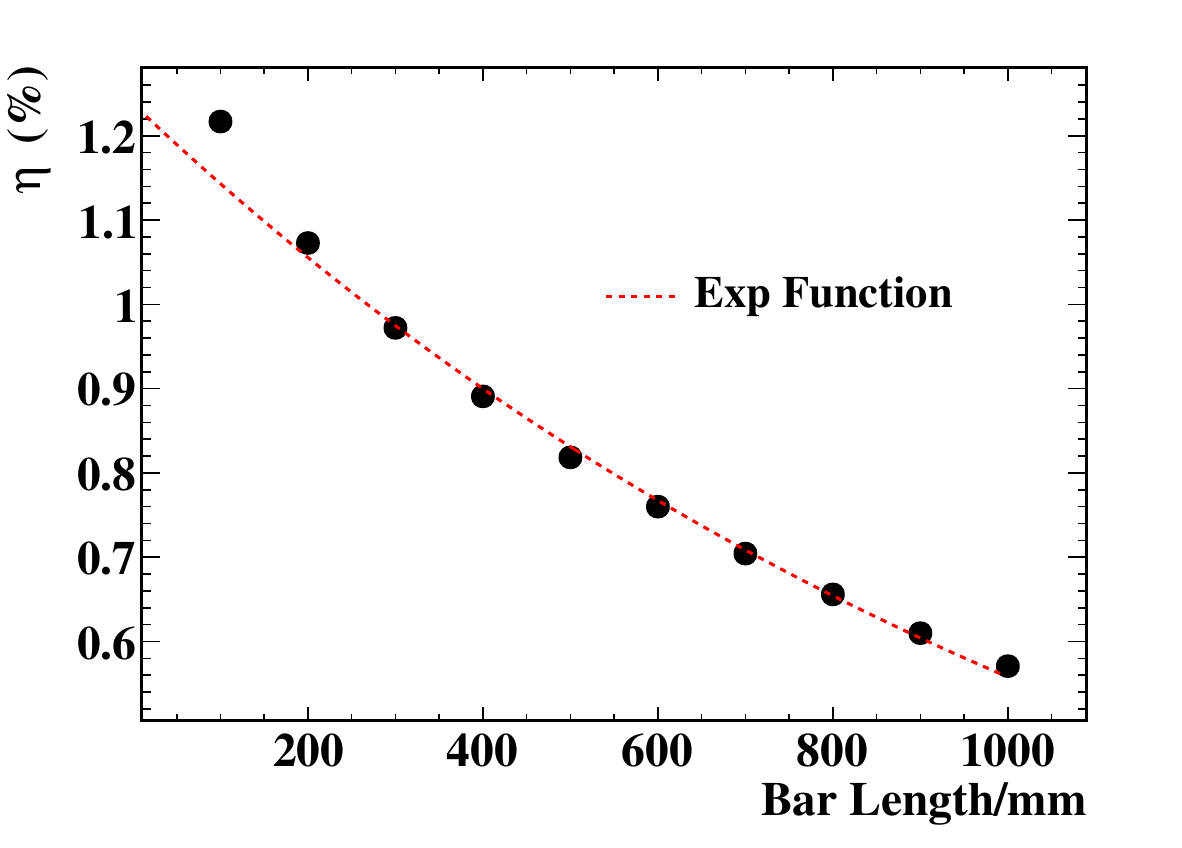}
        \caption{}
        \label{fig:BarLength}
    \end{subfigure}
    \begin{subfigure}{0.49\textwidth}
        \centering
        \includegraphics[width=\textwidth]{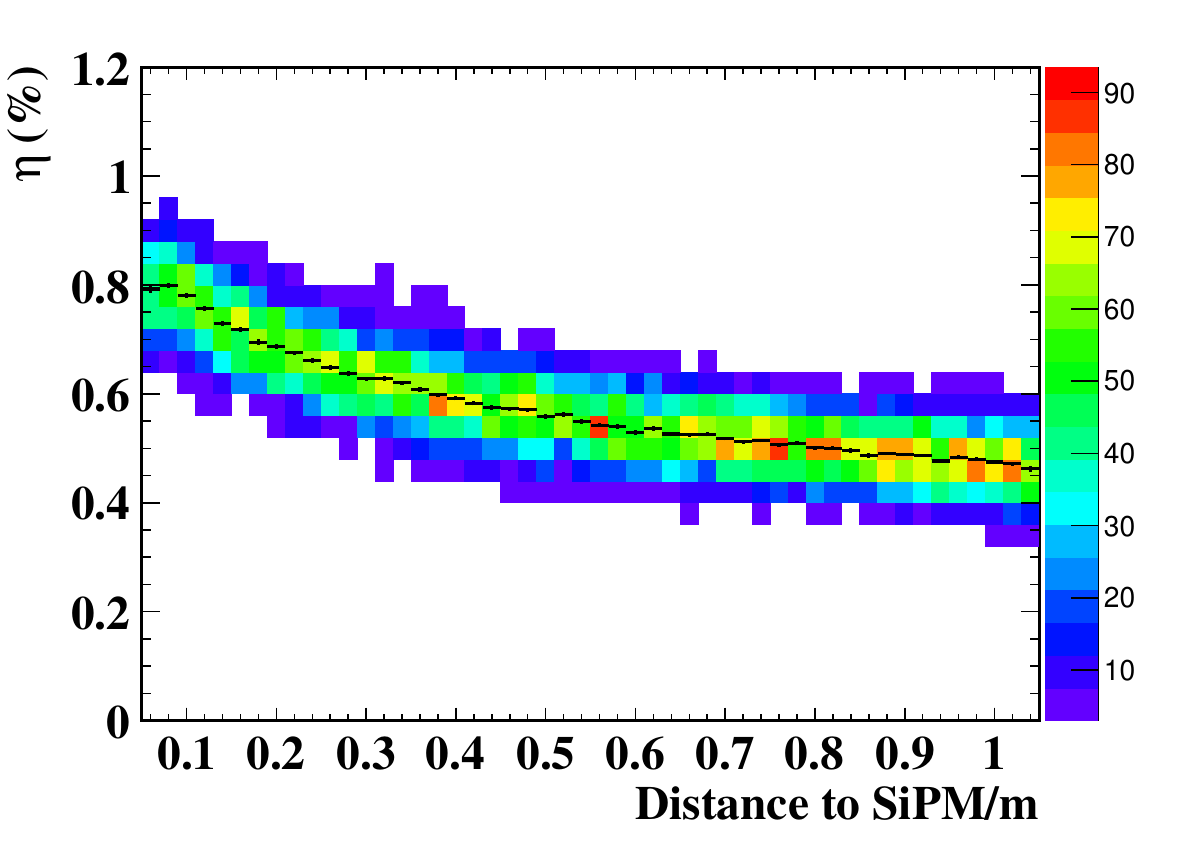}
        \caption{}
        \label{fig:BarLength3}
    \end{subfigure}

    \caption{
    Photon collection efficiency $\eta$ from Geant4 simulations. (a) $\eta$ as a function of groove depth $d$ and offset. (b) $\eta$ as a function of fiber radius $R$. (c) $\eta$ as a function of scintillator bar length $L$. (d) Correlation between $\eta$ and photon propagation distance for a bar length of 1~m. Error bars are smaller than the marker sizes. The default simulation parameters are $d = 1.8$~mm, offset $= 1$~mm, $R = 0.5$~mm, and $L = 500$~mm, with mirror reflection at the fiber end. Two fibers per bar are used in (a) and (b); only one fiber centered at the groove (offset = 0) per bar is used in (c) and (d).
    }
    \label{fig:Detector-Simu2}
\end{figure}

Analysis of the simulation results focus on the photon collection efficiency $\eta$, defined as $ N_\text{det} / N_\text{scint}$, where $N_\text{det}$ is the photoelectron counts detected by the SiPM after accounting for the SiPM photon detection efficiency, and $N_\text{scint}$ is the total photon yield in the scintillator.
Muons are uniformly and randomly incident along the scintillator bar.
Unless otherwise specified, the reported $\eta$ represents an average over all muon interaction positions.
The figure~\ref{fig:DepthOffset} shows the photon collection efficiency $\eta$ as a function of groove offset and depth.
For all considered parameters, including both those plotted and others such as optical coupling and groove shape, the photon collection efficiency $\eta$ exhibits a modest overall variation of approximately 10\%. 
Despite this, the simulation results still define an optimal configuration.
Specifically, the groove offset exhibits a non‑monotonic relationship with $\eta$, peaking at around 3~mm, which was therefore selected as the optimal value;
the depth is set to the minimum that still fully embeds the fiber; and a semicircular groove (the left groove shown in figure~\ref{fig:Detector-structure}) with air coupling is adopted for its manufacturing and installation simplicity.

However, the fiber radius and the scintillator bar length have a significantly greater influence on the efficiency, as shown in figure~\ref{fig:FiberR} and figure~\ref{fig:BarLength}, respectively. 
The fiber radius governs the probability of photon capture and absorption by the WLS fiber, while the bar length determines the propagation distance to the SiPM, thereby affecting the photon transmission efficiency due to attenuation along the fiber.
This effect is further illustrated in figure~\ref{fig:BarLength3}, which shows the variation of $\eta$ with the photon propagation distance for a bar length of 1~m.
In contrast to the average $\eta$ shown in previous figures, here $\eta$ is evaluated at fixed muon interaction positions, with the horizontal axis representing the distance from the interaction point to the SiPM.
As the propagation distance increases, optical losses along the fiber accumulate, leading to a lower photon collection efficiency.
Based on a trade-off between cost and performance, the final design parameters were selected as a scintillator bar length of 600~mm and a fiber radius of 0.75~mm.

\subsubsection{Simulated performance}
\label{sec:simu-detector-sigmax}

\begin{figure}
    \centering
    \begin{subfigure}{0.49\textwidth}
        \includegraphics[width=\linewidth]{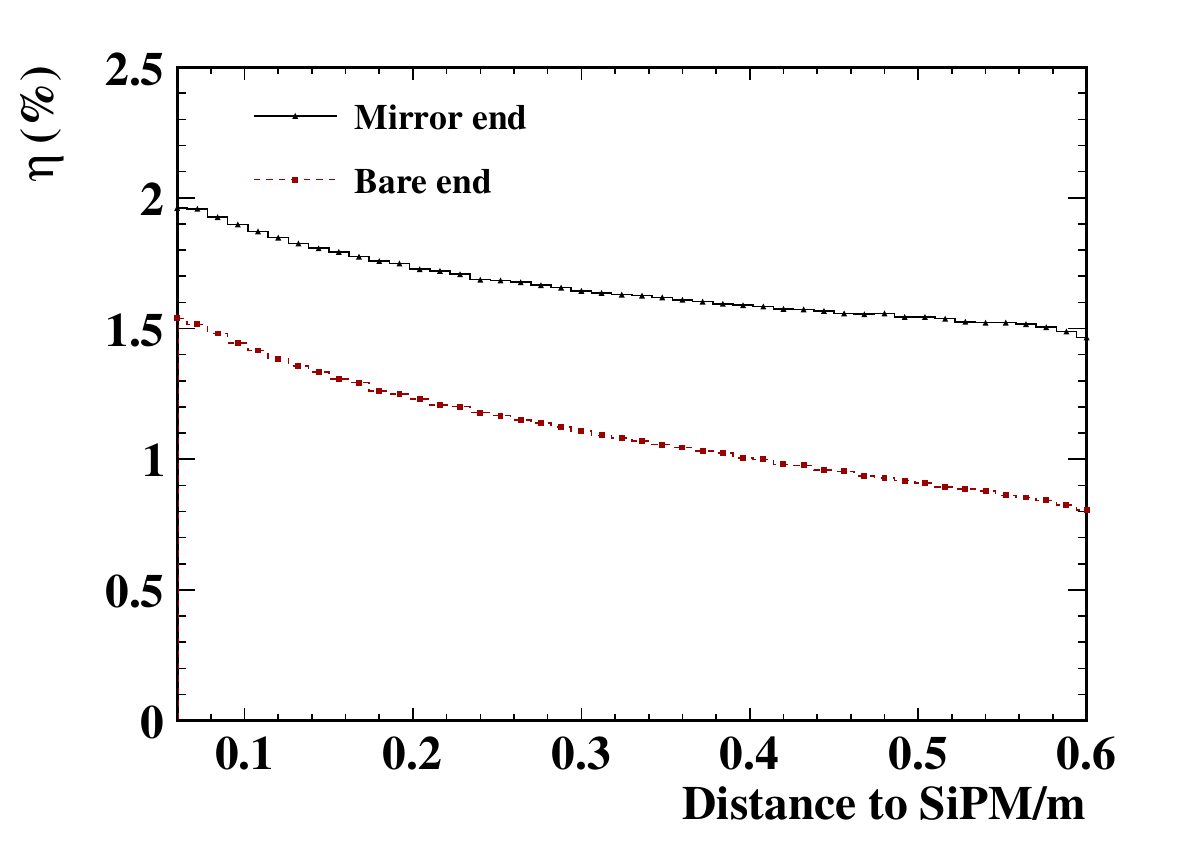}
        \caption{}
        \label{fig:MirrorComparison-Flipped}
    \end{subfigure}
    \begin{subfigure}{0.49\textwidth}
        \includegraphics[width=\linewidth]{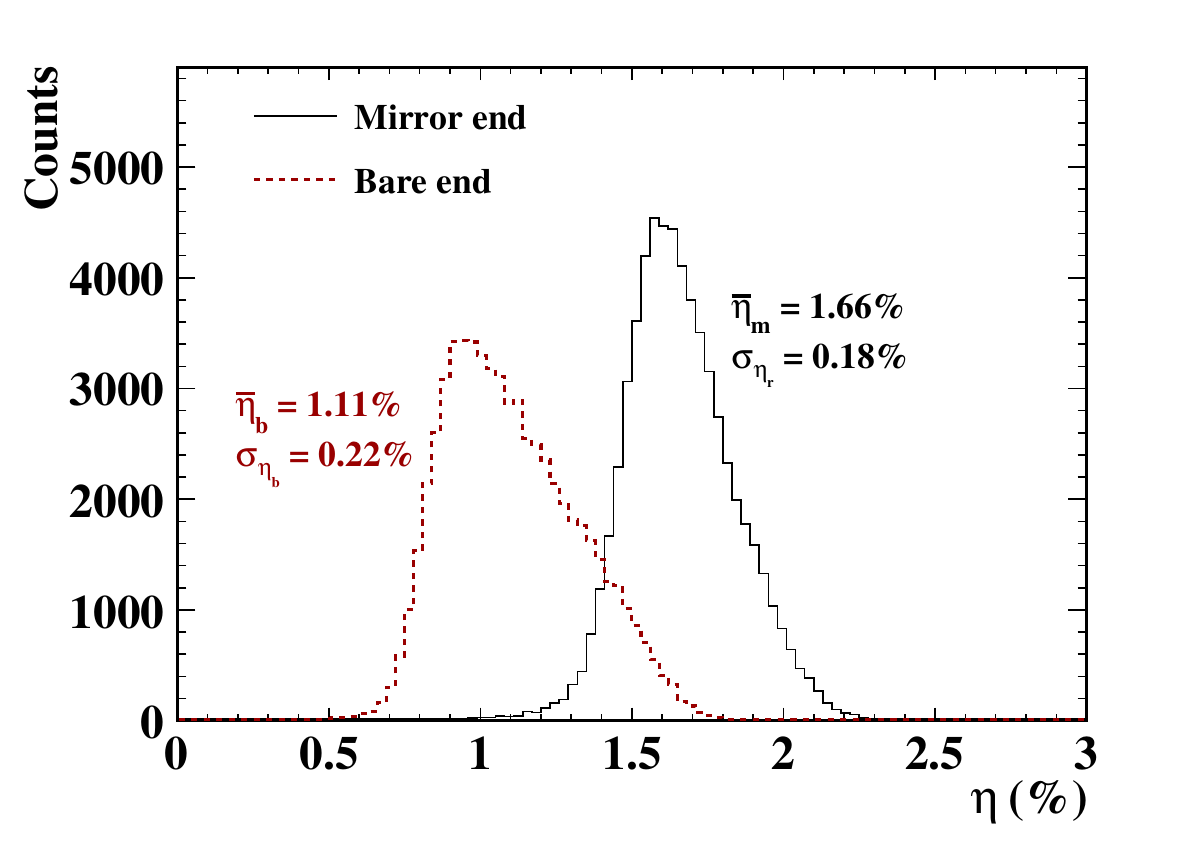}
        \caption{}
        \label{fig:MirrorComparison2}
    \end{subfigure}
    \caption{Effect of mirror coating on photon collection efficiency $\eta$. (a) $\eta$ as a function of distance to the SiPM, with and without a mirror at the distal end. (b) Distribution of $\eta$ with and without a mirror. }
    \label{fig:MirrorComparison}
\end{figure}

With the design parameters finalized, we performed simulations to evaluate the key performance metrics of the detector.
In the simulation, only one end of the fiber was coupled to the SiPM, and a mirror coating was applied to the distal end to reflect uncollected photons back toward the SiPM, following common practice \cite{russoStripDetectorsPortal2014}. 
Figure~\ref{fig:MirrorComparison} shows the effect of the mirror coating on detector performance, presenting $\eta$ as a function of distance to the SiPM and the distribution of $\eta$, with or without the mirror.
The presence of the mirror coating not only significantly increases the average value of $\eta$ but also improves its uniformity along the bar.

\begin{figure}
    \centering
    \includegraphics[width=0.5\linewidth]{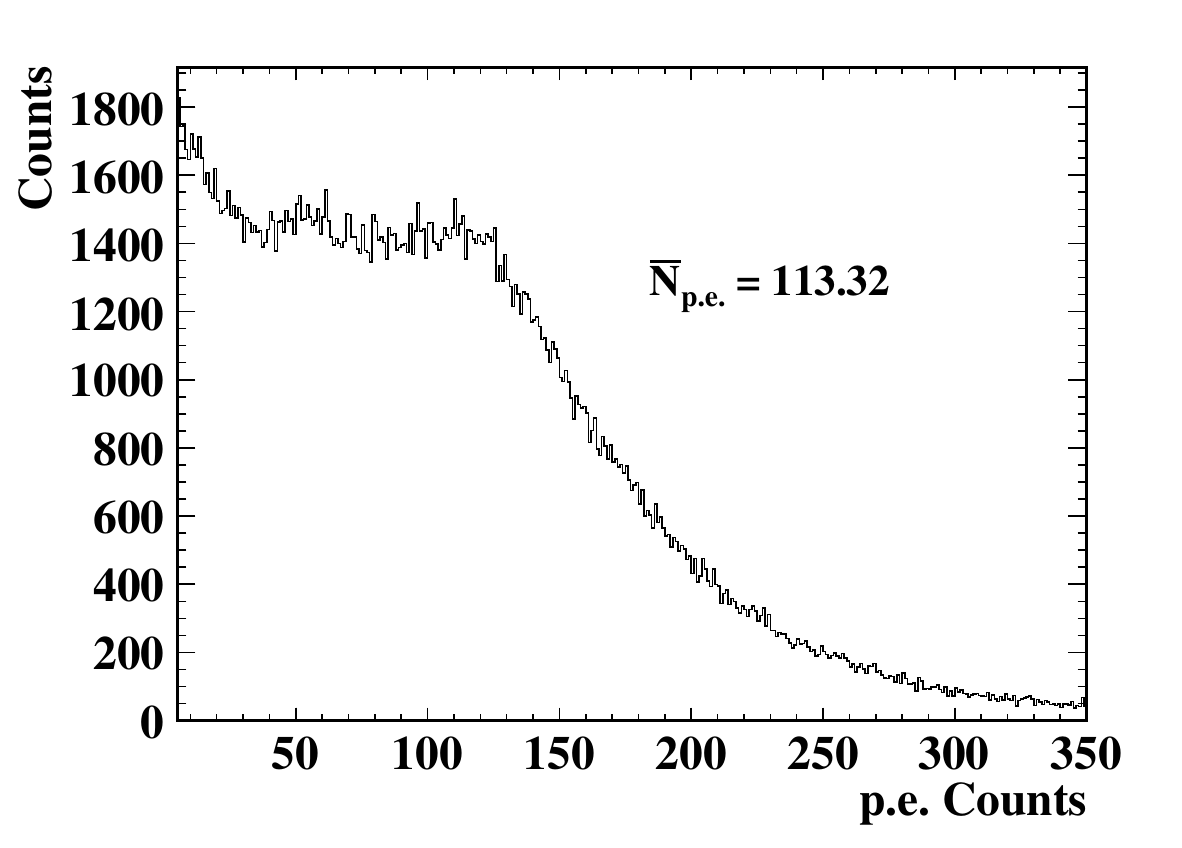}
    \caption{Distribution of the number of photoelectrons collected by a single SiPM with the mirror coating.}
    \label{fig:hpe-ESR}
\end{figure}

The simulation also provides the distribution of the number of photoelectrons converted by a single SiPM with the mirror coating, as shown in figure~\ref{fig:hpe-ESR}. 
The photoelectron count per SiPM ranges from 0 to 350, with an average of approximately 113. 
The nearly plateau region between 0 and 120 p.e. is related to the distribution of muon penetration lengths through the triangular scintillator bars. 
This distribution is used to compare with the photoelectron distribution measured by SiPMs in the experiment and to calibrate the channel-to-channel variations in the experimental setup.

\begin{figure}
    \centering
    \begin{subfigure}{0.45\textwidth}
        \includegraphics[width=\linewidth]{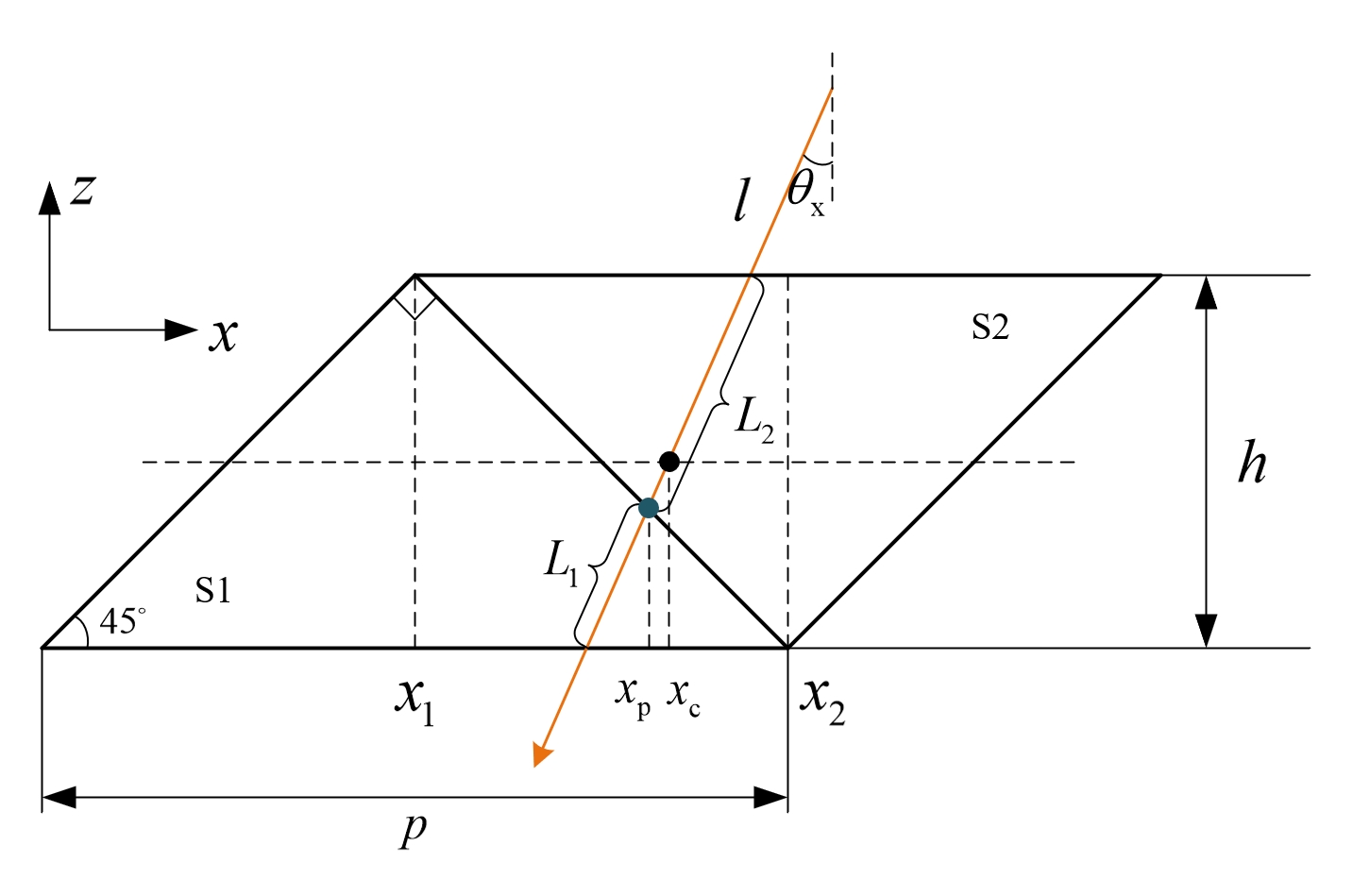}
        \caption{}
        \label{fig:Centroid-positioning}
    \end{subfigure}
    \begin{subfigure}{0.45\textwidth}
        \includegraphics[width=\linewidth]{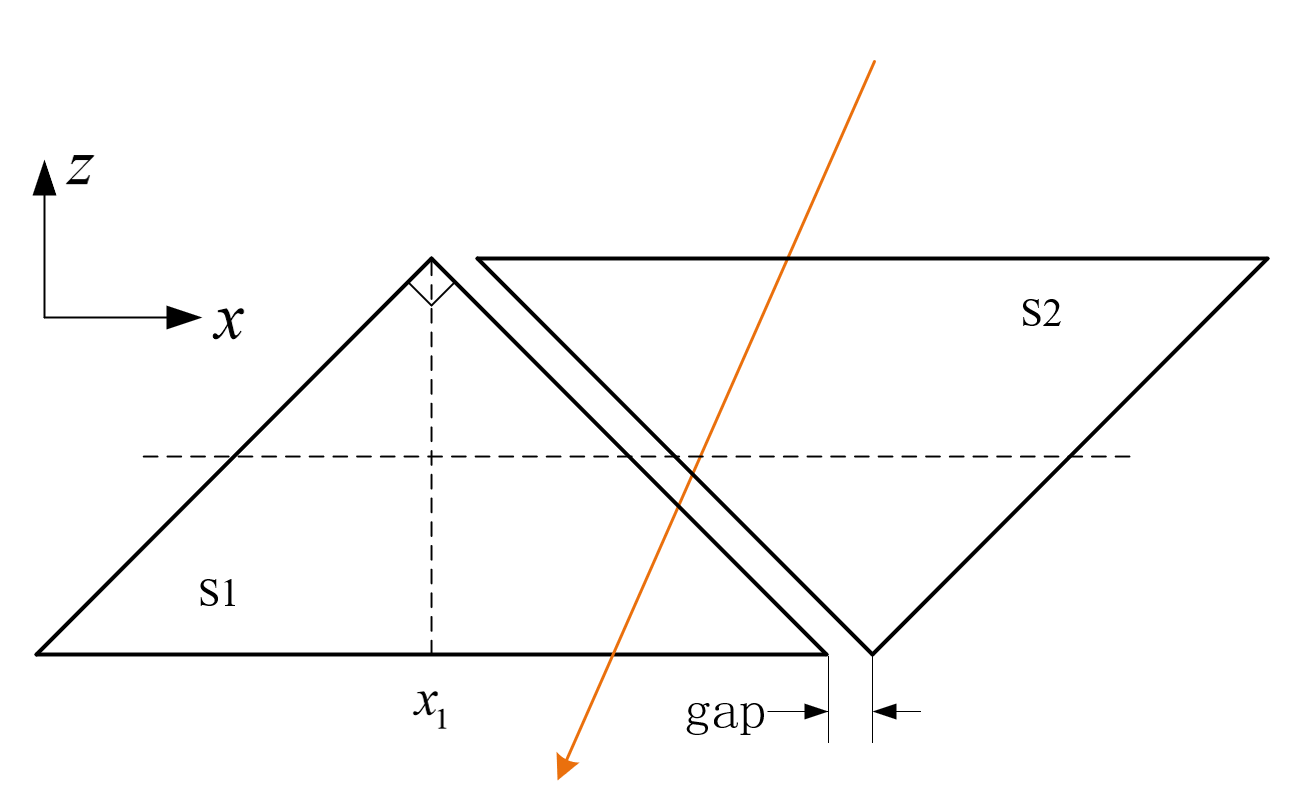}
        \caption{}
        \label{fig:gap-positioning}
    \end{subfigure}
    \caption{
    Schematic diagram of position reconstruction. (a) Ideal case: $x_c$ is determined by the signal ratio (S$_1$/S$_2$) and incident angle $\theta_x$. (b) Actual detector: a gap between adjacent strips necessitates an additional correction.
    }
    \label{fig:scheme-positioning}
\end{figure}

The simulation also enables the evaluation of the detector's spatial resolution.
The charge centroid method, which calculates the hit position of a muon as the weighted centroid of signals from two adjacent strips, is a common algorithm for position reconstruction.
As shown in Figure~\ref{fig:Centroid-positioning}, when the gap between two adjacent strips is neglected, this method determines the intersection point $x_\text{p}$ at the strip boundary. 
However, the muon hit position is conventionally defined as the intercept $x_\text{c}$ at the center of the detector plane. 
Given the incident angle $\theta_x$, $x_\text{c}$ can be derived from $x_\text{p}$ using the following equation:
\begin{eqnarray}
    x_\text{p} &=& \frac{x_1\cdot L_1+x_2\cdot L_2}{L_1+L_2} = x_1 +\lambda \cdot \mathrm{p},\nonumber \\
    x_\text{c} &=& x_\text{p}+(\lambda-0.5)\cdot h\cdot \tan\theta_x, 
    \label{equ:centroid}
\end{eqnarray}
where $x_1$ is the reference strip center, $\mathrm{p}=x_2-x_1$ the strip pitch, 
$\lambda = L_2/(L_1+L_2)$ and $L_i$ is the penetration length in strip $i$,
$h$ the height of scintillator bar, and $\theta_x$ the muon zenith angle.
The photoelectron count $N$ registered by a SiPM is proportional to the energy deposition $E$ of the traversing muon.
For minimum-ionizing particles (MIPs), $E$ is approximately proportional to the path length $L$ within the scintillator bar.
Consequently, the ratio of path lengths $L_1/L_2$ can be estimated from the measured signal ratio $N_1/N_2$ of two adjacent strips.

\begin{figure}
    \centering
    \begin{subfigure}{0.49\textwidth}
        \includegraphics[width=\linewidth]{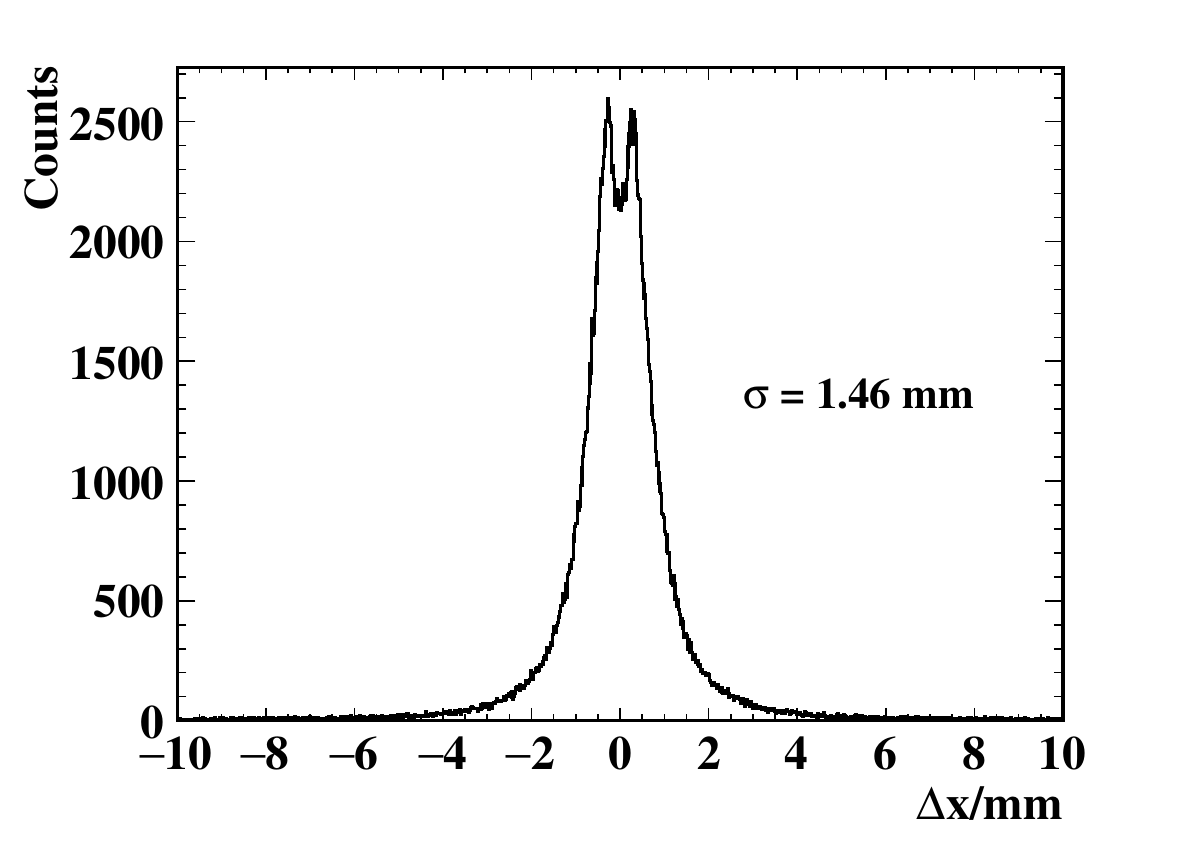}
        \caption{}
        \label{fig:Cor1Resolution-newstyle}
    \end{subfigure}
    \begin{subfigure}{0.49\textwidth}
        \includegraphics[width=\linewidth]{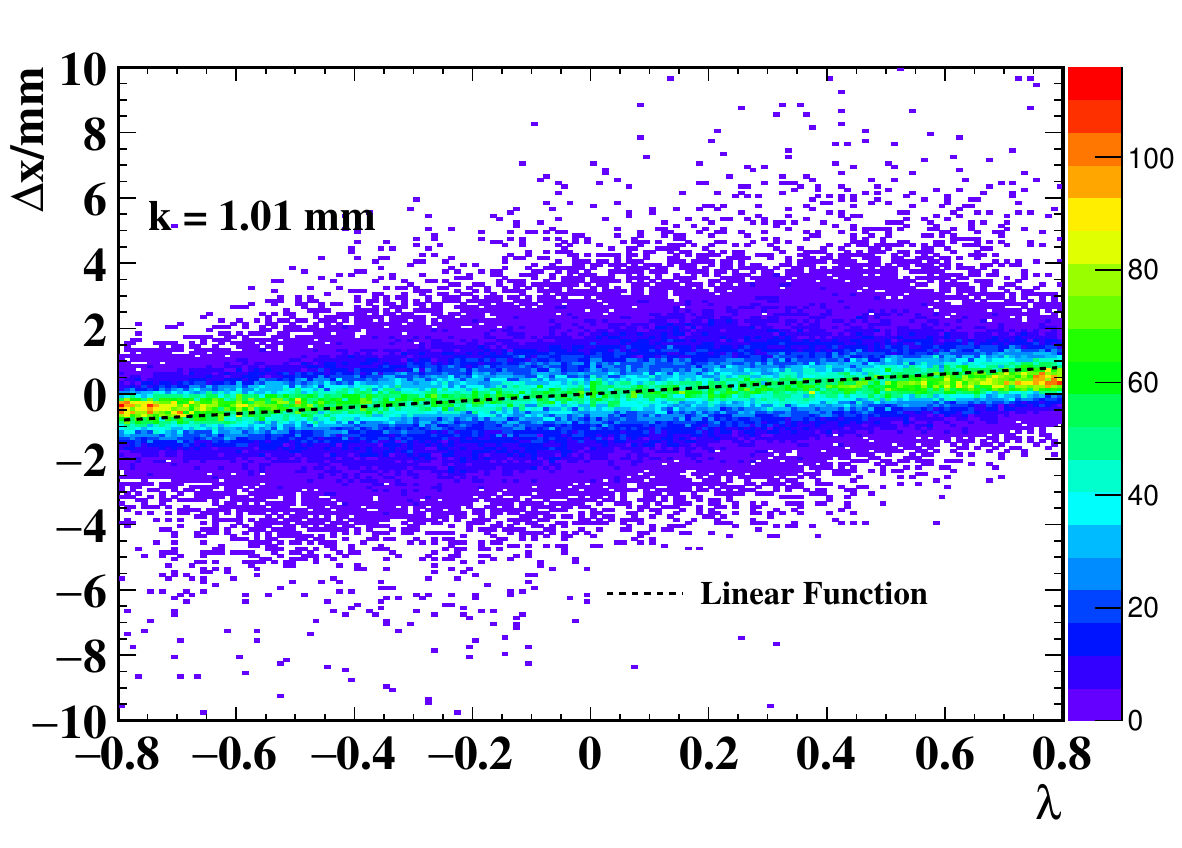}
        \caption{}
        \label{fig:X-EtaLine}
    \end{subfigure}
    \caption{Simulated position residuals and their correlation with $\lambda$. (a) Distribution of $\Delta x$ which splits into two peaks due to the gap between adjacent strips. (b) Correlation between $\Delta x$ and $\lambda$.}
    \label{fig:Residue-NoCor}
\end{figure}

However, the reflective coating on real scintillator strips prevents perfect contact between adjacent strips, creating a gap that affects position reconstruction.
To address this, we analyze the position residuals from the simulation, allowing us to quantify and correct the resulting bias.
Figure~\ref{fig:Residue-NoCor} shows the distribution of the position residual $\Delta x = x_\text{c} - x_{\text{real}}$, where $x_\text{c}$ is the reconstructed position from the simulation and $x_{\text{real}}$ is the true incident position. 
As illustrated in figure~\ref{fig:Cor1Resolution-newstyle}, the $\Delta x$ distribution splits into two distinct peaks.
Figure~\ref{fig:X-EtaLine} further reveals a clear correlation between $\Delta x$ and $\lambda$.
This correlation has been previously reported in \cite{Hu_2020}; however, the correction method we adopt here differs in its implementation.
While previous studies employed a sine function to fit the relationship between $\Delta x$ and $\lambda$, we adopt a more direct linear function, $\Delta x = k \cdot \lambda$, where $k$ is the slope.
Using this correlation, the gap-corrected hit position is given by $x_\text{g} = x_\text{c} - k \cdot \lambda$.
The value of $k$ depends on the detector geometry. 
In our simulation, for scintillator strips with a base length of 20~mm, a height of 10~mm, and a center-to-center pitch of 11~mm, the slope is determined to be $k = 1.01$~mm.

\begin{figure}
    \centering
    \includegraphics[width=0.5\linewidth]{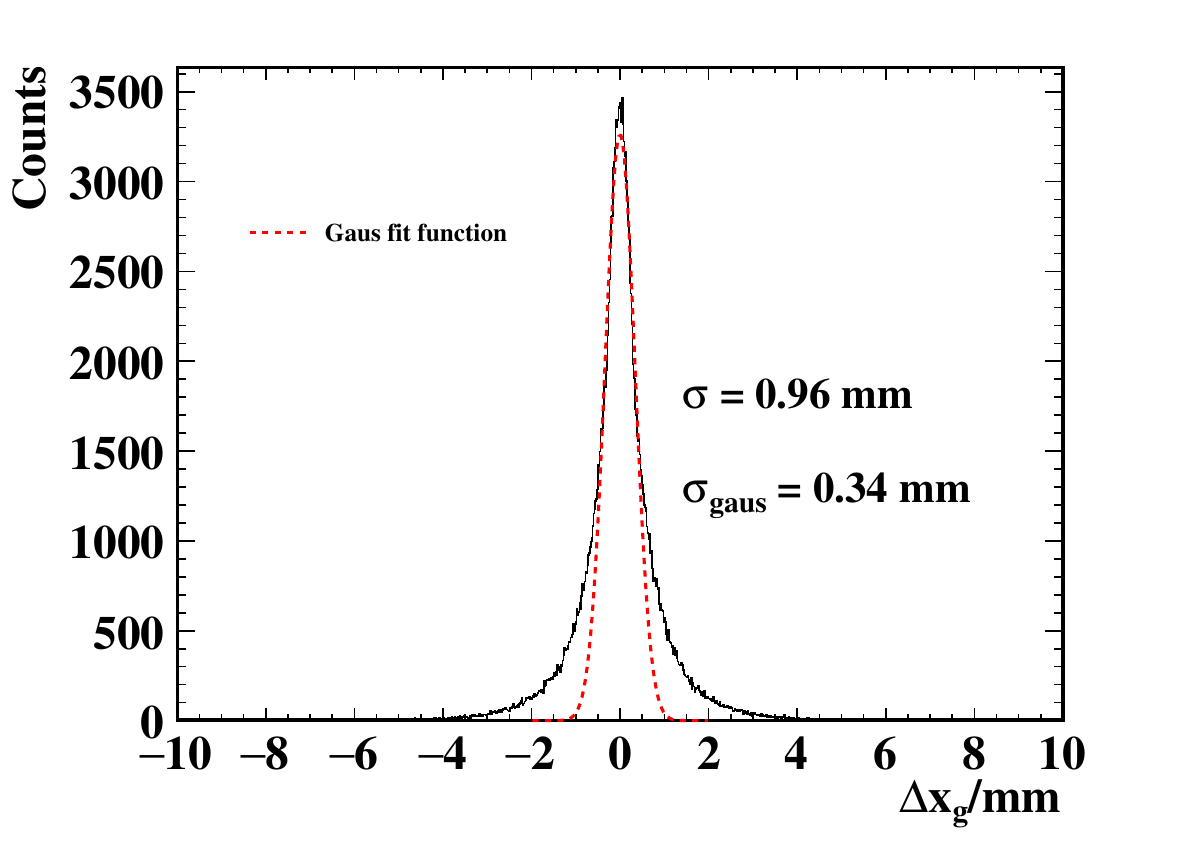} 
    \caption{Distribution of the corrected position residual $\Delta x_\text{g}$ after gap correction.}
    \label{fig:CoatCorLine}
\end{figure}

The distribution of the corrected position residual $\Delta x_\text{g} = x_\text{g} - x_\text{real}$ is presented in figure~\ref{fig:CoatCorLine}.
In contrast to the distribution of $\Delta x$ shown in figure~\ref{fig:Cor1Resolution-newstyle}, the two peaks converge into a single peak. 
Furthermore, the standard deviation is reduced from 1.46~mm to 0.96~mm, demonstrating a marked improvement in spatial resolution.
This value of 0.96~mm represents the best spatial resolution achieved in our simulation.
Unless otherwise stated, the corrected position residual $\Delta x_\text{g}$ is adopted to characterize the spatial resolution performance in all subsequent position reconstruction analyses.

\subsubsection{Readout encoding}
The flux of cosmic ray particles is low, and a single incident particle, in most cases, will hit only two adjacent scintillator bars. 
This characteristic is exploited to implement an encoded signal readout scheme.
Since each scintillator bar incorporates two WLS fibers for light signal output, the signals from these two fibers can be routed to separate SiPMs, thereby enabling the encoding of the optical signal.

In our encoding scheme, 16 scintillator bars and their corresponding 32 WLS fibers are grouped into a single module. 
The signals from every four fibers are input into one SiPM, thus requiring a total of eight SiPMs per encoding module.
Compared to the previous prototype, this design achieves a compression ratio of 4:1.
More specifically, the two fibers in each scintillator bar are designated as $\alpha$ (left) and $\beta$ (right).
Accordingly, the eight SiPMs are also divided into group A and group B, with each group exclusively receiving input from its corresponding set of fibers.
A valid encoding scheme can be readily obtained by populating a $4\times4$ table, defined by the four group A and four group B SiPMs, with the indices of the 16 scintillators.

The encoding scheme implemented in our design is defined by the mapping between scintillator and SiPM numbers given in table~\ref{tab:encoding}, such that the $\alpha$ and $\beta$ fibers of any scintillator are connected to two different SiPMs.
For example, scintillator S$_0$ sends its $\alpha_0$ and $\beta_0$ fiber signals to SiPM~A$_0$ and SiPM~B$_0$, respectively.
The global trigger signal for the encoding module is generated by the coincidence between the OR-ed outputs of group A (A$_0$-A$_3$) and group B (B$_0$-B$_3$).

\begin{table}[htbp]
    \centering
    \smallskip
    \renewcommand\arraystretch{1.6}
    \begin{tabular}{c|cccc}
         & B$_0$ & B$_1 $& B$_2 $& B$_{3}$ \\
        \hline
        A$_0$ & S$_0 $& S$_4 $& S$_{8}$ & S$_{12}$ \\
        A$_1$ & S$_{13}$ & S$_1 $& S$_{5}$ & S$_{9}$ \\
        A$_2$ & S$_{10}$ & S$_7 $& S$_{2}$ & S$_{15}$ \\
        A$_3$ & S$_6 $& S$_{14}$ & S$_{11}$ & S$_{3}$
    \end{tabular}
    \caption{Scheme for encoding 16 scintillator strips to 8 SiPMs.}
    \label{tab:encoding} 
\end{table}

Our encoding scheme satisfies the following two critical requirements:
\begin{enumerate}
    \item Non-adjacency in rows/columns:
    To prevent information loss, adjacent scintillator indices occupy different rows and columns. 
    This arrangement ensures that their four fibers are read by four distinct SiPMs, thereby preserving independent signal paths.
    Merging signals from different scintillator bars would compromise this independence.
    For example, if fibers $\alpha_0$ (from S$_0$) and $\alpha_1$ (from S$_1$) are combined, the independent $\alpha$‑channel information is lost. While the S$_0$/S$_1$ ratio can still be estimated from $\beta_0$ and $\beta_1$, the precision of the measurement is thereby reduced.
    \item Unambiguous pair identification: 
    The encoding scheme guarantees that every four‑fold SiPM coincidence signal maps to a unique adjacent scintillator pair by enforcing that any rectangle in the table has at most one adjacent index pair among its four corners.
    As a concrete example, a coincidence signal on SiPMs A$_0$, A$_1$, B$_0$, and B$_1$ would correspond to several potential scintillator pairs according to table~\ref{tab:encoding}. 
    Since the encoding enforces that any rectangle contains at most one adjacent pair, and among the candidates only S$_0$ and S$_1$ are adjacent, the event is uniquely identified as the S$_0$–S$_1$ pair. 
\end{enumerate}

\begin{figure}
    \centering
    \begin{subfigure}[b]{0.53\textwidth}
        \centering
        \includegraphics[width=\textwidth]{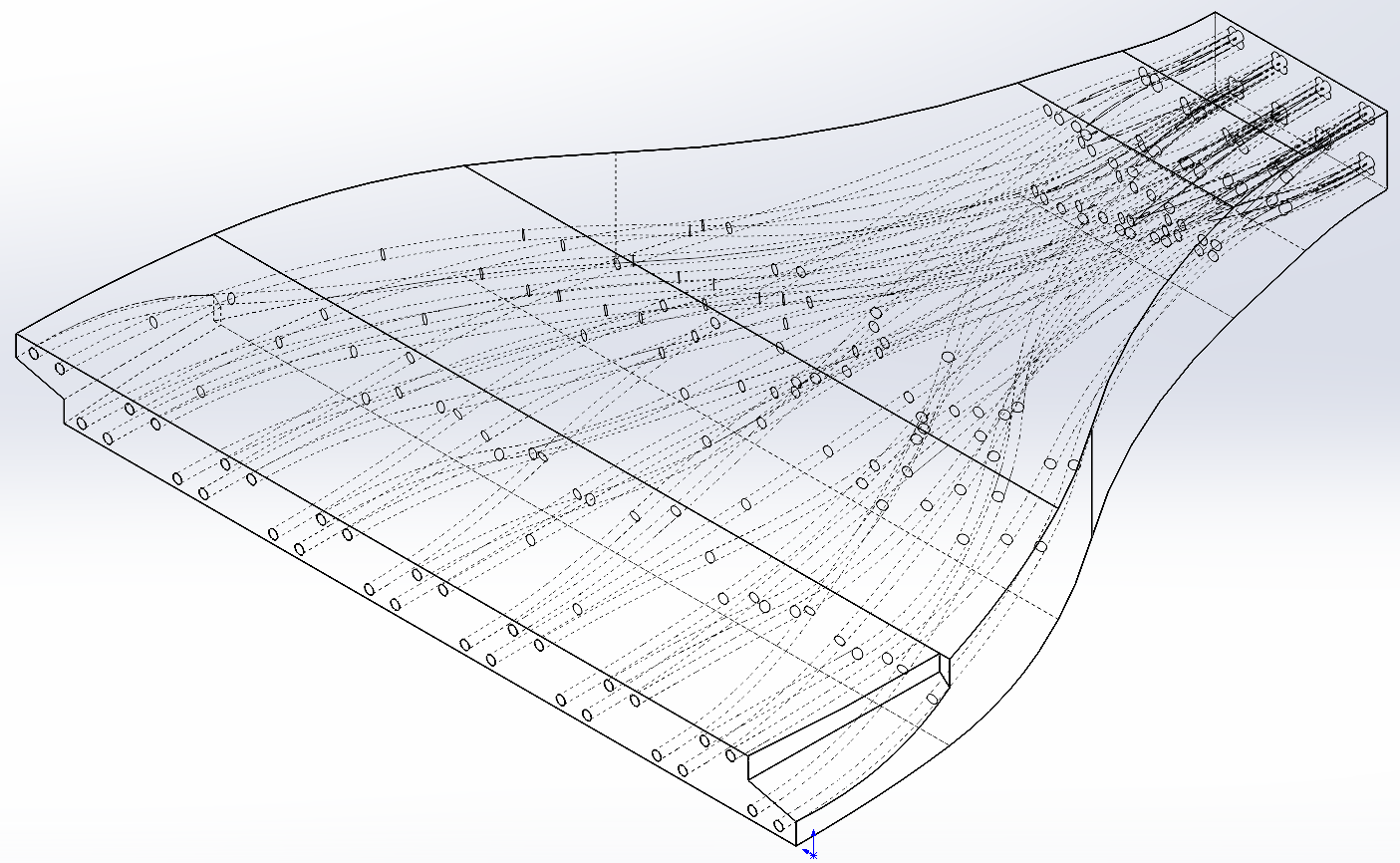}
        \caption{}
		\vspace{10pt}
    \end{subfigure}
    \begin{subfigure}[b]{0.46\textwidth}
        \centering
        \includegraphics[width=\textwidth]{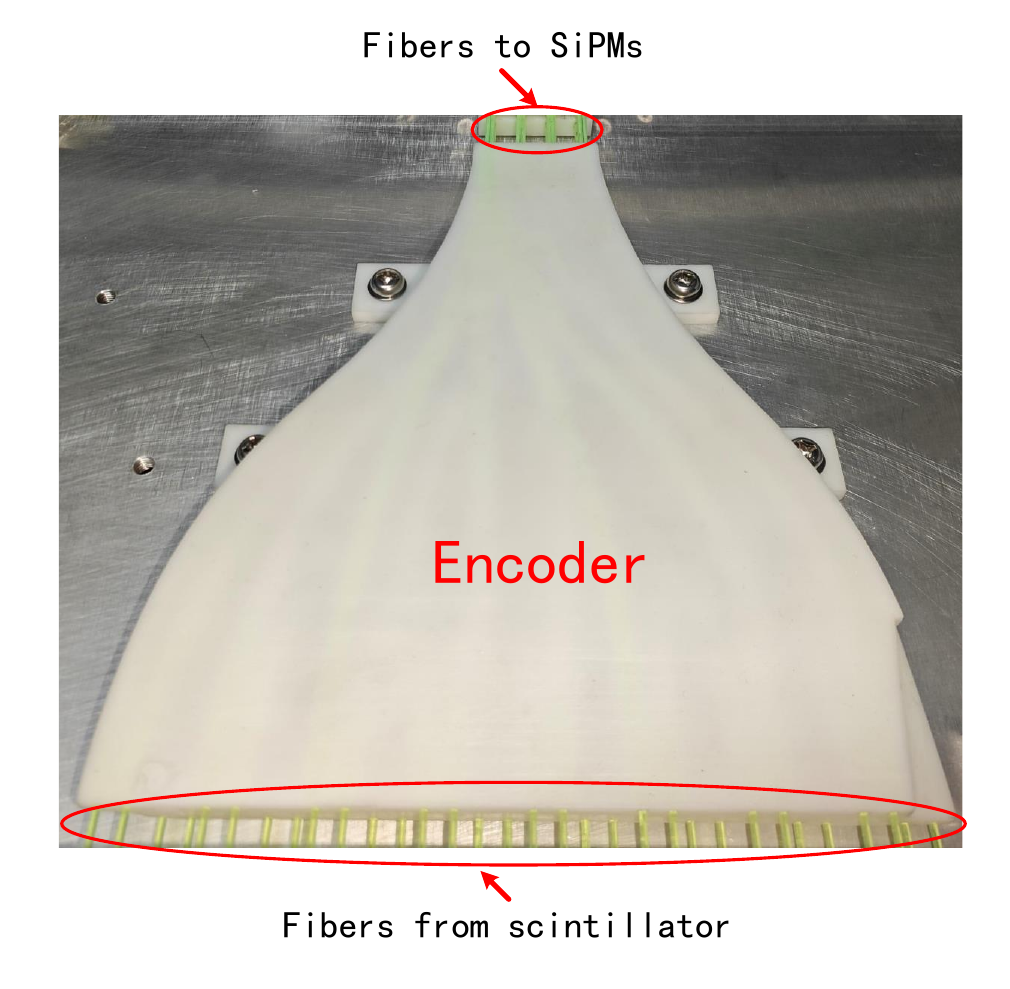}
        \caption{}
    \end{subfigure}
    \caption{Encoder unit design and prototype. (a) Oblique schematic illustrating the internal structure (dashed lines). (b) Photograph of the 3D-printed prototype. }
    \label{fig:Encoder}
\end{figure}

In practical fiber encoding, the tangled fiber arrangement poses a significant challenge for detector assembly. 
Additionally, fibers exposed outside the assembly are highly vulnerable to breakage.
To address these issues, an encoder is necessary to guide the fibers from the scintillators through precisely designed internal paths, thereby organizing them into bundled clusters at the output for interfacing with the SiPM arrays. 
Accordingly, we designed and fabricated the encoder shown in figure~\ref{fig:Encoder}. 
The use of 3D design and printing overcomes the limitations of traditional CNC machining by enabling the production of parts with intricate internal channels that would otherwise be impossible to manufacture.

The SiPMs feature a dark count rate on the order of 1~MHz. 
Furthermore, the use of triangular scintillator bars imposes a requirement to be capable of detecting as low as approximate 10 photoelectrons for higher spatial resolution, which in turn necessitates a low signal threshold.
Consequently, a coincidence measurement scheme using multiple SiPMs must be employed to suppress the false trigger rate effectively. 
This requirement is independent of the encoding scheme and is fundamental for reading out a bar with low photon yield. 
In fact, even a non-encoded design, such as our prototype detector, necessitates at least two SiPMs per bar.
Therefore, the encoding scheme achieves a compression ratio of 1:4.

\section{Construction and test of super layer}

\subsection{Detector fabrication}
The detector was fabricated using the same materials as those modeled in the Geant4 simulation, namely the EJ-200 scintillator, Kuraray Y11 WLS fibers, and Hamamatsu S13360-3075PE SiPMs.
The scintillator slab was first cut into isosceles triangular strips with a base of 20~mm and a height of 10~mm.
The strips were then grooved, polished, and wrapped with Tyvek paper to reflect the fluorescent photons.
The geometric parameters of the slabs were set to the same values as those optimized in the simulation.
During module assembly, the scintillator strips were aligned with a center-to-center pitch of 11~mm, clamped using a fixture, and spaced with a 1~mm inter-strip gap to accommodate the Tyvek wrapper.
The WLS fibers, which have a diameter of 1.5 mm, were then inserted into the grooves.
No optical glue was applied to couple the fibers to the scintillator, simplifying installation and future maintenance.
One end of each fiber was anchored to a piece of Enhanced Specular Reflector (ESR) film cut into small disks of approximately 2~mm in diameter, with a reflectivity of up to 98\%, while the opposite end was passed through the encoder and optically coupled to a SiPM.

\begin{figure}
    \centering
    \begin{subfigure}[b]{0.5\textwidth}
        \centering
        \includegraphics[width=\textwidth]{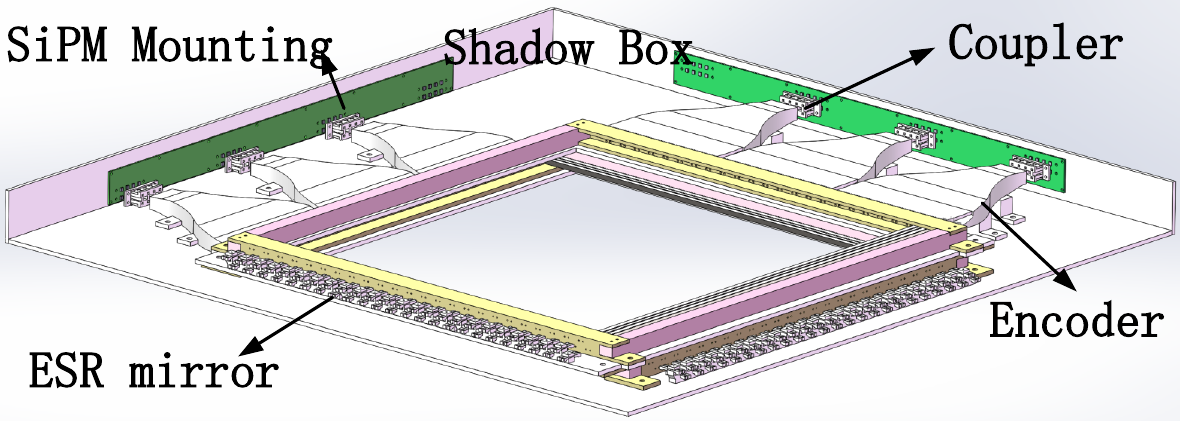}
    \end{subfigure}\\
    \begin{subfigure}[b]{0.55\textwidth}
        \centering
        \includegraphics[width=\textwidth]{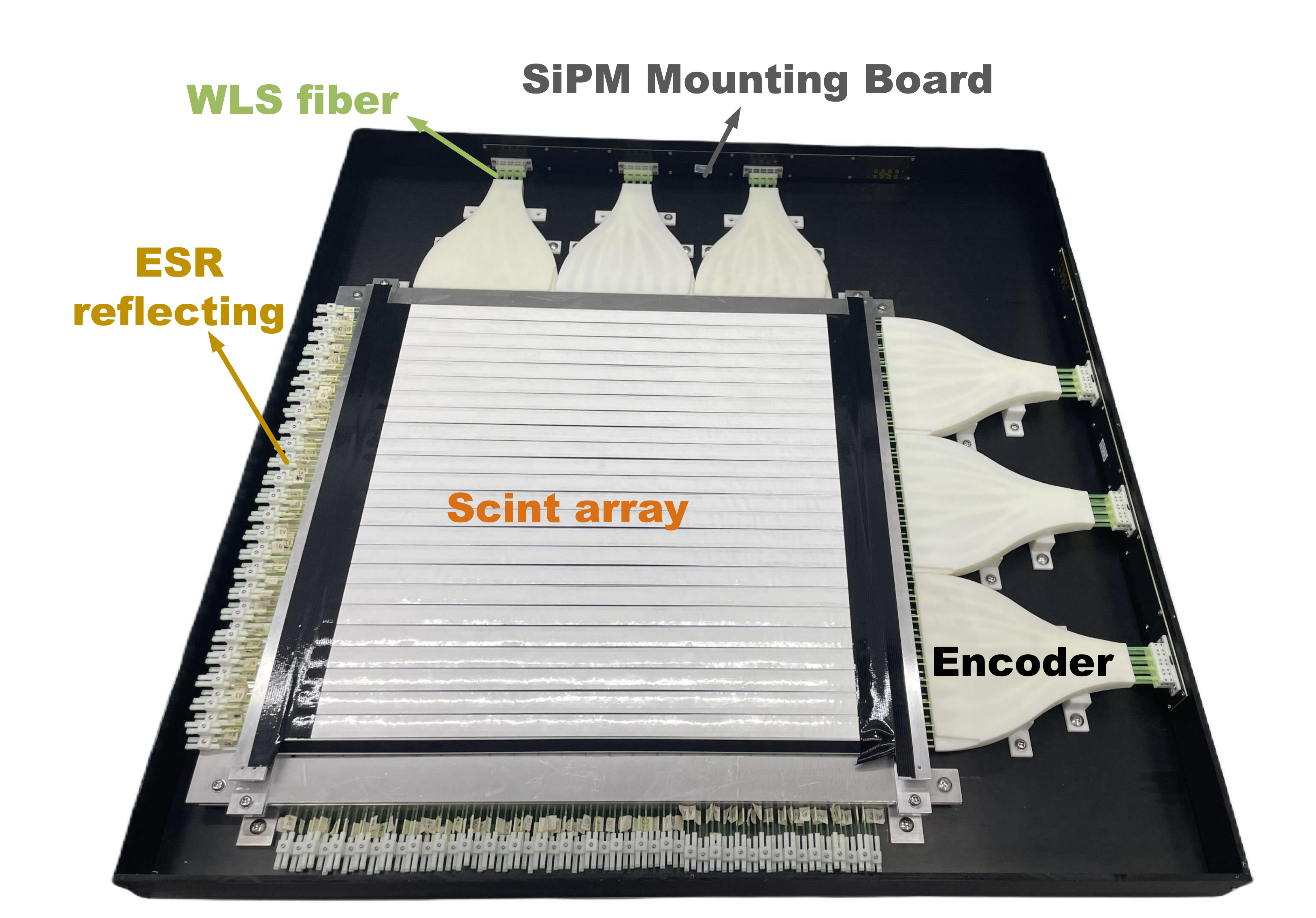}
    \end{subfigure}
    \caption{A super layer within a light-tight box: comprising two detection planes, each constructed from three encoding modules.}
    \label{fig:Detector-Layout}
\end{figure}

Figure~\ref{fig:Detector-Layout} presents the layout and a photograph of a super layer within the light-tight box.
Each box contains a super layer, formed by two orthogonal X and Y detection planes. Each plane contains 48 scintillator strips (60 cm long), giving it an effective area of 53 cm × 60 cm, with the super layer's active area being the 53 cm × 53 cm overlap.
The 48 strips are grouped into three encoding modules, read out by 24 SiPMs and electronic channels. 
Each plane interfaces with one electronic board, which uses 24 of its 32 channels.

\subsection{Readout electronics}
\begin{figure}
    \centering
    \includegraphics[width=0.6\linewidth]{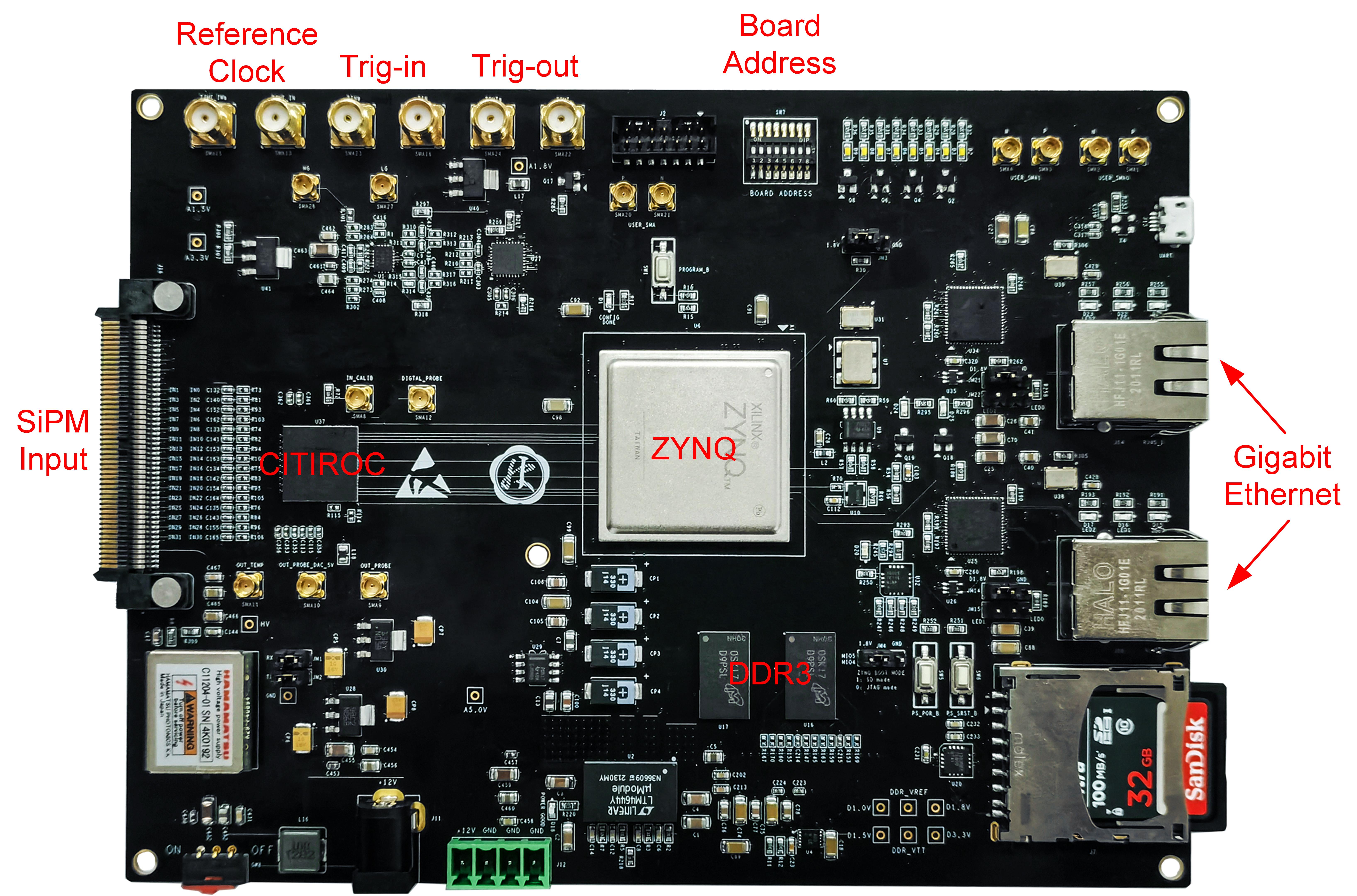}
    \caption{Electronic board designed for SiPM readout.}
    \label{fig:Electronic-board}
\end{figure}

The performance of the readout electronics plays a key role in both the detector and the overall MST system.
The detector requires precise measurement of the photoelectrons count ratio between adjacent scintillator strips.
Furthermore, Geant4 simulations show that the photoelectron count per individual SiPM ranges from 0 to 350, making single photoelectron resolution across this entire range a critical design requirement. 
To fulfill these requirements, we developed the readout electronic board presented in  \cite{Wang_2024}, as shown in figure~\ref{fig:Electronic-board}.

The front-end electronic board integrates a CITIROC 1A \cite{Fleury_2014} for front-end amplification of SiPM signals, an Analog Devices AD9645 dual-channel 14-bit ADC for signal digitization, and a Zynq-7000 SoC, which handles digital logic and serves as the data acquisition interface to the host PC.
The CITIROC chip features 32 input channels, each with adjustable high-gain (HG) and low-gain (LG) charge-sensitive amplifiers and a peak-holding circuit, which extends the dynamic range of the electronic board. 
The held peaks from the 32 channels are sequentially output to the HG and LG pins and then digitized by the dual-channel ADC.
In addition, the board provides temperature-compensated bias voltage to the SiPMs and records the time-of-arrival for each channel. 
The final data read by the host PC from the electronic board include the digitized amplitudes from both the HG and LG ADCs, as well as the time-of-arrival from the TDC for all 32 channels.

\begin{figure}
    \centering
    \includegraphics[width=0.5\linewidth]{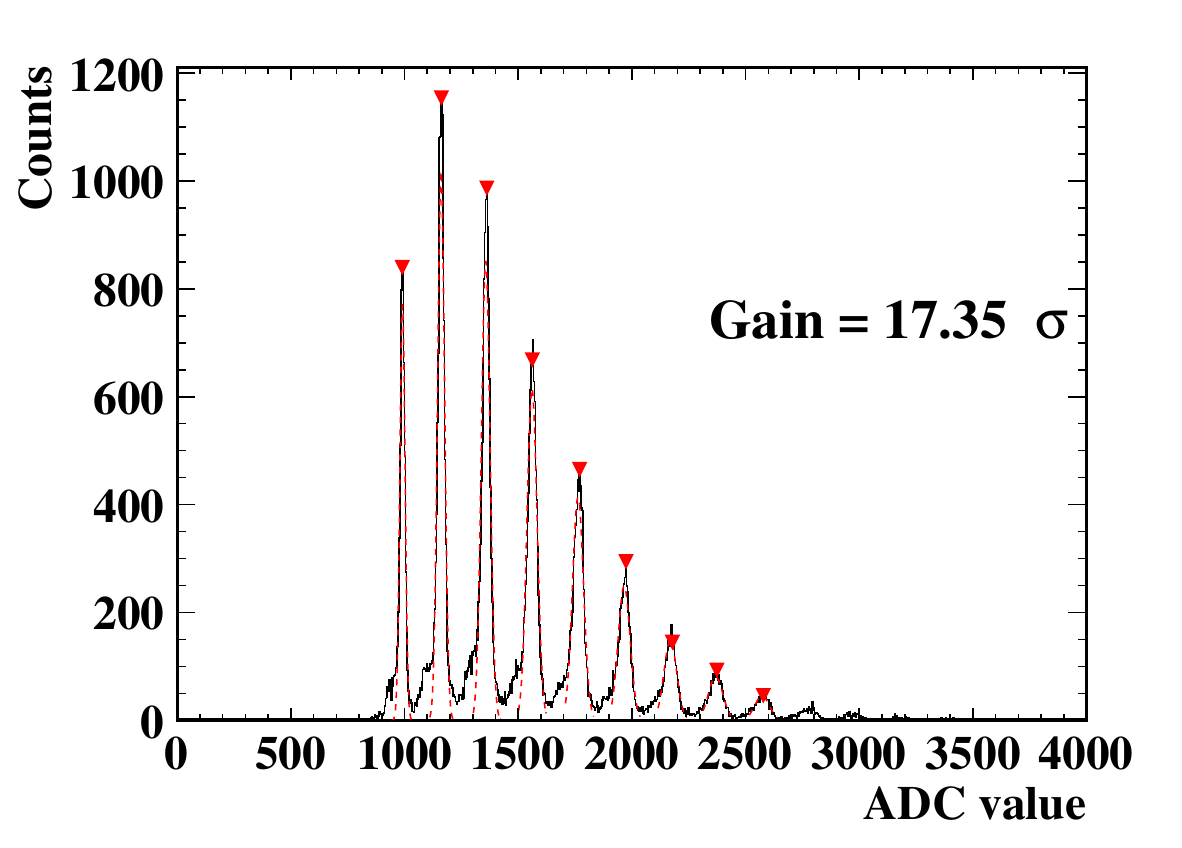}
    \caption{ Characterization of single-photoelectron resolution. Response of a Hamamatsu S13360-3075 SiPM on the readout board to pulsed LED illumination. }
    \label{fig:Electronic-SPE}
\end{figure}

The readout electronics achieve an equivalent noise charge as low as 0.45~fC.
With the SiPM gain set to $4~\times~10^6$, a single photoelectron signal reaches \textasciitilde640~fC, yielding an excellent signal‑to‑noise ratio that enables clear single photoelectron discrimination. 
This performance is verified in figure~\ref{fig:Electronic-SPE}, which shows the ADC spectrum of a Hamamatsu S13360‑3075 SiPM under LED illumination.
A single photoelectron resolution of 17$\sigma$ is achieved, as clearly indicated in the figure.

\begin{figure}
    \centering
    \includegraphics[width=0.8\linewidth]{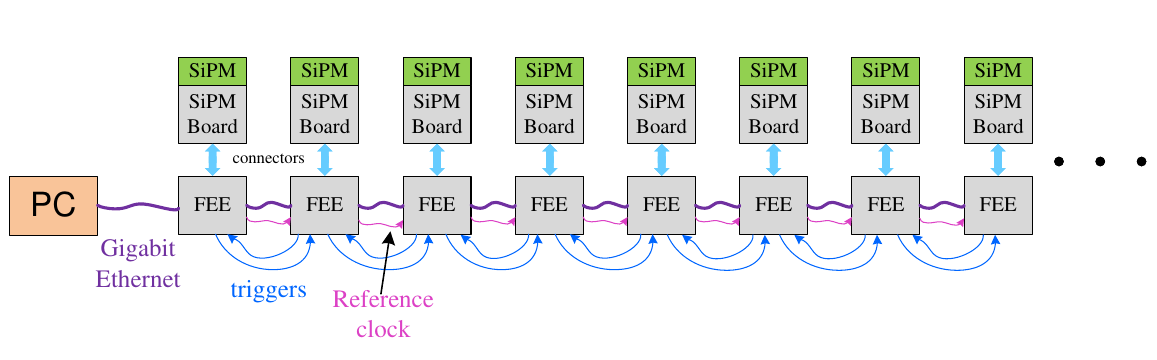}
    \caption{Daisy chain topology across multiple electronic boards.}
    \label{fig:MultiBoards}
\end{figure}

The electronic boards are designed not only to meet the photoelectron detection requirements but also to support system-level functions for the multi-board MST setup, including trigger logic, inter-board communication, and time synchronization.
As shown in figure~\ref{fig:MultiBoards}, all electronic boards are connected in a daisy-chain topology.
Data packets are transferred sequentially between boards through Ethernet cables, while inter-board event handshaking is implemented via the TIN and TOUT signal lines. 
When a board is triggered internally by a cosmic ray event, it sets its TOUT output to a high level. An event is considered confirmed and is recorded by a board only if its TIN input receives a high-level signal within 150 ns after its own internal trigger.
Different test functions can be realized by configuring the logic relationship between TIN and TOUT, which will be described in section~\ref{sec:Detector-performance}.

The distributed nature of the electronics system necessitates precise timing coordination to ensure data consistency. 
To compensate for clock drift between individual boards, a common 1 Hz reference signal is distributed throughout the system.
Each board records the arrival time of this synchronized pulse, enabling offline calibration and alignment of measurement timestamps across all units. 
This synchronization is critical for accurately reconstructing cosmic ray trajectories that involve triggers from multiple detection planes.

In addition to the SiPM FEE board, we developed a mounting board that can accommodate  up to 32 SiPMs, corresponding to four encoding modules.
These are paired with custom 3D-printed couplers to ensure precise alignment between the WLS fibers and the SiPMs.
Each encoding module incorporates a centrally located temperature sensor for real-time monitoring and compensation of its eight SiPMs.
In practice, only three of the four modules are utilized in the current setup.

\subsection{Raw data processing}
With the assembly of multiple detectors completed, we performed a series of processing steps on the raw readout signals, namely time calibration, event matching, channel calibration, signal decoding, and position reconstruction.

\begin{figure}
    \centering
    \includegraphics[width=0.5\textwidth]{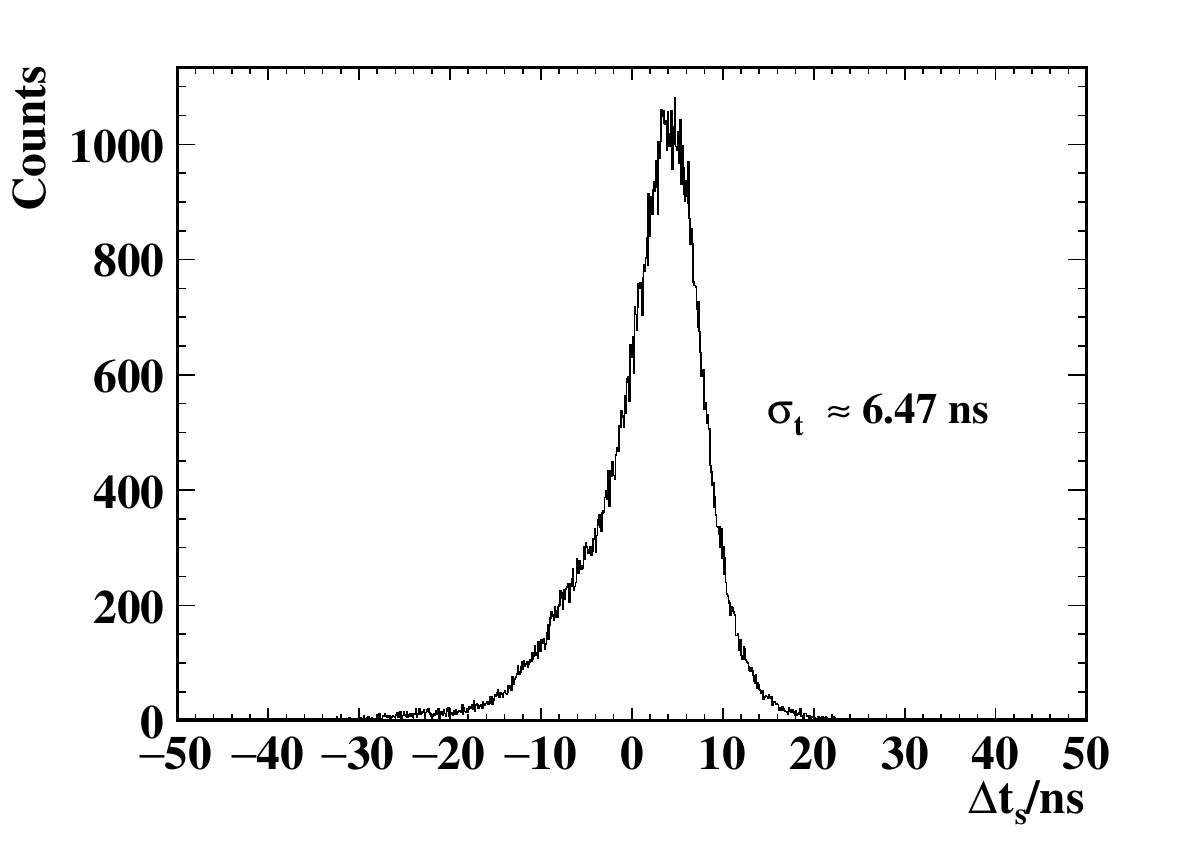}
    \caption{Distribution of corrected timestamp differences. Differences are measured between two boards within the same super-layer for identical muon events after calibration.}
    \label{fig:Time-Cali}
\end{figure}

\paragraph{Time calibration}
System-wide synchronization is based on comparing the arrival times of a common 1 Hz reference pulse across all boards.
For each pulse, each board determines its local arrival time measurement, recorded as a timestamp $\text{TS}_k$, with the index $k$ identifying the pulse sequence number.
Thus, these reference pulses partition the timeline into discrete intervals.
This partition not only enables time calibration based on measurements of the common interval, but also restricts event matching to a limited time window.
The time calibration involves two stages: first determining the board's internal clock frequency within each reference interval ($\text{TS}_k$, $\text{TS}_{k+1}$), then using the frequency to correct the timestamps of muon incident events that occur within that same interval.
Figure~\ref{fig:Time-Cali} presents the typical corrected timestamp difference between two boards in a super-layer, which exhibits a standard deviation of approximately 6.5~ns for the same muon incident. 
This level of precision establishes the essential hardware foundation for accurate event matching in subsequent analysis.

\paragraph{Event matching}

The event reconstruction process is performed in two sequential stages: intra-board and inter-board matching.
In the intra-board stage, the three raw data streams produced by each electronic board ---  high-gain charge, low-gain charge, and timing information ---  are merged into a single coherent data stream. 
This is accomplished by assigning a common trigger identifier that is shared internally across all three measurement channels within the board.
The inter-board stage identifies events from different readout boards that were produced by the same cosmic ray. 
Events are considered coincident if their corrected timestamps differ by less than 100~ns ($\Delta t'<$ 100~ns).

\paragraph{Channel calibration}

\begin{figure}
    \centering
    \begin{subfigure}[b]{0.49\textwidth}
        \centering
        \includegraphics[width=\textwidth]{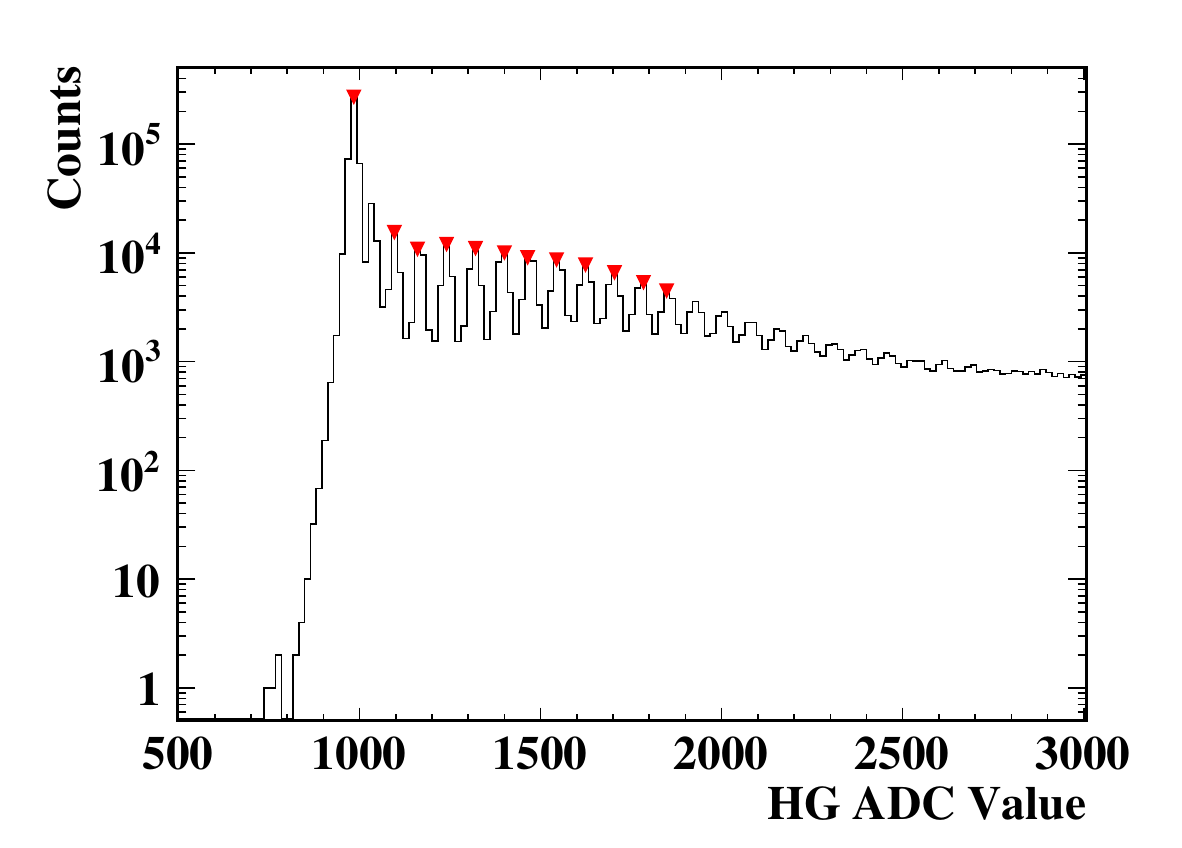}
        \caption{}
    \end{subfigure}
    \begin{subfigure}[b]{0.49\textwidth}
        \centering
        \includegraphics[width=\textwidth]{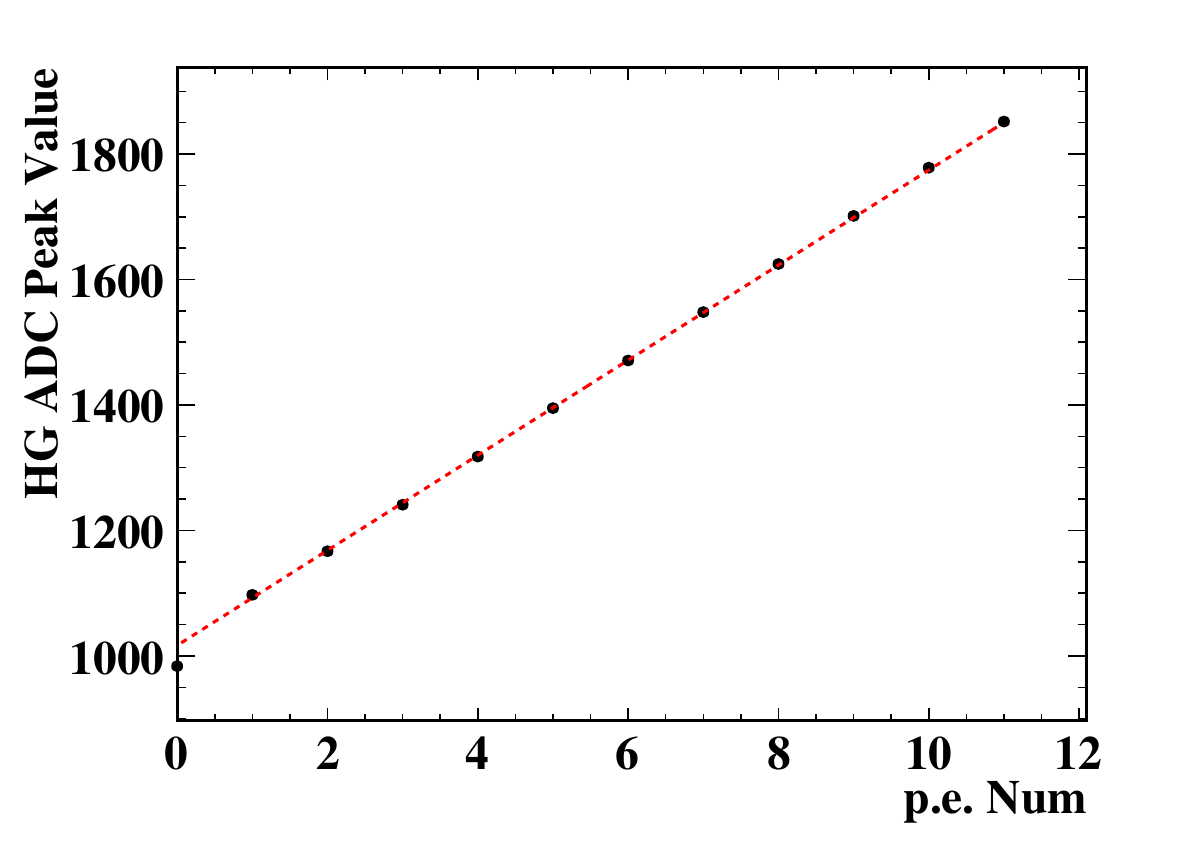}
        \caption{}
    \end{subfigure}\\
    \begin{subfigure}[b]{0.49\textwidth}
        \centering
        \includegraphics[width=\textwidth]{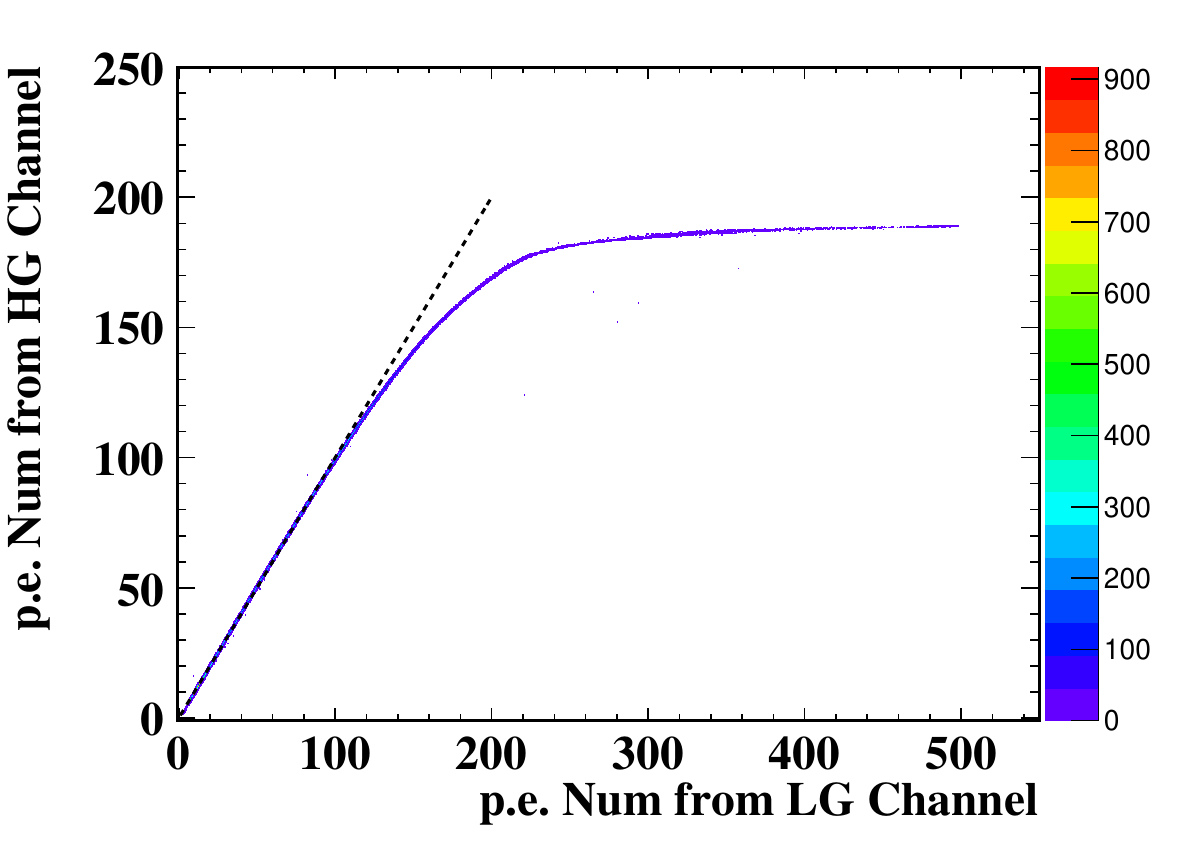}
        \caption{}
    \end{subfigure}
    \begin{subfigure}[b]{0.49\textwidth}
        \centering
        \includegraphics[width=\textwidth]{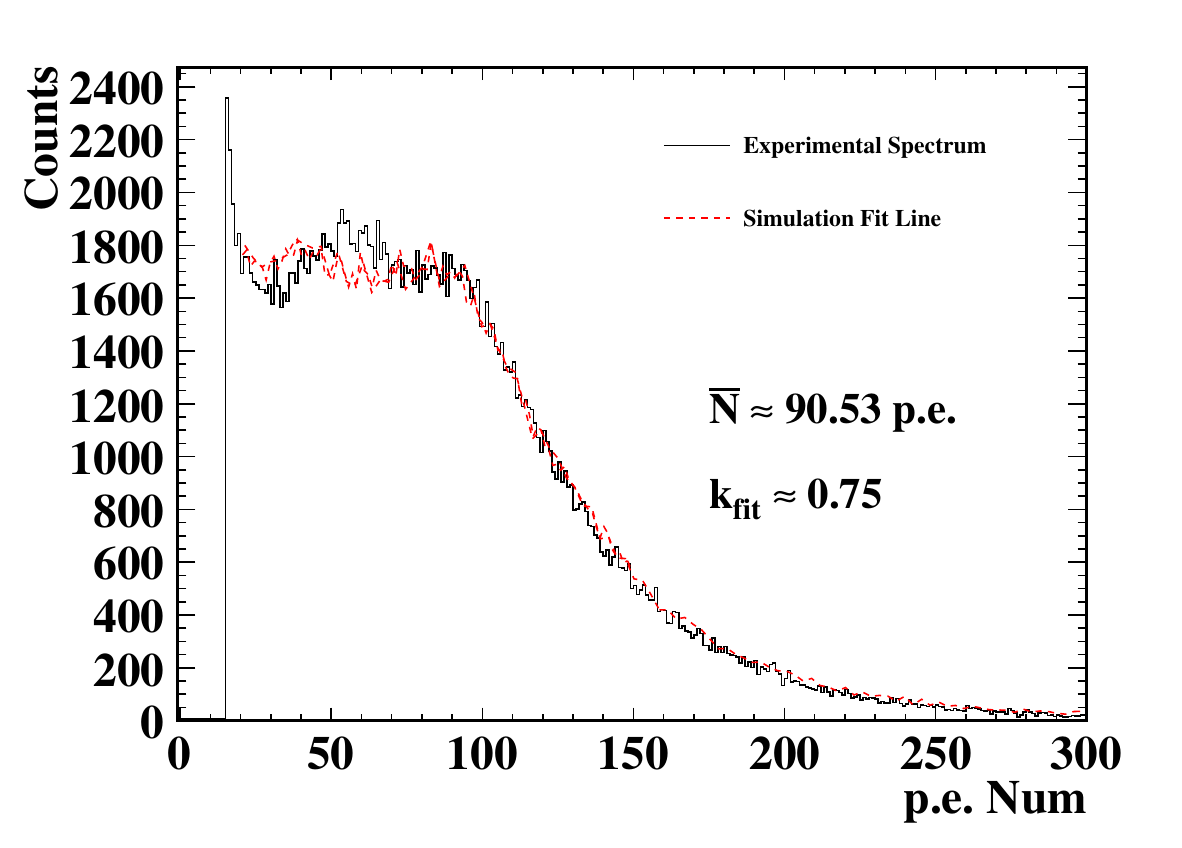}
        \caption{}
    \end{subfigure}
    \caption{Channel Calibration. (a): HG channel spectrum under low intensity. (b): Linear fit to the correlation between HG ADC Values and SiPM photoelectron counts. (c): LG-HG detected photoelectron counts correlation, nonlinearity shown in HG at photoelectron counts > 100. (d): Measured cosmic ray photoelectron spectrum fitted by Geant4 simulation spectrum (red dashed line), with the scale factor $k_\text{fit}$ applied for normalization.}
    \label{fig:Channel-Cali}
\end{figure}

The channel calibration was conducted in two main steps. 
First, the relationship between the HG and LG ADC values and the SiPM photoelectron counts was established.
This correspondence enables the definition of channel-by-channel thresholds and triggers an automatic switch to the LG readout when the HG channel saturates during data analysis.
To establish this correspondence, the distribution of HG ADC values at low photoelectron levels was initially plotted and correlated with SiPM photoelectrons counts, as shown in panels (a) and (b) of figure~\ref{fig:Channel-Cali}.
Subsequently, using high photoelectron intensity data from cosmic ray events, the relationship between LG ADC values and the photoelectron counts derived from HG was plotted, thereby completing HG–LG–photoelectron count calibration, as shown in panel (c).

Second, calibrate the response non-uniformity across different scintillator units.
The non-uniformity arises from two primary sources: differences in electronic components, including amplifiers and ADCs; and variations within the scintillation units themselves, such as the light yield of scintillators and the optical coupling among fibers, scintillators, and SiPMs.
We quantified this non-uniformity by fitting the ADC spectrum from cosmic ray tests to the spectrum from Geant4 simulations, as shown in panel (d) of figure~\ref{fig:Channel-Cali}. 
The derived scale factor $k_\text{fit}$, applied across different scintillators and readout channels, serves this purpose.

\paragraph{Decoding}

The most straightforward decoding approach identifies the four SiPMs with the highest calibrated signal amplitudes and determines the hit position using a pre-defined lookup table presented in table~\ref{tab:encoding}. 
However, this direct decoding scheme is highly susceptible to interference from optical crosstalk between adjacent scintillator strips. 
This vulnerability arises because low photoelectron counts in a SiPM can have two distinct origins: genuine but weak energy deposition from a muon, or spurious signals from crosstalk. 
The algorithm's inability to reliably distinguish between these two cases is the fundamental reason for decoding errors. 
While signals from true weak depositions are valuable for position reconstruction, mistakenly incorporating crosstalk signals can lead to significant errors in the final position determination.

A concrete example illustrates this issue. 
Consider a cosmic-ray event that strikes two adjacent strips, S$_1$ and S$_2$. 
If the energy deposited in S$_2$ is very low, one of its associated SiPMs (e.g., A$_2$ or B$_2$) may fall below the threshold. 
Meanwhile, if crosstalk exists between S$_1$ and S$_0$, spurious signals may appear on the SiPMs corresponding to S$_0$ (A$_0$ and B$_0$).
In such a case, the signal magnitudes typically follow: A$_1$ $\approx$ B$_1$ > A$_2$ $\approx$ B$_2$ $\approx$ A$_0$ $\approx$ B$_0$.
If the decoding algorithm simply selects the four channels with the largest signals, it might incorrectly choose A$_1$, B$_1$, A$_0$, and B$_2$. 
Decoding this selection using table~\ref{tab:encoding} would yield strips S$_4$ and S$_5$, a result that deviates significantly from the true hit position.

To address this issue, we improved the decoding algorithm by exploiting the strong correlation between the two signals generated from the same scintillator strip. 
Using the amplitude relationships among the activated channels, the algorithm can reliably distinguish genuine hit strips from crosstalk-induced signals while maintaining decoding efficiency and positioning precision.
A threshold of approximately 30 photoelectrons is applied to reject crosstalk and noise.
This value is sufficiently high to limit activated SiPMs to no more than four per event, yet low enough to be less than one-quarter of the average total photoelectron count, ensuring that at least two channels from one strip always exceed the threshold.

When four channels exceed the threshold, direct decoding is performed using the lookup table. 
When two or three channels exceed the threshold, the algorithm first decodes the primary hit strip from the SiPMs with the largest signals in the A‑group and B‑group after non‑uniformity calibration.
It then selects the larger of the two adjacent strips as the secondary strip for centroid calculation.
For example, if SiPMs $A_1$ and $B_1$ have the largest signals in the A-group and B-group, respectively, then $S_1$ is identified as the primary strip. 
The signals from the two adjacent strips, $S_0$ ($A_0 + B_0$) and $S_2$ ($A_2 + B_2$), are then calculated.
After comparing magnitudes of S$_0$ and S$_2$, the larger strip is taken as the secondary hit strip for centroid calculation.
If five channels exceed the threshold, the algorithm selects the two largest signals from the A‑group and the two largest from the B‑group, and treats the event as if four channels had been activated.
Single-channel events and events with more than five channels are discarded.

\begin{figure}
    \centering
    \begin{subfigure}[b]{0.49\textwidth}
        \centering
        \includegraphics[width=\textwidth]{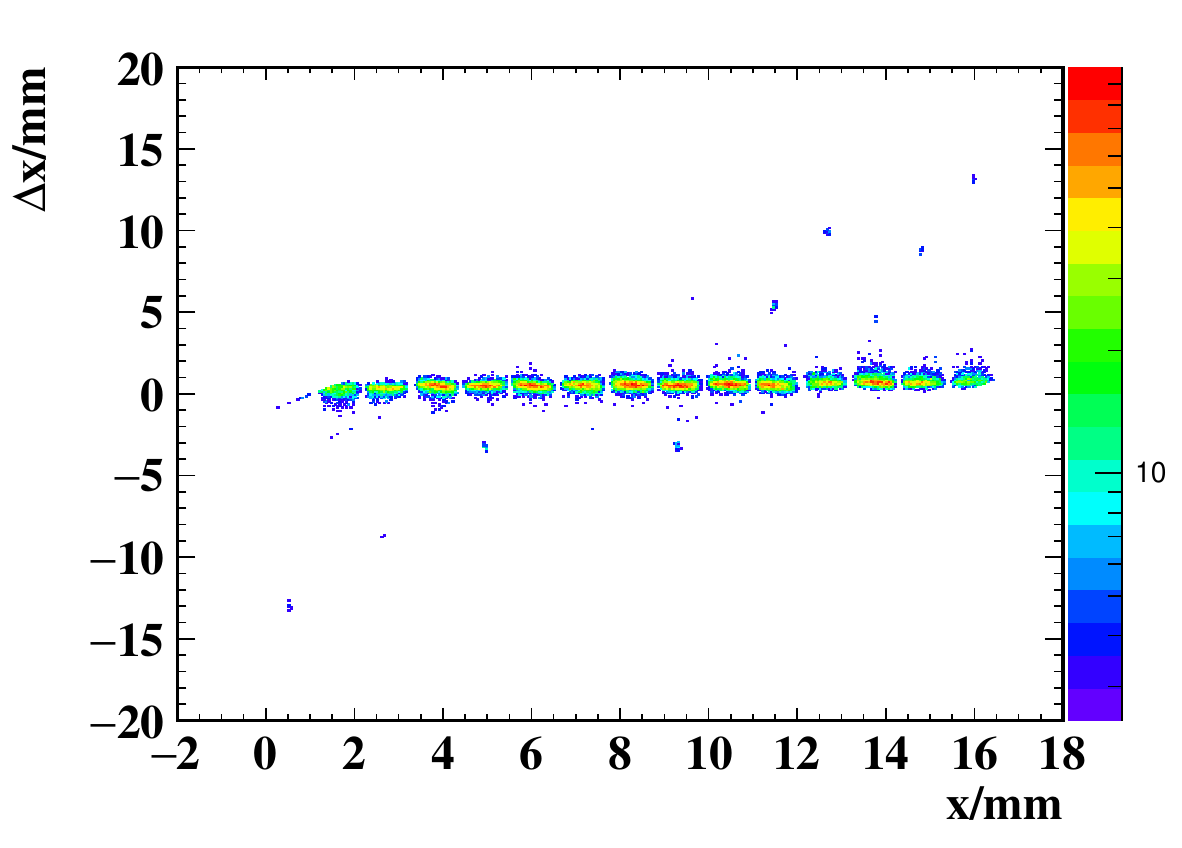}
        \caption{}
    \end{subfigure}
    \begin{subfigure}[b]{0.49\textwidth}
        \centering
        \includegraphics[width=\textwidth]{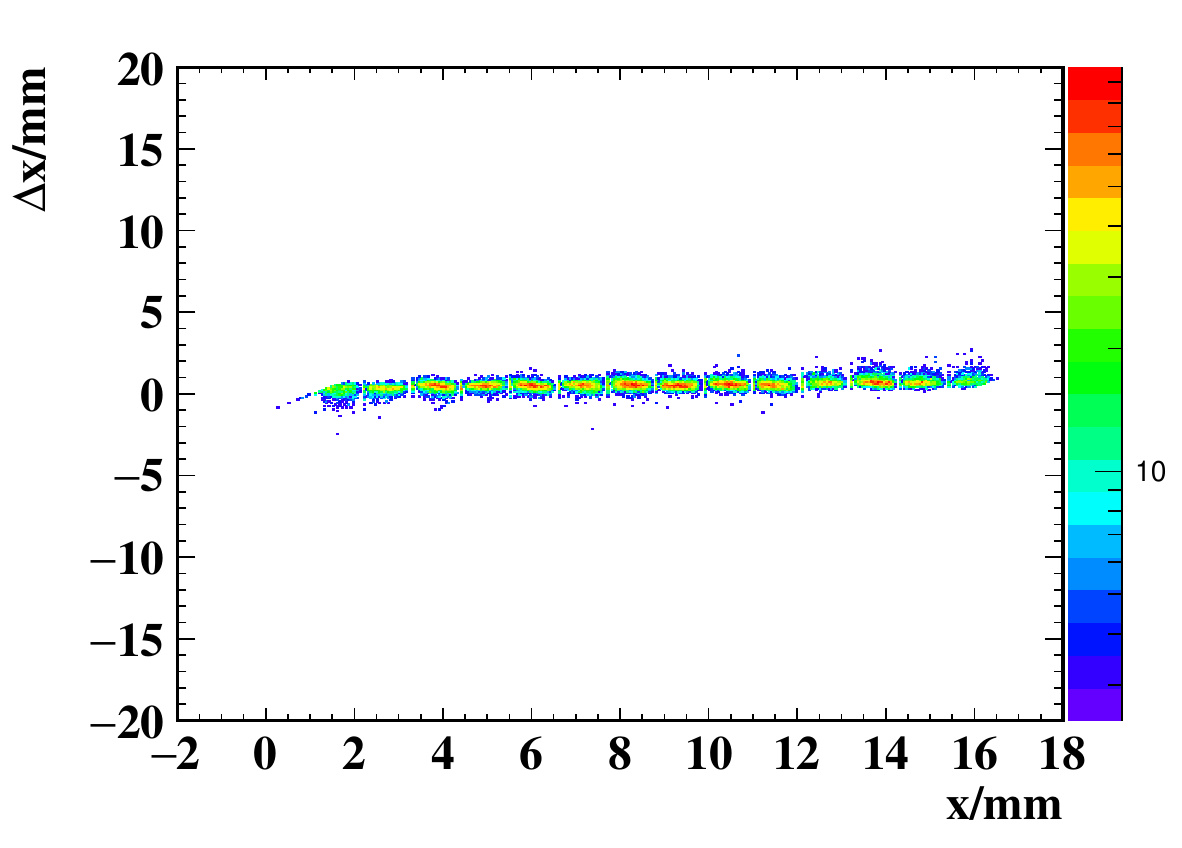}
        \caption{}
    \end{subfigure}
    \begin{subfigure}[b]{0.49\textwidth}
        \centering
        \includegraphics[width=\textwidth]{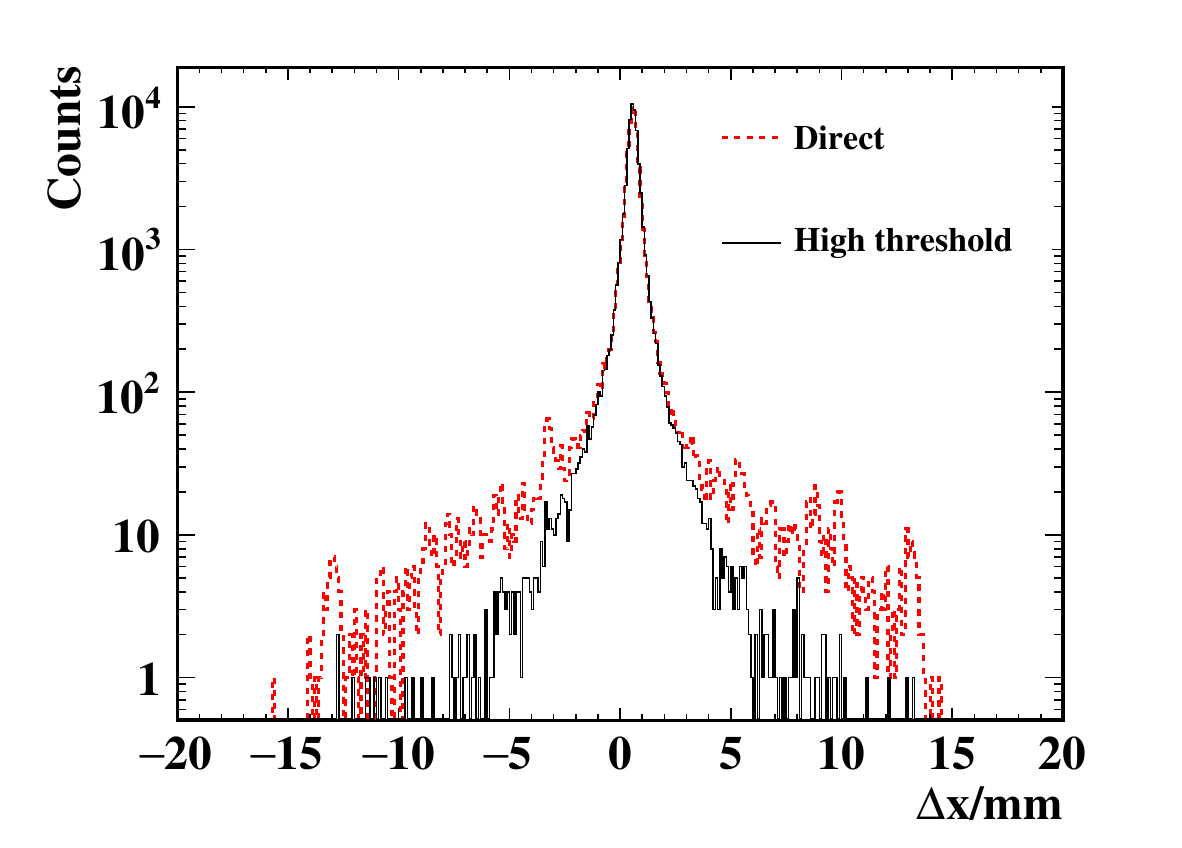}
        \caption{}
    \end{subfigure}
    \caption{Comparison of two decoding schemes. (a, b) Position residual $\Delta x$ as a function of the reference muon incident position $x_\text{ref}$ for the (a) direct (b) new high-threshold methods, respectively. (c) Distribution of the $\Delta x$ for both methods presented on the same plot. }
    \label{fig:Decoding}
\end{figure}

We evaluated the performance of two encoding schemes by integrating the encoding module into a cosmic ray test system. 
Two non-encoding prototype detectors served as a reference, providing triggers and the precise incident position $x_\text{ref}$.
The position residual was defined as $\Delta x=x_\text{dut}-x_\text{ref}$, where $x_\text{dut}$ is the position measured by the module under test.
Figure~\ref{fig:Decoding} compares the $\Delta x$ distributions of the two decoding schemes.
Panel (a), which plots $\Delta x$ against $x_\text{ref}$ for the direct decoding method, shows numerous large outliers indicative of misidentifications. 
These outliers are absent in the corresponding plot for the high-threshold method in panel (b). 
This demonstrates that the new approach effectively eliminates misidentifications, leading to higher spatial resolution —-- an improvement further corroborated by the tighter $\Delta x$ distribution of the new method in panel (c).

\paragraph{Position reconstruction}

As introduced in section~\ref{sec:simu-detector-sigmax}, the charge centroid method relies solely on the signal ratio between adjacent strips.
Therefore, each detection plane independently provides a preliminary hit position $x_\mathrm{p}$ using this method.
The incident angle $\theta_x$ of muons is then reconstructed from the coarse positions $x_\mathrm{p}$ obtained from multiple detection planes.
This angle is subsequently applied as a geometric correction, yielding the refined hit position $x_\mathrm{c}$.
A gap correction, derived from simulation and parameterized as the linear function $x_\text{g} = x_\mathrm{c} - k \cdot \lambda$ (with $k = 1.01$ mm for our detector geometry, as shown in Figure~\ref{fig:X-EtaLine}), is then applied to obtain the final position $x_\text{g}$.
This three-step procedure — centroid positioning, angle correction, and gap correction — thus produces an accurately reconstructed hit position.

\subsection{Detector performance}
\label{sec:Detector-performance}

With the initial processing of the raw signals completed, we can now proceed to assess the performance of the multi-layer detection system.
Its key parameters, including detection efficiency, temporal and spatial resolution, was characterized using a three-super-layer configuration under different coincidence logic.

\begin{figure}
    \centering
    \begin{subfigure}[b]{0.6\textwidth}
        \centering
        \includegraphics[width=\textwidth]{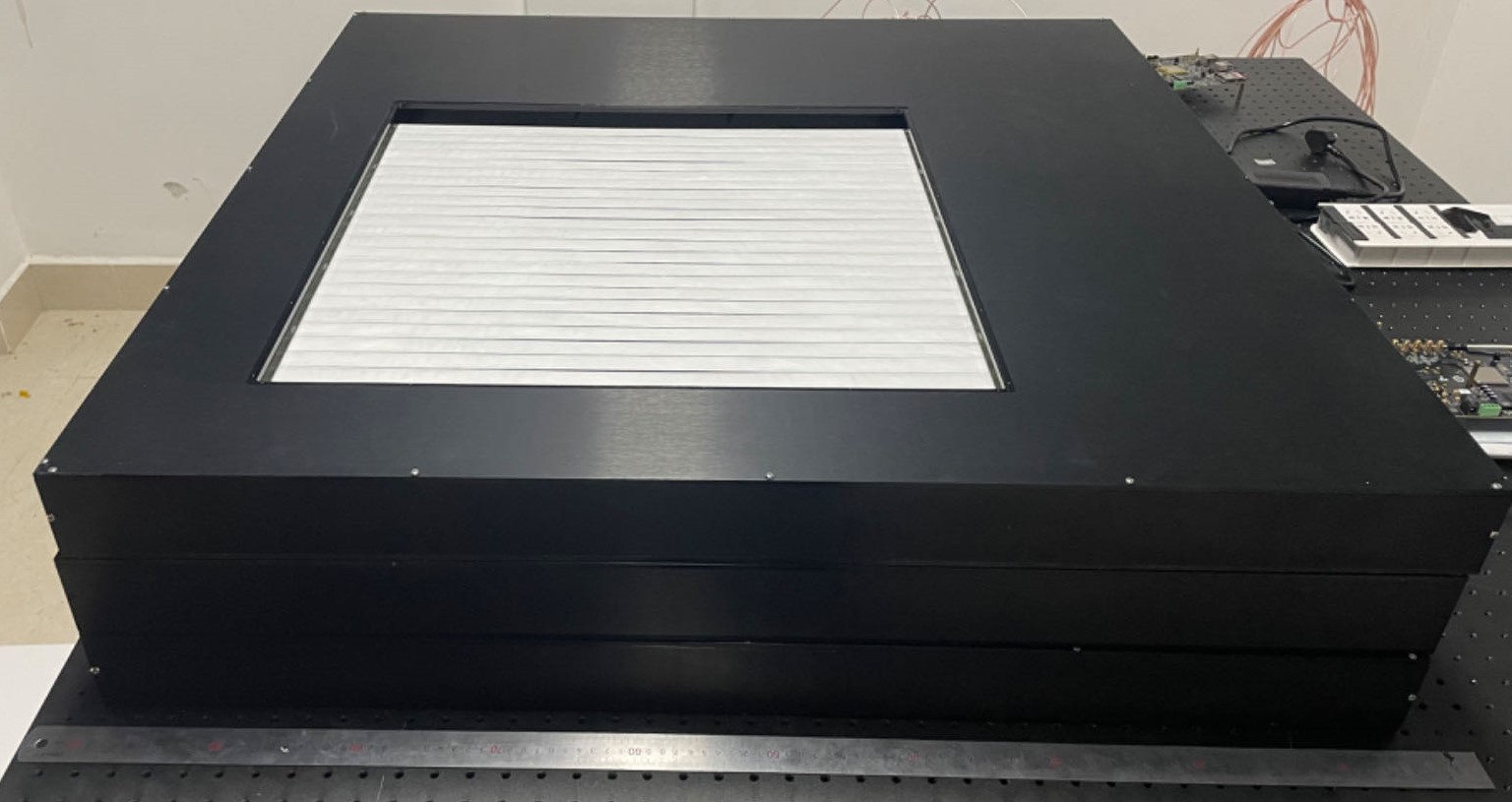}
        \caption{}
    \end{subfigure}\\
    \begin{subfigure}[b]{0.8\textwidth}
        \centering
        \includegraphics[width=\textwidth]{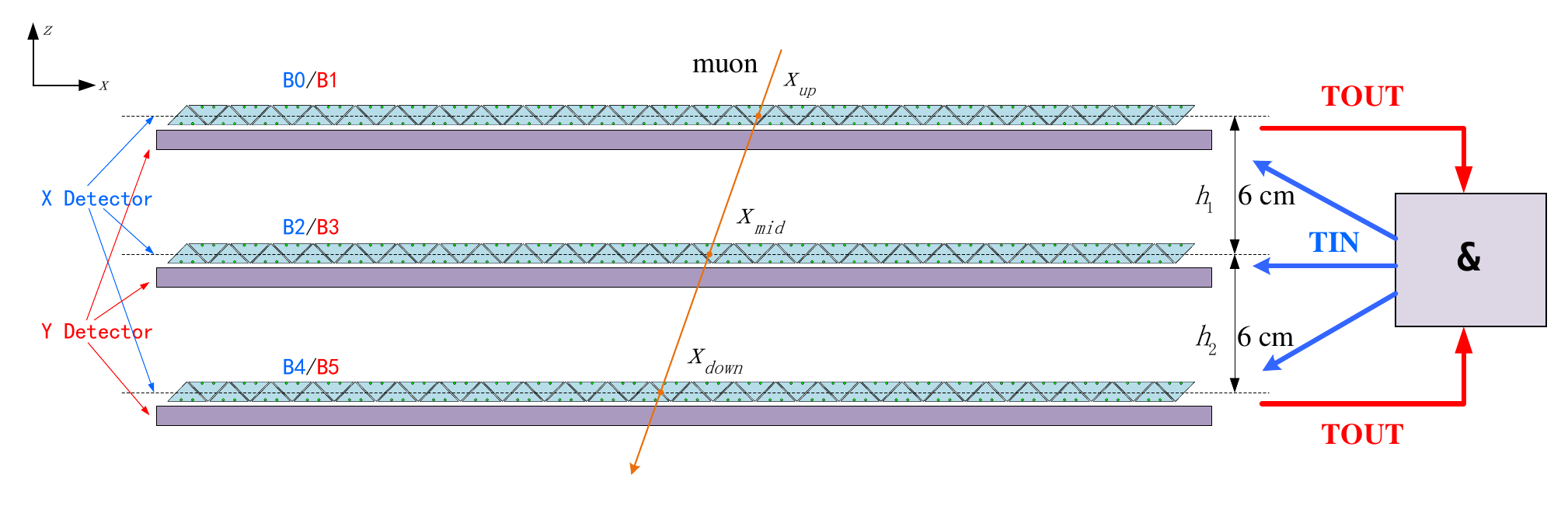}
        \caption{}
        \label{fig:cosmic-ray-efficiency}
    \end{subfigure}\\
    \begin{subfigure}[b]{0.8\textwidth}
        \centering
        \includegraphics[width=\textwidth]{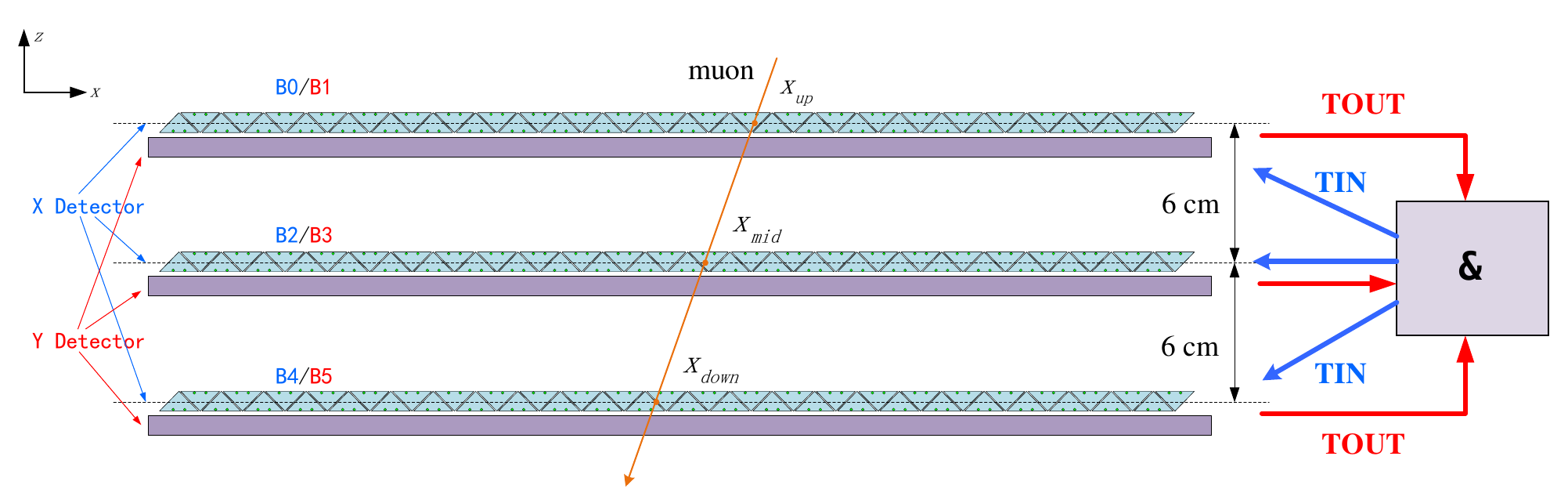}
        \caption{}
        \label{fig:cosmic-ray-spatial}
    \end{subfigure}
    \caption{Configuration for super‑layer performance tests. (a) Photograph showing the three‑layer stack. (b) Connection and coincidence logic for the efficiency test. (c) Connection and coincidence logic for the spatial/temporal resolution test.}
    \label{fig:Cosmic-ray-test}
\end{figure}

\paragraph{Detection efficiency}
For the measurement of detection efficiency, the coincidence signal from the top and bottom layers was used as the trigger input for the middle layer, as shown in figure~\ref{fig:cosmic-ray-efficiency}. 
Thus, the signals recorded in the middle layer represented a three-fold coincidence among all three super-layers, while the top and bottom layers each registered a two-fold coincidence signal.

Let $N_\text{A}$ be the count of coincidence events between the top and bottom layers, and $N_\text{B}$ the total count of three-fold coincidences (top, middle, bottom). The detection efficiency of the middle layer is then $\epsilon_\text{mid}=N_\text{B}/N_\text{A}$.
During a 13-hour measurement period, the recorded counts were  $N_\text{A}$ = 1,195,654 and $N_\text{B}$ = 1,166,606. 
The resulting total detection efficiency for the middle layer $\epsilon_\text{mid}$ was determined to be 97.57\%.
The detection efficiency reported here is a comprehensive value that incorporates the intrinsic detection efficiency of the detectors, the data transmission efficiency of the readout electronics, and the efficiency of the multi-layer event matching process.

\paragraph{Temporal resolution}
The measurement of both temporal and spatial resolution requires events from a three-fold coincidence across the detector layers.
To achieve this, the logic connections between the electronic boards are implemented as illustrated in figure~\ref{fig:cosmic-ray-spatial}.
The TOUT signals from the three layers are combined in coincidence to generate a global validation signal for the entire stack.

\begin{figure}
    \centering
    \includegraphics[width=0.5\linewidth]{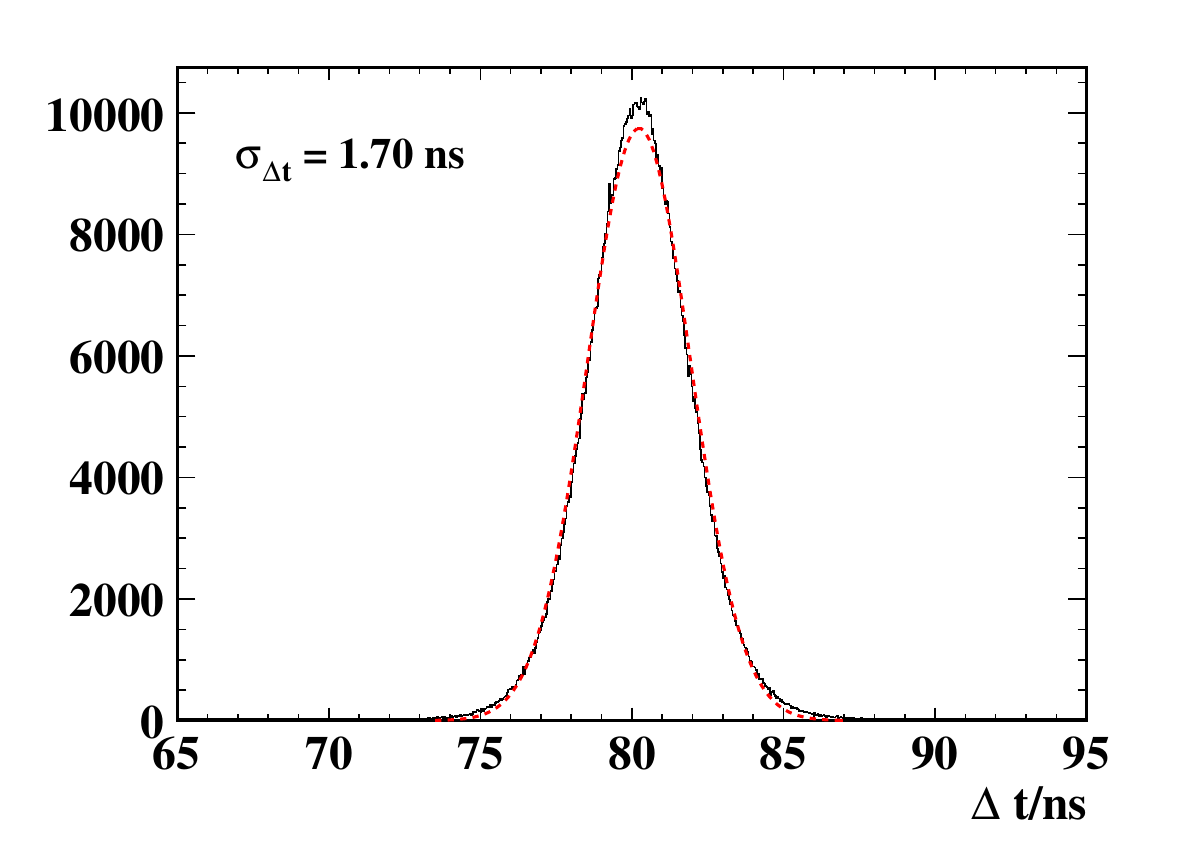}
    \caption{Distribution of the calibrated time difference between the top and bottom detection planes.}
    \label{fig:Sigma-t}
\end{figure}

In order to get precise timing, a final calibration step is necessary after the initial time correction and event matching.
This step incorporates corrections for three key factors: inter-channel timing variations within the front-end electronics, the propagation delay of photons along the WLS fibers (which depends on the hit position), and the time of flight of the muon through the entire detector system.
As shown in the figure~\ref{fig:Sigma-t}, the final time resolution, characterized by the distribution of time differences between the top and bottom detection layers for identical events, $\Delta t=t_\text{top}-t_\text{bottom}$, achieves a $\sigma_{\Delta t}$ of 1.7 ns.
The single detection plane timing resolution is therefore inferred to be $\sigma_t=1.2$ ns.

\paragraph{Spatial resolution}
The same logic configuration from the temporal resolution test was used to measure the detector's spatial resolution.
The test system determines the trajectory and incident angle $\theta$ of muons.
Therefore, equation~\ref{equ:centroid} provides the corrected impact position $x_\text{c}$ at the center of the detection plane.
In addition, the hit position is further corrected for the inter-strip gaps, as determined from Geant4 simulations, resulting in the gap-corrected position $x_\text{g}$.

\begin{figure}
    \centering
    \begin{subfigure}[b]{0.49\textwidth}
        \centering
        \includegraphics[width=\textwidth]{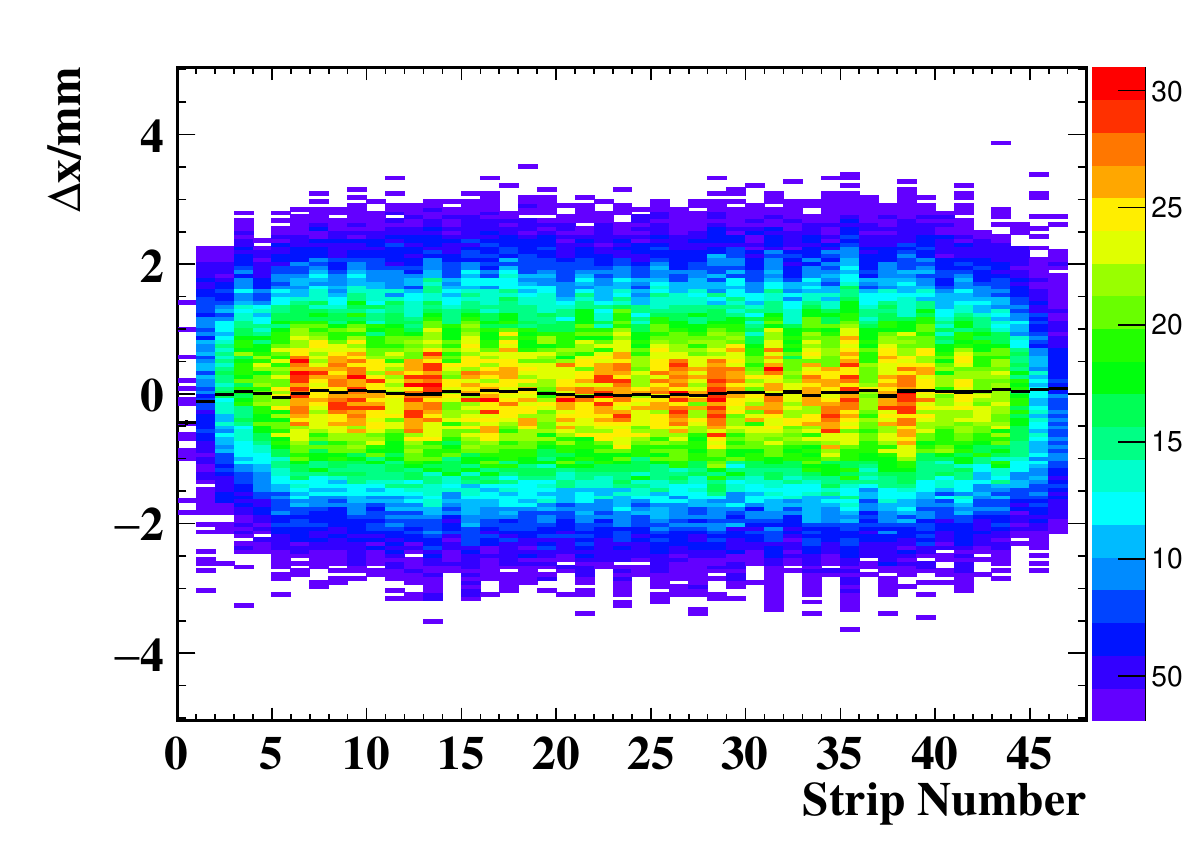}
        \caption{}
    \end{subfigure}
    \begin{subfigure}[b]{0.49\textwidth}
        \centering
        \includegraphics[width=\textwidth]{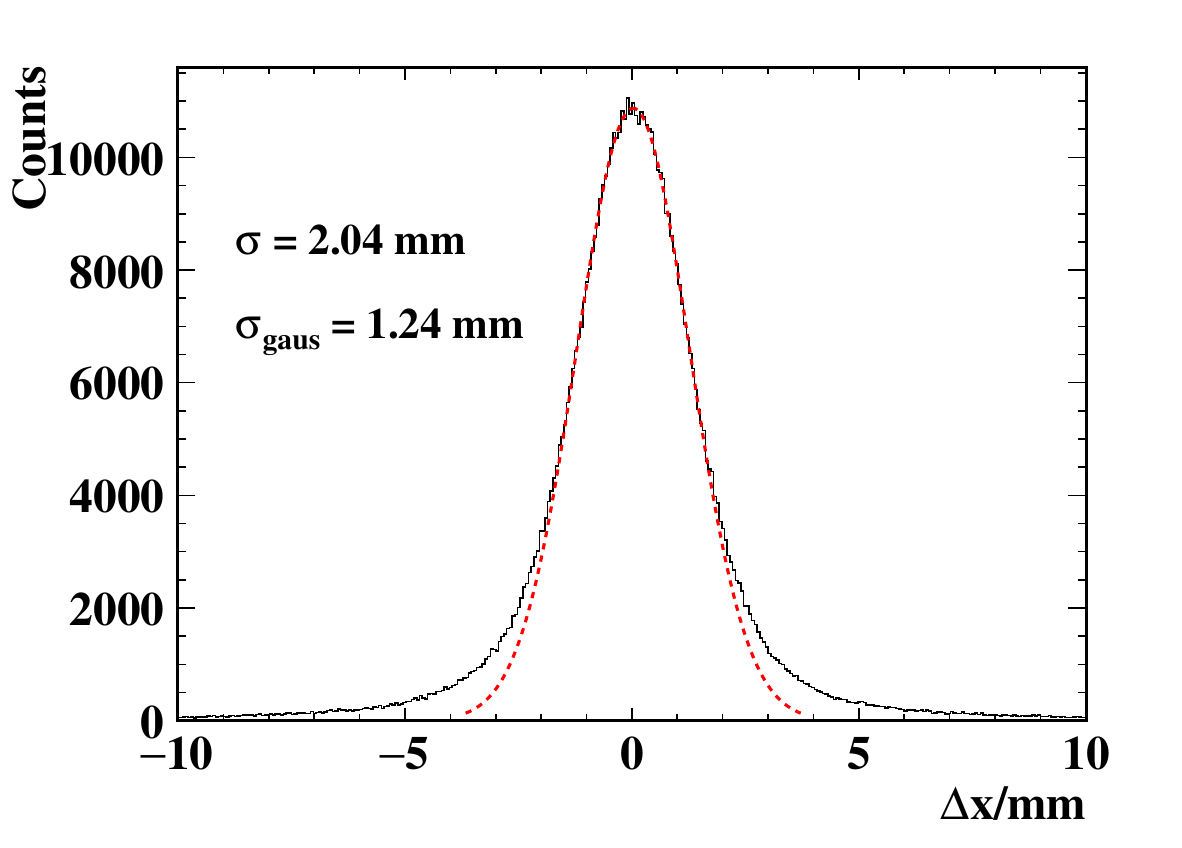}
        \caption{}
    \end{subfigure}
    \caption{Position residual distribution of the middle layer detector. (a): Residual distribution as a function of the scintillator strip index. (b): Residual distribution of all strips.}
    \label{fig:Spatial-Deviation}
\end{figure}

The final residual distribution for the middle-layer position measurement, after incorporating all the aforementioned corrections, is shown in figure~\ref{fig:Spatial-Deviation}.
The standard deviation of the Gaussian peak in the figure is 1.24 mm.
Assuming that each detector layer has an identical spatial resolution of $\sigma_x$, and applying the error propagation formula, the resulting residual $\sigma_{\Delta x}$ for the middle layer is given by:
\begin{eqnarray}
    x_\text{mid, fit} &=& \frac{h_2\cdot x_\text{t}+h_1\cdot x_\text{b}}{h_1+h_2}, \ \Delta x=x_\text{mid,meas}-x_\text{mid,fit},\nonumber\\
    \sigma_{\Delta x} &=& \sqrt{\sigma_x^2\left(\frac{h_1}{h_1+h_2}\right)^2 + \sigma_x^2\left(\frac{h_2}{h_1+h_2}\right)^2 + \sigma_x^2 } ,\nonumber \\
                    &=& \frac{\sqrt{2h_1^2+2h_1h_2+2h_2^2}}{h_1+h_2}\sigma_x = 1.22 \ \sigma_x ,
\end{eqnarray}
where $h_1$ and $h_2$ are the distances between the two detector layers, as labeled in figure~\ref{fig:cosmic-ray-spatial}, $x_\text{t}$ and $x_\text{b}$ denote the position measurements from the top and bottom detector layers, respectively.
Therefore, the measured spatial resolution of the detector is 1.0~mm.
Given that the pitch (center-to-center distance between scintillator bars) is 11~mm, the normalized resolution $\sigma_x$/p is calculated to be 0.09.

Table~\ref{tab:scintillator_comparison} quantitatively compares key performance parameters between our MST system with those of representative scintillator- and gas-based detectors. 
While drift tubes achieve an even smaller $\sigma_x$/p, this requires high-precision timing measurements.
Other gas detectors (e.g., GEM, Micromegas, MRPC) exhibit $\sigma_x$/p values comparable to triangular scintillator detectors ($\sim$0.2) , as seen in systems such as CRIPT, MURAVES, and LUMIS.
In contrast, our detector achieves a significantly lower $\sigma_x$/p of 0.09, demonstrating a clear performance advantage. 
This enhancement is attributed to concurrent improvements in photon collection efficiency, position reconstruction algorithms, and electronic noise suppression.

\begin{table}[htbp]
    \centering
    \caption{Quantitative comparison of key performance parameters between the proposed MST system and representative existing muon imaging systems.}
    \label{tab:scintillator_comparison}
    \begin{tabular}{lcccccc}
        \toprule
        \makecell[l]{Project or \\detector name}  & \makecell{Type}  & \makecell{Detector \\configuration}  & \makecell{Detector \\area (m$^2$)} & \makecell{$\sigma_x$\\ (mm)} & \makecell{Pitch $p$\\  (mm)} & $\sigma_x/p$ \\
        \midrule
        LMT~\cite{MORRIS15102008} &\multirow{4}*{\makecell{Gaseous\\ detector}} & Drift tube   & $3.65\times3.65$ & 0.17 & 50.0 & $3.4\times10^{-3}$\\
        FIT-GEM~\cite{GNANVO201116} & & GEM  & $0.30 \times 0.30$ & 0.13 & 0.40 & 0.33\\
        USTC-Micromegas~\cite{9658560} & & Micromegas  & $0.15 \times 0.15$ & 0.085 & 0.400 & 0.21\\
        TUMUTY~\cite{panExperimentalValidationMaterial2019} & & MRPC  & $0.74 \times 0.74$ & 0.60 & 2.54 & 0.24\\
        
        \addlinespace[0.5em]
        CRIPT~\cite{anghelPlasticScintillatorbasedMuon2015}& \multirow{5}{*}{\makecell{Scintillating \\detector}}  & \makecell{Triangle, single fiber}  & $2.0 \times 2.0$ & 2.5 & 16.5 & 0.15 \\
        MURAVES ~\cite{saracinoMURAVESMuonTelescope2017f} & & \makecell{Triangle, single fiber}  & $1.0 \times 1.0$ &  3.0 & 16.5 &  0.18 \\
        Muon Portal~\cite{RIGGI201816} & & \makecell{Square, dual fibers} &  $1.0 \times 1.0 $ & 3.2 & 10.0 & 0.32 \\
        LUMIS~\cite{luoDevelopmentCommissioningCompact2022} & & \makecell{Triangle, no fiber}  & $0.48 \times 0.48$ & 2.5 & 15.0 & 0.17 \\
        This work & & \makecell{Triangle, dual fibers}  &  $0.53 \times 0.53$ & 1.0 & 11.0 & 0.09 \\
        \bottomrule
    \end{tabular}
\end{table}

\section{Imaging performance of the complete MST system}

The complete MST system consists of an upper tracker and a lower tracker, comprising four super layers in total. 
These super layers collectively contain 8 detection planes, 24 encoding modules, 384 scintillator strips, 768 WLS fibers, 192 SiPMs, and 8 readout electronic boards.
During detector assembly, a quality control system was established to monitor key parameters such as scintillator light yield and fiber transmission efficiency, and to calibrate SiPM temperature compensation and the gain of each electronic channel. 
These efforts ensure the reliable operation of the imaging system.

\begin{figure}
    \centering
    \begin{subfigure}[b]{0.4\textwidth}
        \centering
        \includegraphics[width=\textwidth]{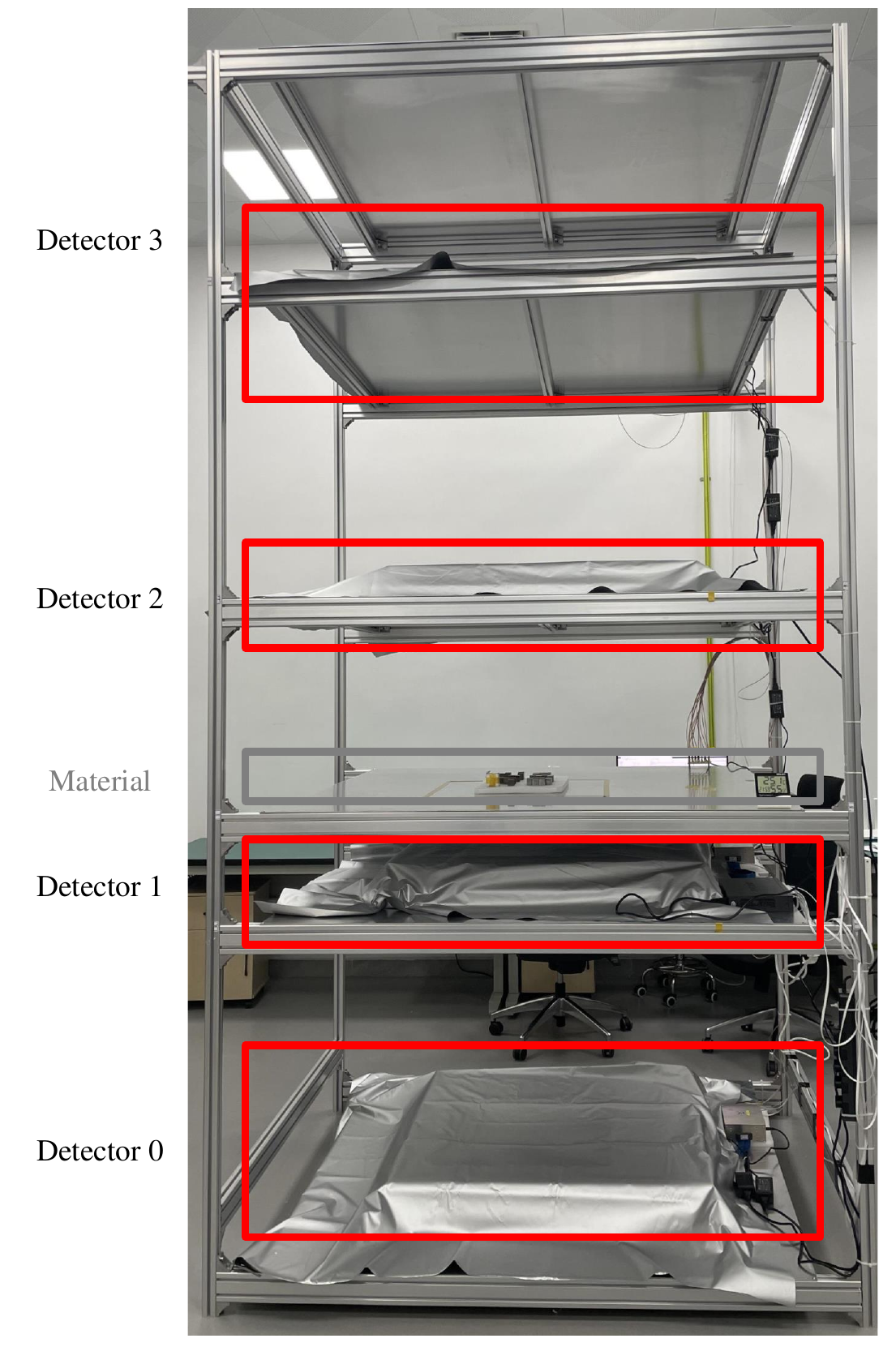}
    \end{subfigure}
    \begin{subfigure}[b]{0.55\textwidth}
        \centering
        \includegraphics[width=\textwidth]{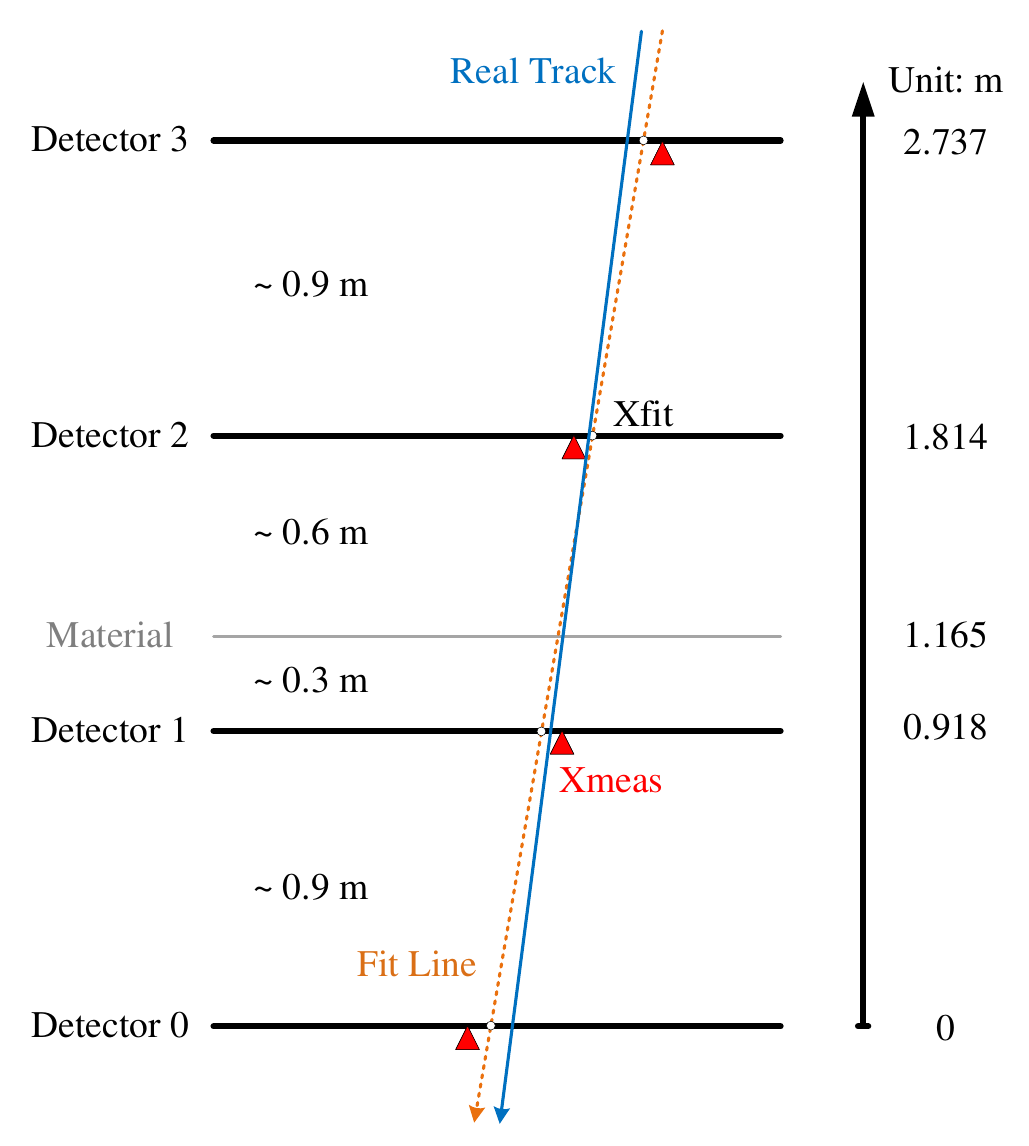}
    \end{subfigure}
    \caption{Muon scattering tomography system. Additionally serves as a cosmic ray telescope, providing cosmic ray trigger and track information.}
    \label{fig:MST-System}
\end{figure}

After assembling the four super-layers, we arranged them at different heights, thus finalizing the construction of the MST system, as shown in figure~\ref{fig:MST-System}.
The four detector layers are arranged with a pairwise separation of around 90 cm. 
The spacing between the two central layers is sufficient to accommodate other detectors under test (DUT).
This configuration allows the system to perform scattering tomography intrinsically, while also enabling it to function as a cosmic ray telescope that provides muon triggers and track information for the DUTs.

\begin{figure}
    \centering
    \begin{subfigure}[b]{0.99\textwidth}
        \includegraphics[width=\textwidth]{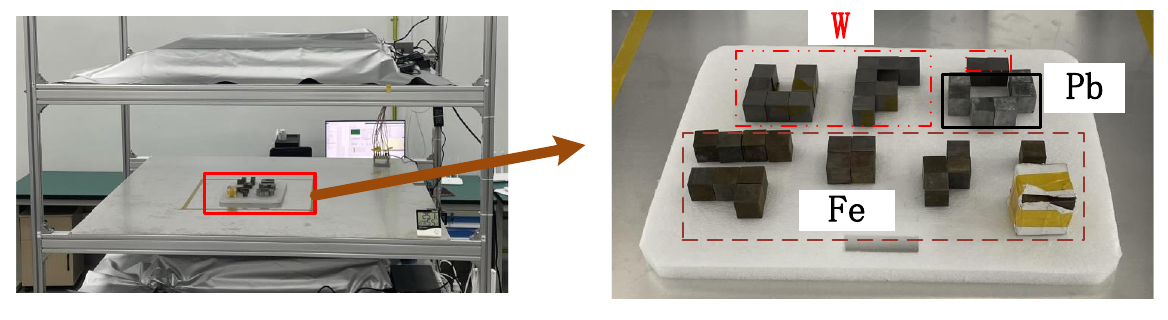}
        \caption{}
        \label{fig:Imaging-Targets-Real}
    \end{subfigure}\\
    \begin{subfigure}[b]{0.99\textwidth}
        \centering
        \includegraphics[width=0.55\textwidth]{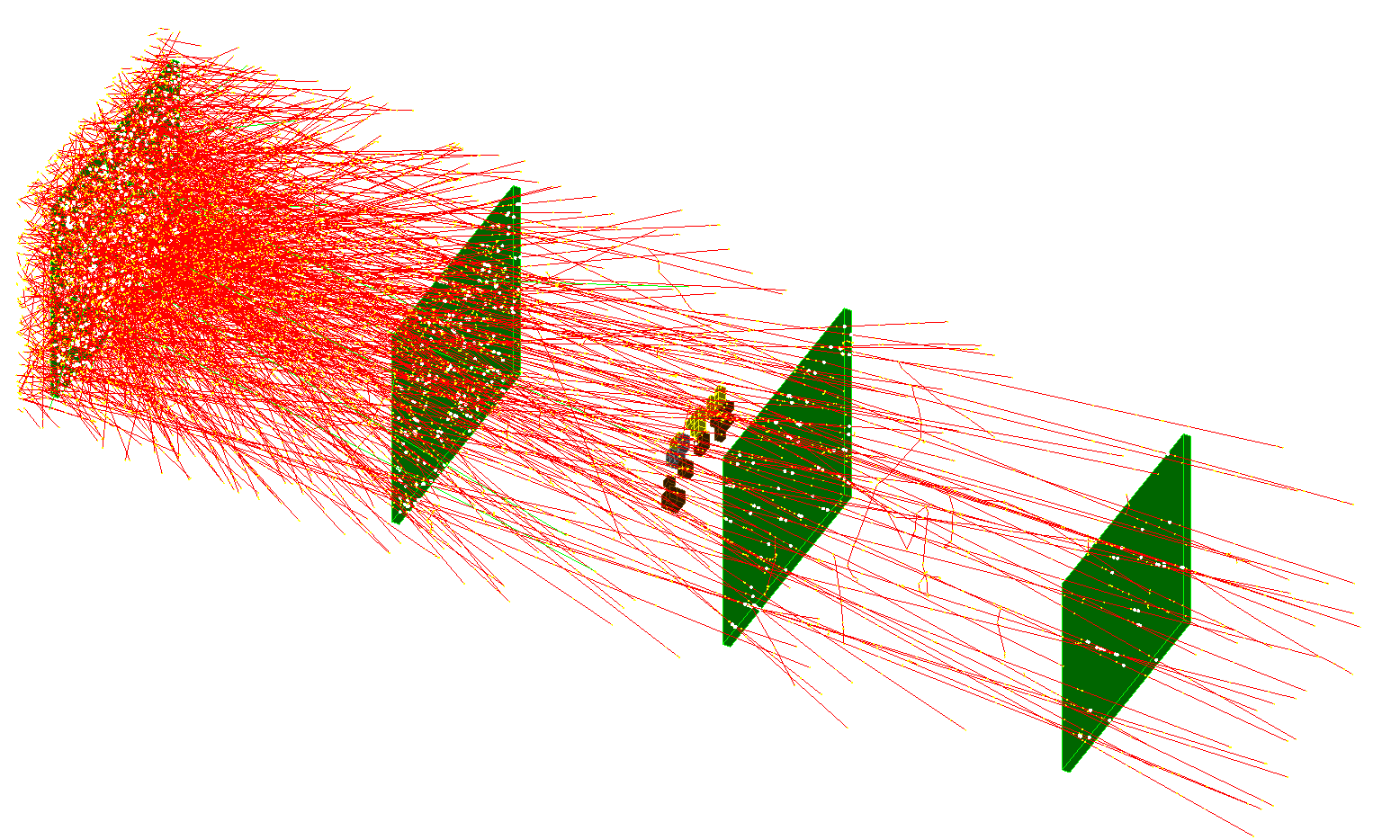}
        \includegraphics[width=0.44\textwidth]{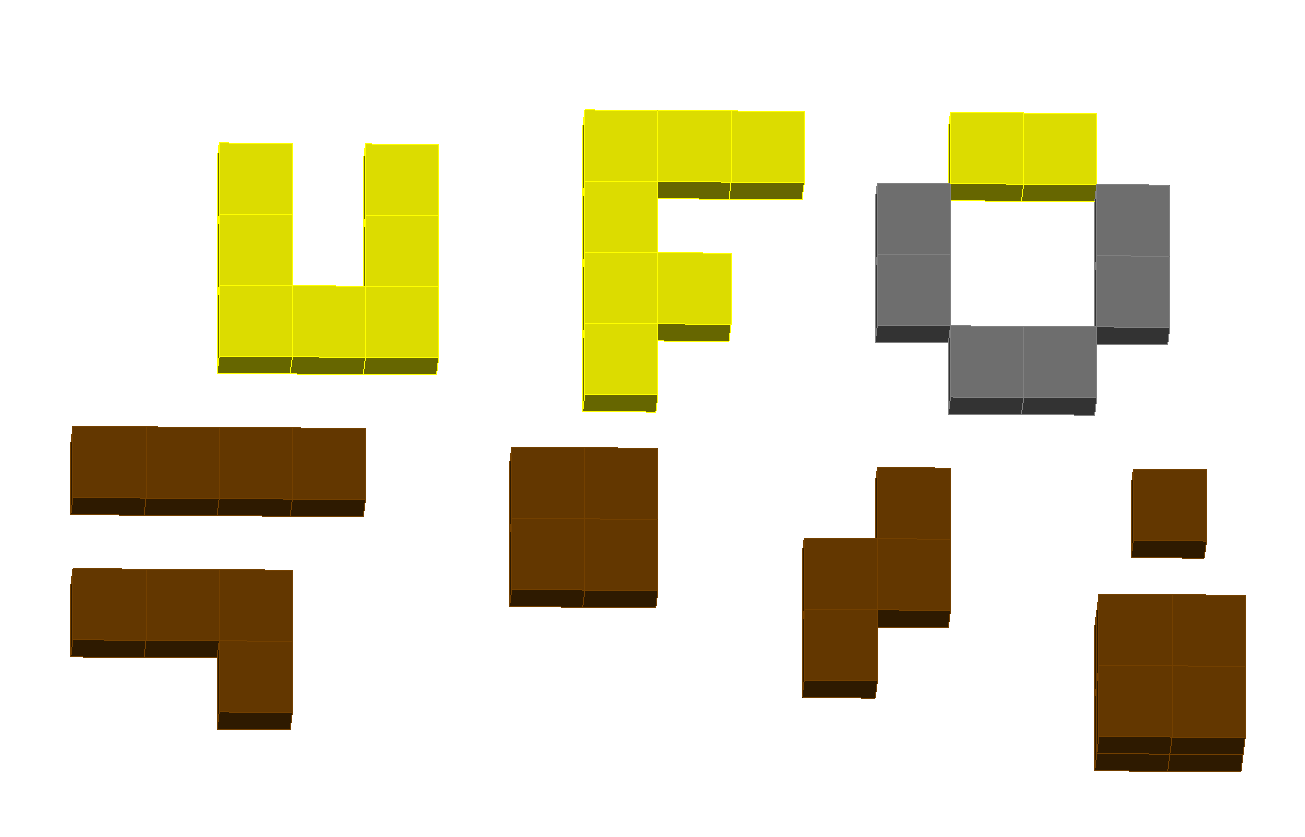}
        \caption{}
        \label{fig:System-Imaging-Simu-Setup}
    \end{subfigure}
    \caption{Imaging setup of the MST system and test objects. (a) Experimental configuration. The letters "UFO" are composed of tungsten and lead blocks, while the Tetris-shaped pattern below is made of iron. Each individual block measures $2 \times 2 \times 2 \text{ cm}^3$. (b) Geant4 simulation setup of the MST system. The left panel shows the relative arrangement of the detection planes and test objects, and the right panel shows the actual placement of the test objects, which are identical to the experimental configuration.}
    \label{fig:Imaging-Targets}
\end{figure}

The imaging performance of the complete MST system was tested and compared against Geant4 simulations.
Within the imaging volume, 2~cm cubic blocks of tungsten, lead, and iron were placed in distinct patterns, as shown in figure~\ref{fig:Imaging-Targets-Real}. 
Data acquisition in the experiment spanned approximately 10 days, accumulating about $2.5\times 10^5$ events for subsequent analysis.
Corresponding Geant4 simulations were performed under identical conditions. 
The modeling of the MST system and target setup in Geant4 is illustrated in Figure~\ref{fig:System-Imaging-Simu-Setup}. 
Muon sampling (energy, angle, position) followed the same procedures described earlier, while detector response was modeled by sampling from the gap-corrected residual distributions, as shown in figure~\ref{fig:CoatCorLine}.
In the Geant4 simulation, the same number of events that passed through all four detection planes was selected for PoCA imaging.

\begin{figure}
    \centering
    \begin{subfigure}[b]{0.49\textwidth}
        \centering
        \includegraphics[width=\textwidth]{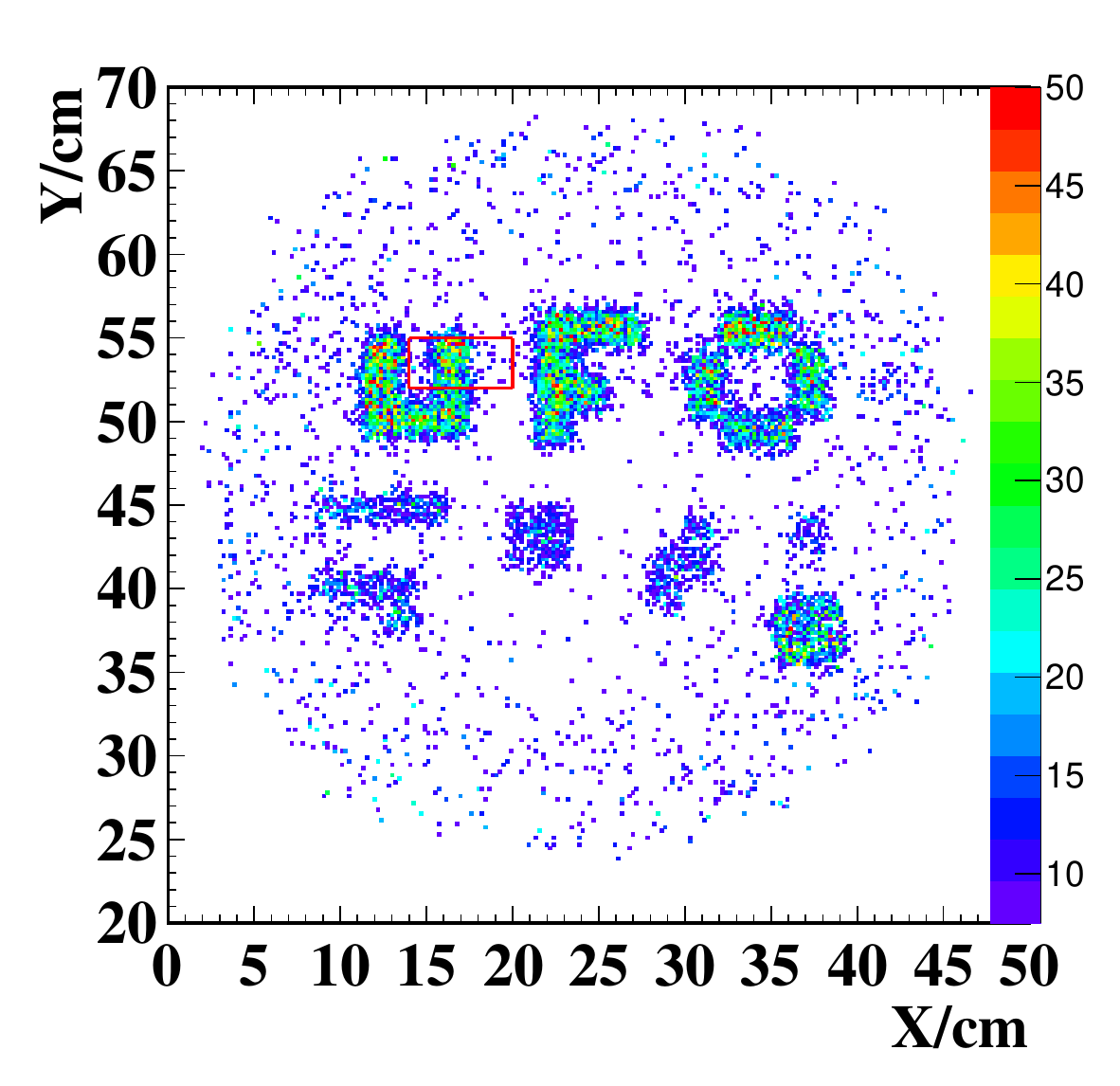}
        \caption{}
        \label{fig:Real-image-2D}
    \end{subfigure}
    \begin{subfigure}[b]{0.49\textwidth}
        \centering
        \includegraphics[width=\textwidth]{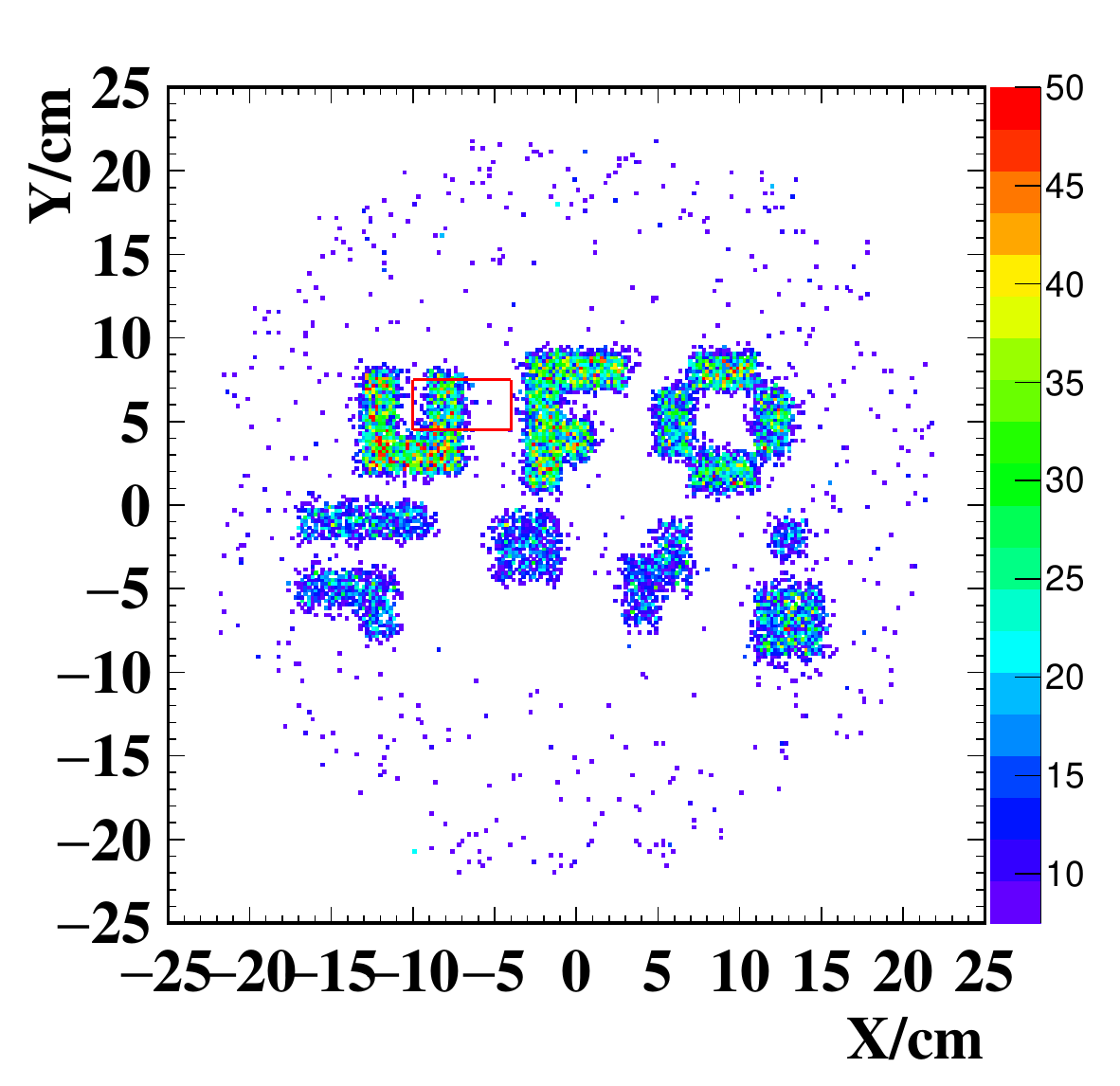}
        \caption{}
        \label{fig:Simu-image-2D}
    \end{subfigure}\\
    \begin{subfigure}[b]{0.49\textwidth}
        \centering
        \includegraphics[width=\textwidth]{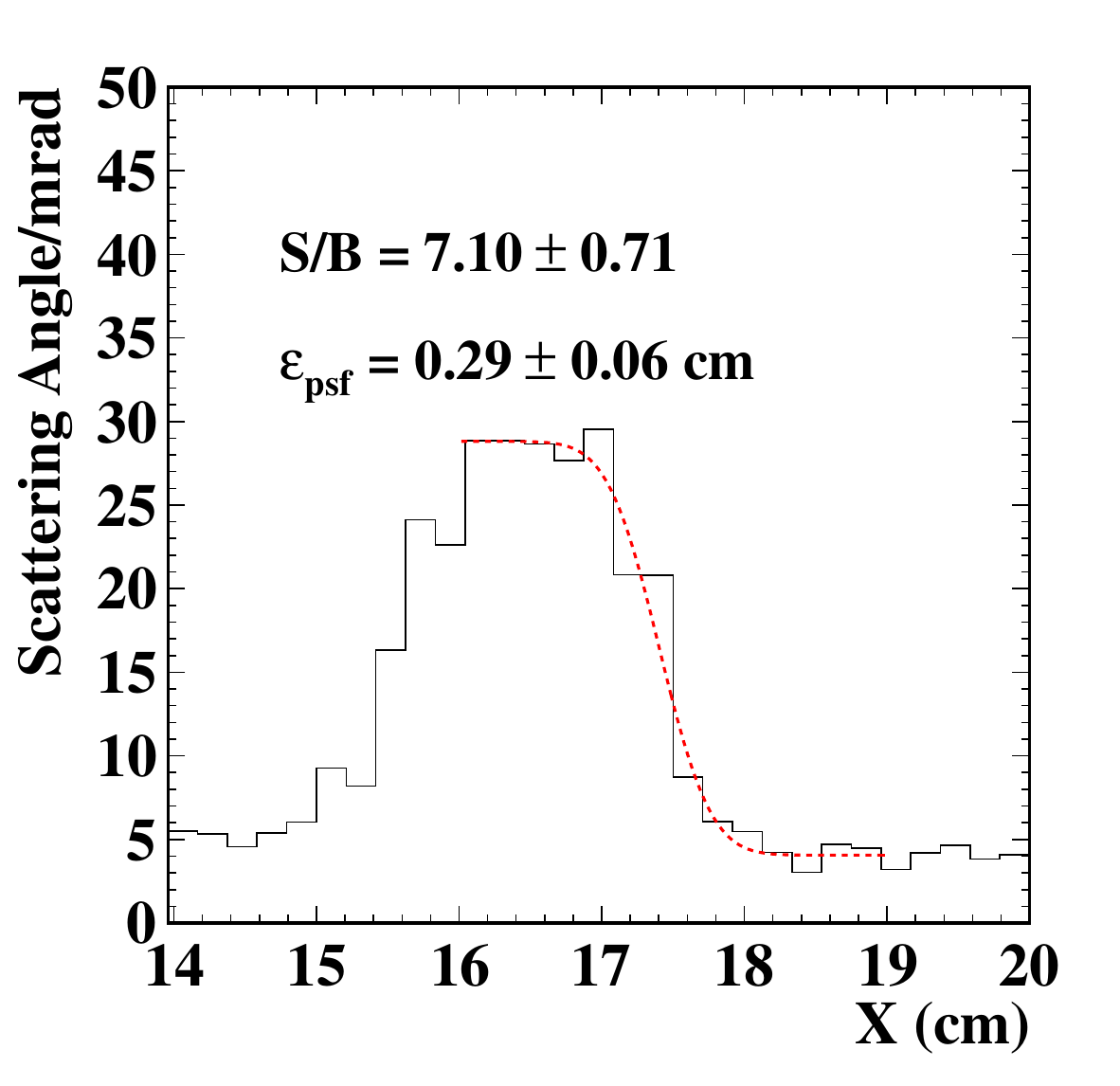}
        \caption{}
        \label{fig:Real-image-Pro}
    \end{subfigure}
    \begin{subfigure}[b]{0.49\textwidth}
        \centering
        \includegraphics[width=\textwidth]{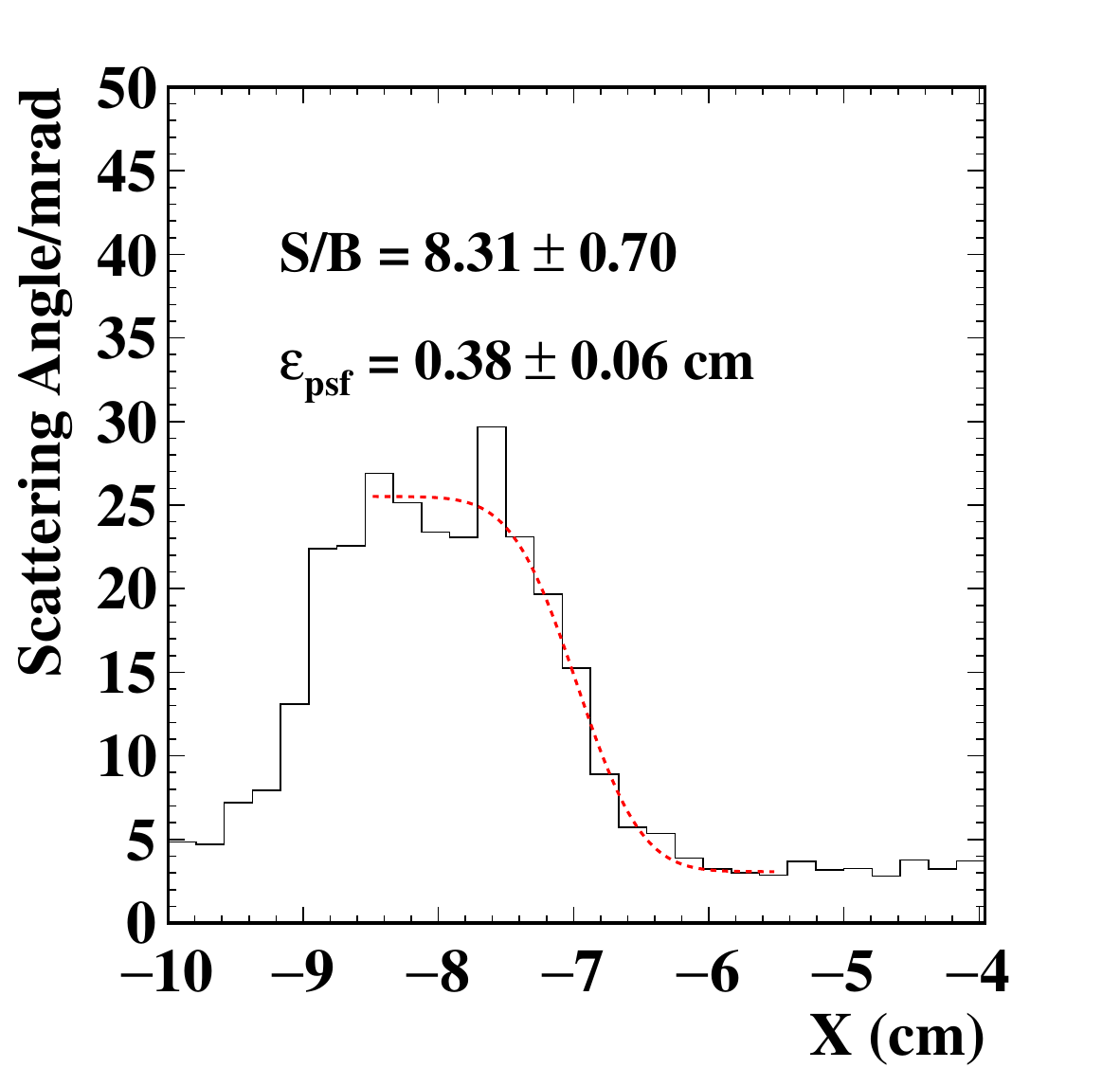}
        \caption{}
        \label{fig:Simu-image-Pro}
    \end{subfigure}
    \caption{
    PoCA imaging results of the MST system: experiment (left) vs. simulation (right),  (a) Experimental 2D image. (b) Simulated 2D image. (c) X-axis profile of the tungsten "U" corner (experiment). (d) X-axis profile of the tungsten "U" corner (simulation). 
    The color axes in (a, b) and the y-axes in (c, d) all represent the mean scattering angle of muons normalized to a 2~cm path length, in units of mrad.
    }
    \label{fig:Imaing-Poca}
\end{figure}

Figure~\ref{fig:Imaing-Poca} presents the PoCA imaging results. 
The left two panels show the experimental results, and the right panels show the simulation results. 
The upper panels display the two-dimensional PoCA images, where the color axis represents the mean scattering angle normalized to a 2~cm path length, in units of mrad. 
The X-axis profiles of the regions indicated by the red boxes are shown in the lower panels, corresponding to the upper-right corner of the letter "U", which is composed of 2~cm wide tungsten blocks.
The fit results using an error function show the imaging signal-to-background ratios of $7.1 \pm 0.7$ (experiment) and $8.3 \pm 0.7$ (simulation), as well as the $\varepsilon_\text{psf} = 0.29 \pm 0.06$ cm (experiment) and $0.38 \pm 0.06$ cm (simulation). 
The experimental and simulated results are in good general agreement.
However, the simulation exhibits a lower background and consequently a higher signal-to-background ratio, while the reconstructed edges are less sharp than those in the experimental results.

\section{Conclusion and outlook}

We present the full development of a cosmic ray muon scattering tomography setup, covering its design, simulation, construction, performance characterization, and imaging validation. 
The core of this setup is a large-area, high-spatial-resolution scintillating detector, which was systematically optimized through Geant4 simulations. 
Key design parameters including a triangular bar geometry with an 11~mm pitch, a fiber encoding scheme, and the low-noise readout electronics, were meticulously engineered to achieve a spatial resolution of 1.0~mm and a detection efficiency of 97.57\% per detection plane.
This results in a large effective detection area of 53~cm~$\times$~53~cm for each super layer. 
As quantitatively compared in table~\ref{tab:scintillator_comparison}, the system achieves a normalized spatial resolution $\sigma_x$/p of 0.09, representing a clear improvement over conventional scintillator-based MST systems (typically $\sim$0.2) and showing competitive performance relative to gas-based detectors such as GEM, Micromegas, and MRPC.

The fully assembled system's imaging capability was successfully demonstrated by clearly reconstructing small test objects ($2\times2\times2\text{ cm}^3$ cubes) made of various materials.
The results confirm that the enhanced spatial resolution directly translates to superior image quality, characterized by a high signal-to-noise ratio and sharp boundaries.
The same imaging conditions were also validated in the simulation, and good agreement was achieved between the experimental and simulated results.

The successful construction and validation of this MST setup open several promising future directions. 
While the PoCA algorithm has proven effective for initial imaging, the high-quality track data from our system is well-suited for more advanced reconstruction algorithms, such as Maximum Likelihood/Expectation Maximization (ML/EM) or machine learning-based methods. 
These algorithms have the potential to further enhance image fidelity and material discrimination capabilities.
Furthermore, the modular design of the plastic scintillator detector offers a clear path towards large-area deployment. The detection area can be cost-effectively expanded to the scale of $2~\text{m} \times 2~\text{m}$ or larger by seamlessly splicing multiple shorter scintillator bars.
This scalability is key to enabling large-area, high-precision muon imaging for practical applications.

\section*{Acknowledgments}
This research work is supported in part  by the  National Natural Science Foundation of China, Nos.~12361141827 and 12375185, and the National Key R\&D Program of China, No.~2023YFA1606903, and the Anhui Provincial Natural Science Foundation, No.~2208085MA12.

We gratefully acknowledge the financial support and the provision of experimental facilities by the Institute of Modern Physics, Chinese Academy of Sciences. 
The scintillators were processed and assembled into detectors at the University of Science and Technology of China (USTC), and the complete MST system was subsequently installed and tested at the Huizhou Campus of the Institute of Modern Physics.

\bibliographystyle{elsarticle-num}
\bibliography{ref}

\end{document}